\documentclass[12pt]{article}

\usepackage{graphicx}
\usepackage{amsmath}
\usepackage{amsfonts}
\usepackage{amssymb}
\usepackage{physics}
\usepackage{microtype}
\usepackage{bbold}
\usepackage{mathtools}
\usepackage{cite}
\usepackage{cancel}
\usepackage{xcolor}
\usepackage{mdframed}
\usepackage{xpatch}
\usepackage{etoolbox}
\usepackage[normalem]{ulem}
\usepackage[most]{tcolorbox}
\usepackage{enumerate}

\definecolor{ForestGreen}{RGB}{34,139,34}

\usepackage{array}
\newcolumntype{P}[1]{>{\centering\arraybackslash}p{#1}}

\allowdisplaybreaks
\numberwithin{equation}{section}

\def\be{\begin{equation}}
\def\ee{\end{equation}}

\def\hd{\frac{d}{2}}
\def\half{\frac{1}{2}}
\def\e{\epsilon}
\def\Gam{\gamma}
\def\ba{\begin{equation}\begin{aligned}}
\def\ea{\end{aligned}\end{equation}}
\newcommand{\reef}[1]{(\ref{#1})}
\definecolor{blueKSV}{RGB}{31, 119, 180}
\definecolor{blueKSV1}{RGB}{141, 206, 247}
\definecolor{blue2}{RGB}{31, 119, 180}
\definecolor{red2}{RGB}{210, 125, 108}
\definecolor{Red}{RGB}{	214, 39, 40}

\makeatletter
\xpatchcmd{\endmdframed}
  {\aftergroup\endmdf@trivlist\color@endgroup}
  {\endmdf@trivlist\color@endgroup\@doendpe}
  {}{}
\makeatother

\definecolor{darkblue}{rgb}{0.15,0.35,0.55}
\definecolor{reddish}{rgb}{0.65, 0.2, 0.2}
\definecolor{darkgreen}{RGB}{50,150,0}
\definecolor{greyish}{rgb}{.90,.90,.90}
\definecolor{greyish2}{rgb}{.96,.96,.96}
\definecolor{greyish3}{rgb}{.37,.37,.37}
\definecolor{darkblue2}{rgb}{0.3,0.4,0.9}
\definecolor{Blue3}{RGB}{31, 119, 180}
\usepackage[linktocpage=true]{hyperref}
\hypersetup{
colorlinks=true,
citecolor=darkblue,
linkcolor=reddish,
urlcolor=darkblue,
pdfauthor={},
pdftitle={},
pdfsubject={}
}

\usepackage{caption}     
\newcommand{\betaB}{B}

\newcommand{\lf}{\text{L}}
\newcommand{\ri}{\text{R}}
\newcommand{\Q}{\Sigma}
\newcommand{\eh}{\text{EH}}
\newcommand{\M}{M}

\usetikzlibrary{decorations.pathmorphing}
\tikzset{snake it/.style={decorate, decoration=snake}}

\begin{document}
\thispagestyle{empty}

\phantom{*}

\begin{center} 

\begin{center}
{\fontsize{20}{16} \bf Microstate counting from defects in de Sitter}
\end{center}

\vskip 25pt
		\begin{center}
			\noindent
			{\fontsize{14}{18}\selectfont 
				Jan de Boer$\,{}^{a}$, Diego Liska$\,{}^{(b+c)/2}$ and Kamran Salehi Vaziri$\,{}^{a}$}
		\end{center}
\vskip 10pt
\normalsize
\small{$\;{}^{a}$\textit{Institute for Theoretical Physics,
University of Amsterdam,}}
\vskip -.22cm
\small{\textit{1098 XH Amsterdam, The Netherlands}}
\vskip .2cm
\small{$\;{}^{b}$\textit{Department of Theoretical Physics, University of Geneva,}}
\vskip -.22cm
\small{\textit{1211 Geneva, Switzerland}}
\vskip .2cm
\small{$\;{}^{c}$\textit{
Laboratory for Theoretical Fundamental Physics, Institute of Physics,}}
\vskip -.22cm
\small{\textit{École Polytechnique Fédérale de Lausanne,}}
\vskip -.22cm
\small{\textit{1015 Lausanne, Switzerland}}
\end{center}

\vspace{5mm}

\begin{abstract}

We explore the microscopic origin of de Sitter entropy using a Lorentzian path-integral approach. We construct a Hilbert space whose states are associated with configurations of thin shells or end-of-the-world branes, with state overlaps defined by the gravitational path integral. By considering states which are indistinguishable to an observer, we find that the variance of microstate overlaps is dominated by Lorentzian wormhole topologies with conical singularities. Evaluating these overlaps, we recover the expected area law for the entropy, relating the dimension of the de Sitter Hilbert space to the area of the cosmological horizon. Extending this analysis to Schwarzschild-de Sitter spacetime, we show that both the cosmological and black hole horizons contribute to the total entropy. Along the way, we present an explicit construction of the shell and brane configurations and examine their compatibility with relevant consistency conditions, including the null energy condition.

\end{abstract}

\small{\vfill \noindent j.deboer@uva.nl\\
diego.liska@epfl.ch (or @unige.ch)\\
k.salehivaziri@uva.nl
}

\newpage

\phantom{a}
\vspace{-4em}
{
\hypersetup{linkcolor=black}
\tableofcontents
}

\newpage

\section{Introduction}
\label{sec:intro}

De Sitter (dS) spacetime possesses a cosmological horizon that, analogous to a black hole horizon, has a temperature and appears to possess an entropy~\cite{Gibbons:1977mu,Gibbons:1976ue}. Despite significant progress over the past few decades in our understanding of quantum gravity, the microscopic origin of this entropy and its associated area law remains an open question. Nevertheless, recent advances in the formulation of the gravitational path integral have offered new perspectives into this longstanding problem. An important insight that has emerged is the reinterpretation of the gravitational path integral as a statistical framework rather than an exact one. This perspective has led to consequential improvements in understanding of the black hole entropy in Anti–de Sitter and Minkowski spacetimes, see, e.g.,  \cite{Balasubramanian:2022gmo,Penington:2019kki,Almheiri:2019qdq,Balasubramanian:2022lnw,Balasubramanian:2025hns}. In this paper, we attempt to extend these techniques to the context of de Sitter space.

Formally, a path integral over a manifold with a single open cut can be understood as defining a state in the Hilbert space of the theory. A well-known example is the Hartle–Hawking no-boundary proposal for the wave function of the universe~\cite{Hartle:1983ai}. The states we consider in this paper are defined using two cuts: one cut is left open, while boundary conditions are imposed on the other. In this framework, the Lorentzian path integral with boundary conditions imposed on the two cuts is naturally interpreted as an inner product in the Hilbert space of the theory.
Schematically, 
\begin{equation}
  \ket{\psi_i}  =  \vcenter{\hbox{
\begin{tikzpicture}
\begin{scope}[scale=1.55]
    \draw[thick, lightgray, line width=1.5pt , fill=gray!10] (-1,0) -- (-1,-2) -- (1,-2) -- (1,0);
    \draw[thick, black,line width=1.5pt] (-1,-2) -- (1,-2);
    \draw[thick, line width=1pt, dashed] (-1,0) -- (1,0);
    \node at (0,-1) {$M$};
     \node at (0,-2.25) {$\partial M = \psi_i$};
    \node at (0,0.25) {$\phantom{(\partial M)_2 = \psi_j}$};
\end{scope}
\end{tikzpicture}
    }}, 
    \qquad 
\bra{\psi_j}\ket{\psi_i}  = \vcenter{\hbox{
\begin{tikzpicture}
\begin{scope}[scale=1.55]
    \draw[thick, lightgray, line width=1.5pt , fill=gray!10] (-1,1) -- (-1,-1) -- (1,-1) -- (1,1);
    \draw[thick, black, line width=1.5pt] (-1,-1) -- (1,-1);
    \draw[thick, black, line width=1.5pt] (-1,1) -- (1,1);
    \node at (0,0) {$M$};
    \node at (0,1.25) {$(\partial M)_2 = \psi_j$};
    \node at (0,-1.25) {$(\partial M)_1 = \psi_i$};
\end{scope}
\end{tikzpicture}}}~.
\end{equation}
where $\psi_i$ denotes the set of asymptotic boundary conditions which render the path integral well defined. This picture is useful for de Sitter space as there are two natural locations for imposing boundary conditions: the asymptotic boundaries at the infinite past and future, denoted by $\mathcal{I}^{^-}$ and $\mathcal{I}^{^+}$, respectively. Boundary conditions specified at $\mathcal{I}^{^-}$ are associated with ‘in’ states, whereas those imposed at $\mathcal{I}^{^+}$ correspond to ‘out’ states. 

One promising way to explicitly construct a family of states $\ket{\psi_i}$, is to choose a modified version of the path integral that includes additional features corresponding to two types of defects: (i) \textit{end-of-the-world branes} and (ii) \textit{Thin shells of matter}. Each type of defect together with its mass parameter and equation of state corresponds to a particular boundary condition $\psi_i$. Each defect will be described using the stress-energy tensor of a perfect fluid. In this effective description, all microscopic degrees of freedom that would otherwise appear in the UV theory have been integrated out. The main advantage of this approach is that these solutions can be described in terms of the geometrical features of the underlying manifold. 

\pagebreak

The Lorentzian path integral over such geometries can be interpreted as defining a (coarse-grained) inner product between two microstates in de Sitter:
\begin{equation}
\label{innerp}
\overline{\bra{\psi_j}\ket{\psi_i}} 
\;\;\approx\;\; \delta_{ij}
\vcenter{\hbox{
\begin{tikzpicture}
\begin{scope}[scale=1.2]
    \draw[black, line width=0.8pt] (-1,1.5) -- (-1.6,1.5) -- (-1.6,-1.5) -- (-1,-1.5);
    \draw[thick, black, line width=0.8pt] (-1,1.5) -- (1,1.5) -- (1,-1.5) -- (-1,-1.5);
    \draw[black!50,dashed,line width=1pt] (1,-1.5) -- (-0.5,0) -- (1,1.5);
    \fill[red,opacity=0.2](1,-1.5) -- (1,1.5) -- (-0.5,0);
    \draw[black, line width=1.5pt]  (-1,-1.5)  to[out = 75, in = -75] (-1,1.5);
    \fill[black] (-1,-1.5) circle (0.075);
    \fill[black] (-1,1.5) circle (0.075);
    \node at (-1.2,-1.75) {$i$};
    \node at (-1.2,1.75) {$j$};
    \draw[thick, red] (1,-1.5) -- (1,1.5);
    \node at (1,1.5) {{\color{red}$\bullet$}};
    \node at (1,-1.5) {{\color{red}$\bullet$}};
    \node at (0.25,-1.75) {$\mathcal{I}^-$};
    \node at (0.25,1.75) {$\mathcal{I}^+$};
\end{scope}
\end{tikzpicture}
}
}\,,
\end{equation}
where the solid black line is the worldvolume of defect and the red line is the worldline of an observer sitting at the center of the static patch, indicated by the red shaded region. The label $i$ denotes a set of continuous or discrete parameters that characterize the properties of the shell. These labels may correspond to quantities appearing in the effective description, such as the mass or the equation-of-state parameter $w$ describing the defect. Alternatively, they could represent internal degrees of freedom that have been integrated out in the effective theory; e.g., labels describing the positions of the individual particles forming the shell, or the flavor of the brane.
The approximate equality sign indicates that this represents only the simplest of the many possible topologies contributing to the amplitude. States with different indices $i, j$ correspond to distinct semiclassical configurations.
We denote \eqref{innerp} as an average (with an overline), rather than an exact equality (without the overline), because the gravitational path integral provides only an approximate evaluation of the inner product. 
This interpretation of the path integral has emerged from the extensive study of solvable models of gravity in two and three dimensions \cite{Saad:2019lba,Belin:2020hea,DiUbaldo:2023qli,Boruch:2025ilr,Altland:2020ccq,Altland:2022xqx,deBoer:2023vsm,Collier:2025lux,Chandra:2022bqq,Collier:2023fwi,deBoer:2025rct,Freivogel:2021ivu,Sasieta:2022ksu,Narovlansky:2023lfz,Hsin:2020mfa,Chandra:2022fwi}, but it is believed to hold in higher dimensions as well. 

A salient feature of the microstates constructed via the gravitational path integral is that they are only approximately orthogonal. The exact inner product between these two states is subject to non-perturbative corrections:
\begin{equation}\label{eq: Rij}
   \bra{\psi_i}\ket{\psi_j} 
   \sim \delta_{ij} + R_{ij}~.
\end{equation}
Here, $R_{ij}$ are a set of pseudorandom variables. In principle, these overlaps can be computed exactly within the UV theory; however, their precise values are not accessible at the semiclassical level. Remarkably, the gravitational path integral nevertheless captures the statistical properties of these variables. In particular, we can compute their variance,  which scales like
\begin{equation}
    \overline{R_{ij}R_{ji}}\sim e^{-S},
\end{equation}
where $S$ is the microcanonical entropy of the Hilbert space of the quantum theory. This variance is consistent with the hypothesis that the states are randomly drawn from a finite dimensional Hilbert space of dimension $e^S$, and gives a method for inferring the dimension of the Hilbert space of the quantum theory. 

\pagebreak

\paragraph{Lorentzian wormholes.} The standard prescription for evaluating $\overline{R_{ij}R_{ji}}$ using the gravitational path integral is to fix asymptotic boundary conditions consistent with the states $\ket{\psi_{i}},\ket{\psi_j}$ and then sum over all topologies and geometries compatible with these conditions.
However, this approach quickly encounters a technical difficulty: topology change is not allowed in Lorentzian signature \cite{Geroch:1967fs,Geroch:1979uc,Hawking:1991nk,Sorkin:1997gi}.
This does not mean that there are no relevant geometries that approximate the gravitational path integral on such topologies. Rather, the dominant contributions to the path integral arise from \textit{complex-valued} spacetime metrics.
The motivation for introducing complex metrics comes from viewing the gravitational path integral as an infinite-dimensional complex integral with a properly chosen integration contour \cite{Gibbons:1978ac}. This contour must pass through an appropriate family of generally complex metrics. 

Louko and Sorkin \cite{Louko:1995jw} showed that the complex-valued metrics responsible for topology change in Lorentzian signature can be effectively described within a purely Lorentzian framework by permitting spacetimes with conical angle singularities. 
Recent work~\cite{Blommaert:2023vbz,Marolf:2022ybi,Held:2024qcl,Colin-Ellerin:2020mva,Colin-Ellerin:2021jev} has argued that these complex geometries provide the correct description of topology change in AdS spacetimes.
For de Sitter space, by allowing such singularities, we identify the following geometry as the leading contribution to the connected moment of de Sitter microstates:
\begin{equation}
\label{wormhole}
\overline{R_{ij}R_{ji}} \,\approx\, 
\vcenter{
\hbox{
\begin{tikzpicture}
\begin{scope}[scale=0.6]
\fill[blueKSV1!70] (3,-3) -- (0,0) -- (3,3) -- (3,-3);
\draw[black,line width=0.8pt] (-2.3,3) -- (3,3) -- (3,-3) -- (-2.3,-3) -- (-2.3,3);
\node at (1.25,-3.4) {$\mathcal{I}^-$};
\node at (1.25,3.5) {$\mathcal{I}^+$};
\node at (-0.5,-3) {{$\bullet$}};
\draw[ line width=1.5pt] (-0.5,-3) to[out=120, in = 270,looseness=1] (-1.4,0);
\node at (-1.5,3) {{$\bullet$}};
\draw[line width=1.5pt] (-0.8,0) to[out=90, in = 300,looseness=1] (-1.5,3);
\draw[black!50,line width=1pt, dashed] (0,0) -- (3,-3);
\draw[black!50,line width=1pt, dashed] (3,3) -- (0,0);
\draw[thick, blueKSV1!200] (3,-3) -- (3,3);
\node at (3,3) {{\color{blueKSV1!200}$\bullet$}};
\node at (3,-3) {{\color{blueKSV1!200}$\bullet$}};
\node at (-1,-3.5) {{$j$}};
\node at (-1.8,3.5) {{$i$}};
\draw[thick,decorate, decoration={zigzag, segment length=4pt, amplitude=1pt}](-2.3,0)--(-0.2,0);
\node at (-0.2,0) {{$\bullet$}};
\end{scope}
\end{tikzpicture}
}
}\,
\vcenter{
\hbox{
\begin{tikzpicture}
\begin{scope}[ scale=0.6]
\fill[red!20] (3,-3) -- (0,0) -- (3,3) -- (3,-3);
\draw[ thick] (-2.3,3) -- (3,3) -- (3,-3) -- (-2.3,-3) -- (-2.3,3);
\node at (1.25,-3.5) {$\mathcal{I}^-$};
\node at (1.25,3.5) {$\mathcal{I}^+$};
\draw[black!50,line width=1pt , dashed] (0,0) -- (3,-3);
\draw[ black!50,line width=1pt , dashed] (3,3) -- (0,0);
\draw[thick, red] (3,-3) -- (3,3);
\node at (-1.5,-3) {{$\bullet$}};
\draw[black ,line width=1.5pt] (-1.5,-3) to[out=60, in = 270,looseness=1] (-0.6,0);
\node at (-0.5,3) {{$\bullet$}};
\draw[ black ,line width=1.5pt] (-1.4,0) to[out=90, in = 240,looseness=1] (-0.5,3);
\node at (3,3) {{\color{red}$\bullet$}};
\node at (3,-3) {{\color{red}$\bullet$}};
\node at (-1.8,-3.5) {{$i$}};
\node at (-0.8,3.5) {{$j$}};
\draw[thick,decorate, decoration={zigzag, segment length=4pt, amplitude=1pt}] (-2.3,0)--(-0.2,0);
\node at (-0.2,0) {{$\bullet$}};
\end{scope}
\end{tikzpicture}
}
}
\end{equation}
This topology corresponds to two de Sitter spacetimes joined along a cut (zigzag line) with the black lines representing the defects. The cut introduces a conical singularity with a (Lorentzian) surplus angle of $4\pi$ at its endpoint, which is situated at the cosmological horizon of the static patches in blue and red.  This geometry has two important features. First, there are continuous timelike curves (shown in blue and red) connecting kets to their corresponding bras. Second,  all non-trivial features lie outside the causal diamond of these curves. 
These properties make such topologies compatible with the idea of an ``observer" existing within the spacetime, a concept that has received renewed interest in the study of closed universes, see e.g. \cite{Abdalla:2025gzn,Blommaert:2025bgd,Harlow:2025pvj,Witten:2023xze,Chandrasekaran:2022cip,Arias:2019pzy}.

The presence of the conical singularity in \eqref{wormhole} modifies the Lorentzian path integral in a non-trivial way, leading to the estimate 
\begin{equation}
   \overline{R_{ij}R_{ji}} \sim \exp(-\frac{A_{\text{c}}}{4G}) ~,
\end{equation}
where $A_{\text{c}}$ is the area of the cosmological horizon. The approximate equality arises because this expression is obtained through a saddle-point approximation without taking into account  higher-order loop corrections. This is one of the main results of the paper, as the formula reproduces the Gibbons-Hawking entropy \cite{Gibbons:1976ue}, relating area of the cosmological horizon to the dimensionality of the Hilbert space associated with de Sitter spacetime:\footnote{More precisely, this entropy counts the number of states that are indistinguishable to an observer inside the de Sitter static patch.}
\begin{equation}
   \log(\dim \mathcal{H}_{\text{dS}}) = \frac{A_{\text{c}}}{4G} + \order{\log(G)}~.
\end{equation}

\paragraph{Schwarzschild-de Sitter.}

One advantage of working with the Lorentzian path integral is that it enables the analysis of Schwarzschild–de Sitter (SdS) spacetimes. 
By considering configurations of cuts with two defects placed within the SdS geometry, we obtain the following estimate for the variance
\begin{equation}
\overline{R_{ij}R_{ji}} 
\sim \; \exp(-\frac{A_{\text{c}} +A_{\text{b}}}{4G})~, 
\end{equation}
where $A_{\text{c}}$ and $A_{\text{b}}$ correspond to the area of the cosmological and black hole horizons of the SdS observer. As a result, the microcanonical entropy associated with the Schwarzschild-de Sitter spacetime is given by 
\begin{equation}
    \log( \dim \mathcal{H}_{\text{SdS}} ) = \frac{A_{\text{c}}+A_{\text{b}}}{4G} + \order{\log(G)}~.
\end{equation}

\paragraph{Microscopic conditions.} 
This paper adopts a semiclassical approach to microstate counting, which inherently limits our access to the ultraviolet degrees of freedom of the theory. Nevertheless, we search for solutions that satisfy certain microscopic conditions, focusing on two in particular: the Null Energy Condition (NEC) and \emph{the matching background condition.} The latter requires that all states be defined within a single, fixed value of the cosmological constant. This restriction is consistent with a UV-complete theory of de Sitter in which there is a single de Sitter vacuum. The NEC is imposed to ensure causality.

In this work, we explore a broad class of de Sitter and Schwarzschild–de Sitter geometries to find solutions meeting these two conditions. However, we do \emph{not} find configurations meeting both requirements simultaneously. Instead, we find the following configurations which have the potential to admit a UV completion. 

The first corresponds to a family of de Sitter solutions in which a positive-tension brane separates two de Sitter spacetimes with different cosmological constants. As discussed above, the cost of finding such configurations is that the UV theory must include multiple de Sitter vacua. Nonetheless, this requirement may not be too severe, since similar constructions can be realized in AdS/CFT setups, and the landscape may be obtained through mechanisms like those introduced by Bousso and Polchinski \cite{Bousso:2000xa}. Requiring many values of the cosmological constant is a reasonable condition when considering only empty de Sitter solutions (with no additional matter fields), as the entropy naturally arises from bubble nucleation processes. These bubbles, whose interiors have a different value of the cosmological constant, are examples of the well-known Coleman–De Luccia instantons \cite{Coleman:1980aw}.

If one insists on having a single cosmological constant but allows for matter in the theory, the entropy can be accounted for by considering small black holes in de Sitter space. To reproduce the de Sitter entropy, one can then take the limit where the black hole disappears.
One of the configurations that we identify in this paper corresponds to an SdS geometry with two shell insertions: one shell escapes to infinity, while the other falls into the black hole. The infalling shell does not satisfy the NEC,  but we find evidence suggesting that the overall spacetime satisfies the Achronal Averaged Null Energy Condition (AANEC), which is a sufficient condition to prove many important causality theorems\cite{Graham:2007va}. In the AdS literature, there are examples of infalling shells which lower the energy of black hole microstates \cite{Papadodimas:2017qit}. The CFT operators needed to create these shells are fine-tuned, but they exist in the dual CFT and are generally present in other quantum mechanical systems. This suggests that similar shells in de Sitter may also have a UV description. 
We leave a more detailed investigation of these solutions for future work.

\subsection*{Outline}

This paper is organized as follows.

In Section~\ref{sec:setup}, we review the various qualitative features of our geometries. This includes a discussion of the role of the observer, the placement of defects and conical singularities, and the microscopic consistency conditions that our geometries should ideally satisfy—namely, the energy and background matching conditions.

In Section~\ref{sec:shells}, we derive the equations of motion for the defects by varying the Gibbons–Hawking\\–York boundary term in a generic spherically symmetric spacetime. We then present solutions consistent with the criteria outlined in Section~\ref{sec:setup}, first for the de Sitter case (Subsection~\ref{sec:puredS}) and subsequently for the Schwarzschild–de Sitter case (Subsection~\ref{sec: SdS}). Finally, in Subsection~\ref{sec: explicit sol}, we provide explicit examples of a typical defect solution obtained by numerically solving the equations of motion.

In Section~\ref{sec:wormholes}, we discuss how to compute the moments of the state overlaps, and hence the entropy, using the Lorentzian gravitational path integral with complex geometries. In particular, Subsection~\ref{crotches} is devoted to the “crotch” geometry in both flat and curved spacetimes. We then use Lorentzian wormholes geometries in Subsection~\ref{sec:Lwormholes} to show that the leading correction to the connected microstate overlap is governed by the area of the extremal surface. Finally, in Subsection~\ref{microcounting}, we complete the microstate count and derive the entropy for de Sitter and Schwarzschild–de Sitter spacetimes.

We finish the paper with a discussion and open questions in Section~\ref{sec:discussion}. Appendix~\ref{eq:conventions} summarizes notations and conventions. Appendix~\ref{sec:GHY} provides a detailed derivation of the Gibbons–Hawking–York terms in the gravity action. The NEC for codimension-one defects is reviewed in Appendix~\ref{sec: NEC}. Finally,  Appendix~\ref{app:crotches} contains an explicit computation of the Einstein–Hilbert term in the crotch geometry. 

\pagebreak

\section{Set up}\label{sec:setup}

Before turning to the technical analysis of branes and shells in Sections~\ref{sec:shells} and \ref{sec:wormholes}, in this section we outline the relevant qualitative features of these geometries. In particular, we specify the role and placement of the observer, the defects, the conical singularities together with their associated cut and, finally, the microscopic consistency conditions that the spacetime is expected to satisfy after the defect is introduced.

\paragraph{The role of the observer.} 
An important aspect in the construction of gravitational microstates is the choice of an observer.
Both the entropy we calculate and the topologies we consider in the path integral depend crucially on the existence and placement of this observer inside the spacetime.
In the case of de Sitter space, a natural choice for the observer is the south pole (or, equivalently, the north pole) of the global geometry. Similarly, there is a natural location for the defects that define the microstates: outside the causal diamond of the observer’s static patch.
This is the setting that we focus on in this paper where the static path of the observer is left untouched. 

The reason for considering a region of spacetime where there is no back and forth communication between the perturbation and the observer comes from analogous computations performed in AdS black holes. There, it is natural to place the observer at the asymptotic boundary, and the microstates are defined by perturbations hidden behind black hole horizons \cite{Climent:2024trz,Balasubramanian:2022gmo,Balasubramanian:2024rek}. In both cases, the intuition is that the relevant microstates should be indistinguishable for the single semiclassical observer. Therefore, in de Sitter, we are going to consider overlaps of the form 
\begin{equation}
\label{innerp2}
\overline{\bra{\psi_j}\ket{\psi_i}} 
\;\;\approx\;\; \delta_{ij}
     \underbrace{
    \vcenter{\hbox{
    \begin{tikzpicture}
    \begin{scope}[scale=1.2]
    \draw[thick, black] (-1,1.5) -- (1,1.5) -- (1,-1.5) -- (-1,-1.5);
    \draw[black!50,dashed,line width=1pt] (1,-1.5) -- (-0.5,0) -- (1,1.5);
    \fill[red,opacity=0.2](1,-1.5) -- (1,1.5) -- (-0.5,0);
    \draw[thick, black, line width=1.5 pt] (-1,-1.5)  to[out = 75, in = -75] (-1,1.5);
    \fill[black] (-1,-1.5) circle (0.075);
    \fill[black] (-1,1.5) circle (0.075);
    \node at (-1.2,-1.2) {$i$};
    \node at (-1.2,1.2) {$j$};
    \draw[thick, red] (1,-1.5) -- (1,1.5);
    \node at (1,1.5) {{\color{red}$\bullet$}};
    \node at (1,-1.5) {{\color{red}$\bullet$}};
    \end{scope}
    \end{tikzpicture}
    }}}_{\text{End-of-the-world brane}}
    \quad \text{or} \qquad 
\underbrace{
\vcenter{\hbox{
\begin{tikzpicture}
\begin{scope}[scale=1.2]
    \draw[thick, black,fill=white] (-1,1.5) -- (-1.6,1.5) -- (-1.6,-1.5) -- (-1,-1.5);
    \draw[thick, black,fill=white] (-1,1.5) -- (1,1.5) -- (1,-1.5) -- (-1,-1.5);
    \draw[black!50,dashed,line width=1pt] (1,-1.5) -- (-0.5,0) -- (1,1.5);
    \fill[red,opacity=0.2](1,-1.5) -- (1,1.5) -- (-0.5,0);
    \draw[line width=1.5, black] (-1,-1.5)  to[out = 75, in = -75] (-1,1.5);
    \fill[black] (-1,-1.5) circle (0.075);
    \fill[black] (-1,1.5) circle (0.075);
    \node at (-1.2,-1.2) {$i$};
    \node at (-1.2,1.2) {$j$};
    \draw[thick, red] (1,-1.5) -- (1,1.5);
    \node at (1,1.5) {{\color{red}$\bullet$}};
    \node at (1,-1.5) {{\color{red}$\bullet$}};
\end{scope}
\end{tikzpicture}
}
}}_{{\text{Thin shell of matter}}}
\end{equation}
where the solid red line is the observer worldline, the shaded red region is its causal diamond, and the solid black line is the defect which is outside the observer's causal diamond.

The role of the observer is not limited to determining where the non-trivial features of the geometry are located. It also constrains the set of topologies that should be considered in the gravitational path integral. As argued in \cite{Abdalla:2025gzn,Blommaert:2025bgd,Harlow:2025pvj}, we should only include topologies in which the observer’s worldline connects their corresponding bra and ket. Topologies that fail to meet this condition, such as the one illustrated below, are \emph{excluded} from the sum and do not contribute to the moments of the overlap between microstates:
\begin{equation}
\overline{R_{ij}R_{ji}}\;\not\supset\;
\vcenter{
\hbox{
\begin{tikzpicture}
\begin{scope}[ scale=0.6]
        \fill[gray!20] (3,-3) -- (0,0) -- (3,3) -- (3,-3);
        \draw[thick] (-2.2,3) -- (3,3) -- (3,-3) -- (-2.2,-3) -- (-2.2,3);
        \node at (1.5,-3.4) {$\mathcal{I}^-$};
        \node at (1.5,3.5) {$\mathcal{I}^+$};
        \draw[ line width=1.5] (-0.5,-3) to[out=120, in = 240,looseness=1] (-0.5,3);
        \draw[black!50,dashed,line width=1pt]  (0,0) -- (3,-3);
        \draw[black!50,dashed,line width=1pt]  (3,3) -- (0,0);
        \node at (3,3) {{\color{red}$\bullet$}};
        \node at (3,-3) {{\color{blueKSV1!200}$\bullet$}};
        \node at (-0.5,-3) {{$\bullet$}};
        \node at (-0.5,3) {{$\bullet$}};
        \node at (-0.8,-3.6) {{$j$}};
        \node at (-0.8,3.6) {{$j$}};
\end{scope}
\end{tikzpicture}
}
}\quad 
\vcenter{
\hbox{
\begin{tikzpicture}
\begin{scope}[ scale=0.6]
        \fill[gray!20] (3,-3) -- (0,0) -- (3,3) -- (3,-3);
        \draw[ thick] (-2.2,3) -- (3,3) -- (3,-3) -- (-2.2,-3) -- (-2.2,3);
        \node at (1.5,-3.4) {$\mathcal{I}^-$};
        \node at (1.5,3.5) {$\mathcal{I}^+$};
        \draw[ line width=1.5] (-0.5,-3) to[out=120, in = 240,looseness=1] (-0.5,3);
        \draw[black!50,dashed,line width=1pt]  (0,0) -- (3,-3);
        \draw[black!50,dashed,line width=1pt] (3,3) -- (0,0);
        \node at (3,3) {{\color{blueKSV1!200}$\bullet$}};
        \node at (3,-3) {{\color{red}$\bullet$}};
        \node at (-0.5,-3) {{$\bullet$}};
        \node at (-0.5,3) {{$\bullet$}};
        \node at (-0.8,-3.5) {{$i$}};
        \node at (-0.8,3.5) {{$i$}};
\end{scope}
\end{tikzpicture}
}
}
\end{equation}
The boundary conditions in this setup are the same as in \eqref{wormhole}, but the spacetime lacks a worldline connecting the observer's in-state with its out-state. As explained in \cite{Blommaert:2025bgd,Harlow:2025pvj}, the inclusion of these topologies in the path integral collapses the Hilbert space to a single state, $\dim \mathcal{H}_{\text{dS}} = 1$. Hence, for closed universes like de Sitter, introducing an observer is required to find a nontrivial entropy.

\paragraph{Placement of the conical singularities.}
As we briefly mentioned in Section~\ref{sec:intro} and discuss in more detail in Section~\ref{sec:wormholes}, we consider spacetimes with conical singularities as the leading contributions to the connected moment of the microstate overlap. Following the logic that was discussed above, these singularities and the cut attached to them must be placed outside the casual diamond of the observer. The cuts, that are represented by the zigzag lines, are used to construct the wormhole geometry connecting the two copies of spacetime:
\begin{equation}
\vcenter{
\hbox{
\begin{tikzpicture}
\begin{scope}[ scale=0.6]
\fill[blueKSV1!70] (3,-3) -- (0,0) -- (3,3) -- (3,-3);
\draw[ thick] (-2.3,3) -- (3,3) -- (3,-3) -- (-2.3,-3) -- (-2.3,3);
\node at (1.5,-3.4) {$\mathcal{I}^-$};
\node at (1.5,3.5) {$\mathcal{I}^+$};
\node at (-0.5,-3) {{$\bullet$}};
\draw[ line width=1.5pt] (-0.5,-3) to[out=120, in = 270,looseness=1] (-1.4,0);
\node at (-1.5,3) {{$\bullet$}};
\draw[ line width=1.5pt] (-0.8,0) to[out=90, in = 300,looseness=1] (-1.5,3);
\draw[black!50,dashed,line width=1pt] (0,0) -- (3,-3);
\draw[black!50,dashed,line width=1pt] (3,3) -- (0,0);
\draw[blueKSV1!200,thick] (3,-3) -- (3,3);
\node at (3,3) {{\color{blueKSV!!200}$\bullet$}};
\node at (3,-3) {{\color{blueKSV!!200}$\bullet$}};
\node at (-1,-3.5) {{$j$}};
\node at (-1.8,3.5) {{$i$}};
\draw[thick,decorate, decoration={zigzag, segment length=4pt, amplitude=1pt}](-2.3,0)--(-0.2,0);
\node at (-0.2,0) {{$\bullet$}};
\end{scope}
\end{tikzpicture}
}
}
\qquad
\vcenter{
\hbox{
\begin{tikzpicture}
\begin{scope}[ scale=0.6]
\fill[red!20] (3,-3) -- (0,0) -- (3,3) -- (3,-3);
\draw[ thick] (-2.3,3) -- (3,3) -- (3,-3) -- (-2.3,-3) -- (-2.3,3);
\node at (1.5,-3.4) {$\mathcal{I}^-$};
\node at (1.5,3.5) {$\mathcal{I}^+$};
\draw[black!50,dashed,line width=1pt] (0,0) -- (3,-3);
\draw[black!50,dashed,line width=1pt] (3,3) -- (0,0);
\draw[thick, red] (3,-3) -- (3,3);
\node at (-1.5,-3) {{$\bullet$}};
\draw[ line width=1.5pt] (-1.5,-3) to[out=60, in = 270,looseness=1] (-0.6,0);
\node at (-0.5,3) {{$\bullet$}};
\draw[ line width=1.5pt] (-1.4,0) to[out=90, in = 240,looseness=1] (-0.5,3);
\node at (3,3) {{\color{red}$\bullet$}};
\node at (3,-3) {{\color{red}$\bullet$}};
\node at (-1.8,-3.5) {{$i$}};
\node at (-0.8,3.5) {{$j$}};
\draw[thick,decorate, decoration={zigzag, segment length=4pt, amplitude=1pt}] (-2.3,0)--(-0.2,0);
\node at (-0.2,0) {{$\bullet$}};
\end{scope}
\end{tikzpicture}
}
}
\end{equation}
In this setting, the shells originate at $\mathcal{I}^-$ in one copy, pass through the cut and exit on the other copy, ending at $\mathcal{I}^+$. There exists a similar picture for the case of an end-of-the-world brane where spacetime terminates at the brane.

\paragraph{Schwarzschild-de Sitter.}
Compared to de Sitter, the situation differs slightly in SdS spacetimes because the Penrose diagram of the maximally extended SdS geometry corresponds to an infinite collection of causally disconnected static patches, separated by alternating black-hole and cosmological horizons,
\begin{equation}
\label{identifiedPRS}
\vcenter{\hbox{
    \begin{tikzpicture}
    \begin{scope}[scale=1.4]
        \draw[thick,decorate,decoration={zigzag,amplitude=0.3mm,segment length=2mm}] (-1,1) -- (1,1);
        \draw[thick, decorate,decoration={zigzag,amplitude=0.3mm,segment length=2mm}] (-1,-1) -- (1,-1);
        \draw[black!50,dashed,line width=1pt] (0,0) -- (1,1) -- (2,0) -- (1,-1) -- (0,0);
        \draw[black!50,dashed,line width=1pt] (0,0) -- (-1,1) -- (-2,0) -- (-1,-1) -- (0,0);
        \draw[thick] (1,1) -- (2.5,1);
        \draw[thick] (2.5,-1) -- (1,-1);
        \draw[thick] (-1,1) -- (-2.5,1);
        \draw[thick] (-2.5,-1) -- (-1,-1);
        \node at (2.4,0) {$\cdots$};
        \node at (-2.4,0) {$\cdots$};
        \node at (0,1.25) {$r = 0$};
        \node at (0,-1.25) {$r = 0$};
        \node at (2,1.25) {$\mathcal{I}^+$};
        \node at (2,-1.2) {$\mathcal{I}^-$};
        \node at (-2,1.25) {$\mathcal{I}^+$};
        \node at (-2,-1.2) {$\mathcal{I}^-$};
        \node[rotate = 45,black!50] at (0.65,0.35) {$r = r_\text{b}$};
         \node[rotate = 45,black!50] at (1.42,-0.37) {$r = r_\text{c}$};
        \end{scope}
    \end{tikzpicture}
    }}.
\end{equation}
Here, $r_\text{b}$ and $r_{\text{c}}$ denote the radii of the black hole and cosmological horizons, respectively. Due to this structure, when considering SdS microstates, we are required to place two end-of-the-world branes in the geometry
\begin{equation}
\overline{\bra{\psi_i}\ket{\psi_j}}  \approx \delta_{ij} 
\vcenter{\hbox{
    \begin{tikzpicture}
    \begin{scope}[scale=1.5, yscale = -1]
        \draw[thick,decorate,decoration={zigzag,amplitude=0.3mm,segment length=2mm}] (-0.45,1) -- (1,1);
        \draw[thick, decorate,decoration={zigzag,amplitude=0.3mm,segment length=2mm}] (-0.45,-1) -- (1,-1);
        \draw[line width=1,dashed,black!50, fill=red!20] (0,0) -- (1,1) -- (2,0) -- (1,-1) -- (0,0);
        \draw[line width=1,dashed,black!50] (-0.6,-0.6) -- (0,0) -- (-0.6,0.6);
        \draw[thick] (1,1) -- (2.3,1);
        \draw[thick] (2.34,-1) -- (1,-1);
        \draw[line width=1.5] (-0.5,1) to[out = -110, in = 110]  (-0.5,-1);
        \draw[line width=1.5] ({-1.3+3.6},1) to[out = -70, in = 70] ({-1.3+3.6},-1);
        \draw[line width=1,dashed,black!50] ({2+0.42},-0.42)--(2,0)--({2+0.42},0.42);
        \fill[black] ({-0.5},-1) circle (0.04);
        \fill[black] ({-1.3+3.6},-1) circle (0.04);
        \fill[black] ({-0.5},1) circle (0.04);
        \fill[black] ({-1.3+3.6},1) circle (0.04);
        \node at (1,1.2) {$i$};
        \node at (1,-1.2) {$j$};
        \end{scope}
    \end{tikzpicture}
    }}
\end{equation}
For the case of thin shells, we can use a standard trick where we identify the left and right regions of the spacetime
\begin{equation}
\label{blackhole}
\vcenter{\hbox{
\begin{tikzpicture}
    \begin{scope}[scale=1.75]
    \draw[line width=1,dashed,black!50,fill=blueKSV1!60] ({-0.8+0.22},{1-0.22})-- (-0.8,1) -- ({-0.8-0.44},{1-0.44})
        to[out=-82, in = 82] ({-0.8-0.44},{-1+0.44})
        -- (-0.8,-1)
        -- ({-0.8+0.22},{-1+0.22})
          to[out=105, in = -105] ({-0.8+0.22},{1-0.22});
        \draw[thick,decorate,decoration={zigzag,amplitude=0.3mm,segment length=2mm}] (-0.8,1) -- (1,1);
        \draw[thick, decorate,decoration={zigzag,amplitude=0.3mm,segment length=2mm}] (-0.8,-1) -- (1,-1);
        \draw[line width=1,dashed,black!50, fill=red!20] (0,0) -- (1,1) -- (2,0) -- (1,-1) -- (0,0);
        \draw[line width=1,dashed,black!50] (-0.6,-0.6) -- (0,0) -- (-0.6,0.6);
        \draw[thick] (1,1) -- (2,1) -- (2,-1) -- (1,-1);
        \draw[thick,black] (-0.8,1) -- (-1.6,1) -- (-1.6,-1) -- (-0.8,-1);
        \draw[line width=1.5] (-0.5,-1) to[out = 110, in = -110] (-0.5,1);
        \draw[line width=1.5] (-1.3,-1) to[out = 80, in = -80] (-1.3,1);
        \draw[line width=1,dashed,black!50] ({-1.6+0.37},-0.37)--(-1.6,0)--({-1.6+0.37},0.37);
        \fill[black] (-0.5,-1) circle (0.05);
        \fill[black] (-1.3,-1) circle (0.05);
        \fill[black] (-0.5,1) circle (0.05);
        \fill[black] (-1.3,1) circle (0.05);
        \draw[thick, ->] (-1.6,-0.55) -- (-1.6,-0.5);
        \draw[thick, ->] (-1.6,0.45) -- (-1.6,0.5);
        \draw[thick, ->] (2,-0.55) -- (2,-0.5);
        \draw[thick, ->] (2,0.45) -- (2,0.5);
        \node at (-0.8,1.2) {$i$};
        \node at (-0.8,-1.2) {$j$};
        \node[rotate = 90] at (-2,0) {identified};
        \node[rotate = 90] at (2.4,0) {identified};
        \end{scope}
    \end{tikzpicture}
    }}
\end{equation}
In this picture, the SdS microstate for an observer in the right static patch is prepared by inserting a pair of shells in the left region of the Penrose diagram. At least two shells are required to be able to make the required identification; we will mostly look at the case with two shells. The first shell transitions from the (red) SdS spacetime characterized by the parameters ($\ell_1, M_1$) to a (blue) SdS spacetime with $(\ell_2, M_2)$. 
The second shell then transitions back from the second to the first set of parameters, ensuring a consistent gluing at the boundaries of the diagram. Here $M$ and $\ell$ are the mass parameter and de Sitter radius of the geometry.

\paragraph{Energy and Matching background conditions.} Although we do not have access to an explicit construction of the UV theory, we can still infer certain microscopic consistency conditions that we expect a viable solution to satisfy. Therefore, on top of the properties described above, we are looking for solutions subject to the following two conditions:
\begin{itemize}
    \item \textit{Null Energy Condition:} From causality considerations, we expect the perturbed geometry to satisfy the Null Energy Condition (NEC). Although the NEC can be locally violated in the quantum theory (for instance, due to Casimir energy), one generally expects that an averaged version should hold in any consistent semiclassical regime. This is typically ensured by the Achronal Averaged Null Energy Condition (AANEC)~\cite{Hartman:2016lgu,Kelly:2014mra,Graham:2007va}, which imposes that 
    \begin{equation}
        \int_\gamma T_{\mu\nu}k^\mu k^\nu \geq 0,
    \end{equation}
    over all \emph{achronal} null geodesics $\gamma$ with tangent vector $k^\mu$. Achronal null geodesics are those that do not contain any pair of timelike-separated points.

    \item \textit{Matching Background Condition:} 
    We seek microstates corresponding to distinct configurations of the \emph{same} underlying theory, all constructed on a shared asymptotic vacuum. Translated to de Sitter,  the prediction would be that for a thin shell of matter, both sides of the shell must share the same background, i.e. the same cosmological constant. We refer to this requirement as the {matching background condition}. 
\end{itemize}
As it will become apparent, finding solutions that satisfy all the conditions outlined above is a non-trivial task. Nevertheless, in Section~\ref{sec:shells} we take an exploratory approach, systematically classifying the possible scenarios and identifying viable solutions.

\section{Defects in de Sitter}
\label{sec:shells}

This section is  dedicated to the construction of  spacetimes that contain a codimension-one timelike defect $\Sigma$ in the bulk. As discussed in Section~\ref{sec:setup}, we are looking for configurations in which (i) the observers' worldline connects their bra and ket states, and (ii) defects are outside the causal diamond of the observer. We focus on two scenarios: 
\vspace{-0.15cm}
\begin{itemize}
    \item[] \textbf{Scenario I:} \textit{End-of-the-world branes.} The spacetime is truncated beyond the defect. 
    \item[] \textbf{Scenario II:} \textit{Thin shells.} The defect divides spacetime into two regions, which we refer to as left $\M_{\lf}$  and right $M_{\ri}$ manifolds with metrics $g^\lf_{\mu \nu}$ and $g^\ri_{\mu \nu}$ respectively.
\end{itemize}
\vspace{-0.15cm}
To support these solutions, we introduce a matter source at the hypersurface $\Q$ where the defect is located, characterized by a perfect-fluid stress–energy tensor $T_\Q^{ij}$.

\begin{figure}[h!]
   \centering
\begin{tabular}{cc}
\raisebox{-33pt}{
\begin{tikzpicture}[line width=1. pt, scale=0.75]
 \fill[gray!20] (-5,-3) to[out=90, in = 270,looseness=1]
    (-5,3) -- (0,3) to[out=-120, in = -290,looseness=1] (0,-3) -- (-5,-3);
    \node at (-2.5,0) {$\M$};
    \node at (0,1.8) {$\Q$};
    \draw[thick, line width=1.5pt] (0,3) to[out=-120, in = -290,looseness=1] (0,-3);
    (-4,3) -- (4,3) to[out=-70, in = 120,looseness=1] (4,-3) -- (-4,-3);
\end{tikzpicture}
}& ~~
\raisebox{-33pt}{
\begin{tikzpicture}[line width=1. pt, scale=0.75]
 \fill[red!20] (-4,-3) to[out=90, in = 270,looseness=1]
    (-4,3) -- (0,3) to[out=-120, in = -290,looseness=1] (0,-3) -- (-4,-3);
    \fill[fill = blueKSV1!70] (4,-3) to[out=90, in = 270,looseness=1]
    (4,3) -- (0,3) to[out=-120, in = -290,looseness=1] (0,-3) -- (4,-3);
   \draw[thick, ->] (4.5,-2.5) -- (4.5,2.5);
    \node[anchor = west] at (4.6,0) {time};
    \node at (-2,0) {$\M_{\lf}$};
    \node at (2,0) {$\M_{\ri}$};
    \node at (0,1.8) {$\Q$};
    \draw[thick, line width=1.5pt] (0,3) to[out=-120, in = -290,looseness=1] (0,-3);
\end{tikzpicture}
}
\end{tabular}
   \caption{Left, a Lorentzian manifold $\M$ with boundary $\Sigma$ where the space ends (Scenario~I). Right, two Lorentzian manifolds $M_{\lf}$ and $M_\ri$ joined along a timelike codimension-one defect $\Q$ (Scenario~II).}
    \label{fig:1}
\end{figure}
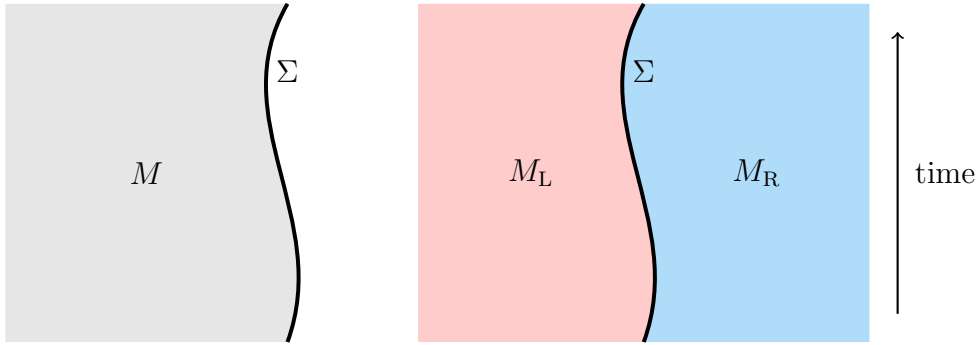

The section proceeds as follows.
In Subsection~\ref{sec3:setup}, we introduce the gravitational action and the equations of motion describing defects. In Subsection~\ref{sec:Equation of motion}, we focus on the specific case of spherically symmetric shells and branes; we extract the energy density and pressure of these solutions. In Subsections~\ref{sec:puredS}, \ref{sec:gaowald}, and~\ref{sec: SdS}, we assess the viability of these configurations as potential de Sitter and SdS microstates. This is done by checking whether they satisfy the null energy and matching background conditions. 
A summary of the results is as follows:
\vspace{-0.15cm}
\begin{itemize}
    \item \emph{Branes in dS and SdS:} for branes outside the cosmological horizon of the observer, the NEC is violated but AANEC is satisfied if one imposes reflective boundary conditions on the brane. 
    \item \emph{Shells in dS:} solutions require a jump in the cosmological constant across the shell. For the right sign of this jump, the energy density of the shell is positive and the NEC is satisfied. For the wrong sign both NEC and AANEC are violated. 
    \item \emph{Shells in SdS:} These solutions require a jump in the cosmological constant and/or a jump in the mass of the black hole. Solutions that satisfy the NEC have both: different masses and cosmological constants. We also consider a class of solutions with a fixed cosmological constant. These solutions violate the NEC, but we find evidence suggesting that they satisfy the AANEC.
\end{itemize}
\vspace{-0.15cm}
\noindent To end this section, we establish the existence of solutions to the equations of motion, discuss their qualitative properties, and present explicit numerical construction of such solutions in Subsection~\ref{sec: explicit sol}.

\subsection{Boundary field equations}
\label{sec3:setup}

For the gravity action, we consider the Einstein-Hilbert action, with a cosmological constant, along with a Gibbons–Hawking–York boundary term~\cite{York:1972sj,Gibbons:1976ue} for each region. In appendix~\ref{sec:GHY}, we provide a pedagogical derivation of the Gibbons–Hawking–York terms where we set the conventions. The results derived in the next few sections hold for a generic manifold until in Section~\ref{sec:puredS}, where we specialize to de Sitter or Schwarzschild-de sitter.  

In \textit{Scenario I}, the full action reads 
\be\label{eq: scenario I action}
S = S_{\text{EH}}[g_{\mu\nu}] +  S_{\text{GHY}}[h_{ij}] + S_{\Sigma}~,
\ee
where the Einstein-Hilbert term $S_{\text{EH}}$, the Gibbons–Hawking–York boundary term $S_{\text{GHY}}$ and the hypersurface matter-field action $ S_{\Sigma}$ are given by
\ba
S_{\eh}[g_{\mu \nu}] &= \frac{1}{16 \pi G} \int_{\M} \sqrt{|g|}(R-2\Lambda)~,\\
S_{\text{GHY}}[h_{ij}] &=  \frac{1}{8 \pi G}\int_{\Sigma} \sqrt{|h|} \,K~,\\
S_\Sigma&= \int_{\Sigma} \sqrt{|h|}\, \mathcal{L}^\Sigma_{\text{matter}}~.
\ea
$R$ is the Ricci scalar, $\Lambda$ is the cosmological constant, $h_{ij}$ is the induced metric on $\Sigma$, $K_{ij}$ is the extrinsic curvature and $K$ is its trace which is given by $\nabla_\mu n^\mu$ where $n^\mu$ is the outward-pointing normal vector to $\Sigma$. See appendix~\ref{eq:conventions} for more details on our conventions\footnote{Note that for a timelike hypersurface, this implies that $n_\mu$ is inward pointing}. We use Greek letters $\mu,\nu,\dots$ for the $(d+1)$-dimensional indices and Latin letters $i,j,\dots$ for the $d$-dimensional indices that describe the hypersurface $\Q$. We also attribute a stress tensor to the boundary matter field: 
\be
T_{ij}^{\Sigma}\equiv\frac {-2}{\sqrt{|h|}}\frac {\delta ({\sqrt{|h|}}\mathcal {L}^{\Sigma}_{\text {matter} })}{\delta h^{ij}}~.
\ee
In this work, we consider the Neumann boundary condition at the defect in which there is no flux of energy in or out of the defect. By requiring that the variation of action vanishes, as reviewed in appendix~\ref{sec:GHY}, this boundary condition implies certain field equations for location of the brane. In particular, we have:
\be\label{eq: Israel II a}
K_{ij}- K\, h_{ij}=8\pi\e G \, T^{\Sigma}_{ij}~.
\ee

The action for the \textit{scenario II} is given by
\be
\label{eq: scenario II action}
    S = S_{\text{EH}}[g^\lf_{\mu\nu}] + S_{\text{EH}}[g^\ri_{\mu\nu}]+ S_{\text{GHY}}[h^\lf_{ij}]+ S_{\text{GHY}}[h^\ri_{ij}] + S_{\Sigma}~,
\ee
where $h^{\lf/\ri}_{ij}$ is the induced metric of $\Sigma$ for the left/right manifold. The continuity of the metric implies the first Israel-Lanczos junction condition\cite{Israel:1966rt,Lanczos:1924bgi}:
\be\label{eq: first Israel}
h^\lf_{ij} = h^\ri_{ij}~,
\ee
therefore, from now on, we drop the superscript for the left and right manifolds and simply use $h_{ij}$ for the induced metric at the defect. 
Similar to scenario I, the matter field at the defect, sources the extrinsic curvature, leading to the second junction condition:
\be\label{eq: Israel II b}
\kappa_{ij}- \kappa\, h_{ij}=8\pi\e G \, T^{\Sigma}_{ij}~,
\ee
where $\epsilon = n_\mu n^\mu = \pm 1$, and we defined the total extrinsic curvature 
\be\label{eq: kappa def}
\kappa_{ij} \equiv K^{\lf}_{ij} +K^{\ri}_{ij}~. 
\ee
For a more detailed derivation of this condition see appendix~\ref{sec:GHY}.
One can see scenario I as a limit of scenario II in which one of the manifolds is set to have zero extrinsic curvature. This can be easily verified by comparing two actions~\reef{eq: scenario I action} and \reef{eq: scenario II action}. 

\textit{A note of caution}: $\kappa_{ij}$ is the sum of the extrinsic curvatures of the left and right manifolds because our convention for the normal vector at the junction always assumes that it is \emph{outward-pointing} for each manifold ($n^\lf_{ij}=-n^\ri_{ij}$). Some sources, take the normal vector of the left and right manifolds to be the same. In that case, $\kappa_{ij}$ would be the difference of the extrinsic curvature of the two manifolds following the definition of the extrinsic curvature in~\reef{eq: K def}.

\subsection{Equations of motion}
\label{sec:Equation of motion}
We now turn to a concrete problem: the evolution of a spherically symmetric thin shell $\Sigma$. In the following, we employ the defect field equations~\reef{eq: Israel II a} and~\reef{eq: Israel II b} to derive the equation of motion for a perfect-fluid thin defect in both scenarios.
We first focus on scenario I and later recover the solution for scenario II.

\subsection*{Scenario I}
Using the spherical symmetry, the metric of the bulk can be written as
\begin{equation}\label{eq: metric spherical}
ds^2 = -f(r) \dd t^2+ \frac{dr^2}{f(r)} + r^2 \dd\Omega^2_{d-1}~, 
\end{equation}
where $\dd\Omega^2_{d-1}$ is the metric of the unit sphere $S^{d-1}$. 
We parametrize the evolution of the brane with an auxiliary variable $\tau$. More concretely, we take points on the spherical symmetric brane to have the worldvolume represented by $r=R(\tau)$ and $t=T(\tau)$:
\begin{equation}
    \Sigma = \left(T(\tau), R(\tau), \theta_1,\theta_2,\dots,\theta_{d-1}\right)~,
\end{equation}
where we parametrize the points on the hypersurface $\Sigma$ as $\xi^i =(\tau,\theta_1,\cdots,\theta_{d-1})$.

To find the extrinsic curvature, we first construct the tangent  and normal vectors.
A set of convenient tangent vectors $e^\mu_i=\frac{\partial x^\mu}{\partial\xi^i}$  on $\Sigma$ is:
\ba\label{eq:tangent vec}
    e_\tau&=\left(\dot T(\tau)\,,\,\dot R(\tau)\,,\,0\,,\,0\,,\,\cdots\right)~,\\
e_{\theta_a} &= \left(0\,,\,0\,,\,\cdots\,,\, {1}\,,\, \cdots \right)~,
\ea
where we define $\dot A \equiv \partial_\tau A$. These are perpendicular to the outward-pointing normal vector
\be
n^{\mu} =  \Gam\left(\frac{\dot R}{f(R)}\,,\,f(R)\,\dot T\,,\,0\,,\,0\,,\,\cdots\right)~,
\ee
where we keep the dependence on $\tau$ implicit to avoid clutter. Here, we are assuming that variations in $\tau$ is such that $f(R)\dot T>0$.

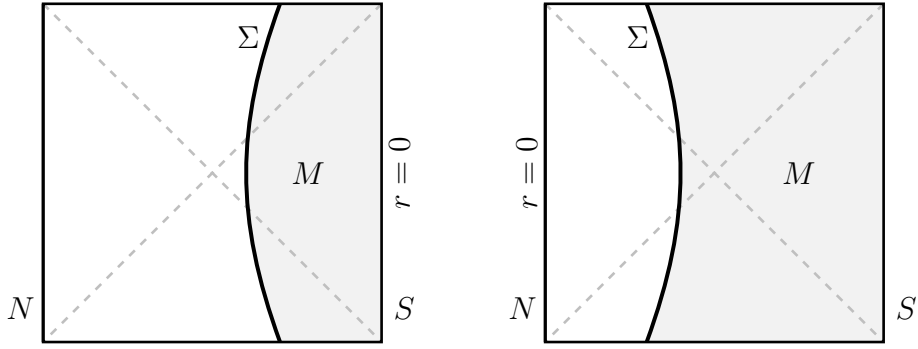
\begin{figure}[t!]
   \centering
\begin{tabular}{cc}
\scalebox{1}{
\raisebox{-33pt}{
\begin{tikzpicture}[line width=1. pt, scale=1.5]
\fill[gray,opacity=0.1](1.5,-1.5) -- (1.5,1.5) -- (0.6,1.5) to[out=180+70, in = 180-70,looseness=1] (0.6,-1.5) --  (1.5,-1.5) -- (1.5,1.5);
\node[scale=1,black] at (0.85,0) {$\M$};
\draw[lightgray,dashed, line width=1pt]  (1.5,1.5)-- (0,0) -- (1.5,-1.5);
\draw[lightgray,dashed, line width=1pt]  (-1.5,1.5)-- (0,0) -- (-1.5,-1.5);
\draw[black, line width=1.5pt] (0.6,1.5) to[out=180+70, in =180- 70,looseness=1] (0.6,-1.5) ;
\node[scale=1,black] at (0.32,1.2) {$\Sigma$};
\node[scale=1,black, rotate=90] at (1.66,0) {$r=0$};
\draw[black, line width=1pt]  (-1.5,1.5)-- (-1.5,-1.5) -- (1.5,-1.5) -- (1.5,1.5)-- (-1.5,1.5)--(-1.5,-1.5);
\node[scale=1,black] at (1.7,-1.2) {$S$};
\node[scale=1,black] at (-1.7,-1.2) {$N$};
\end{tikzpicture}
}
}&
\scalebox{1}{
\raisebox{-33pt}{
\begin{tikzpicture}[line width=1. pt, scale=1.5]
\fill[gray,opacity=0.1](1.5,-1.5) -- (1.5,1.5) -- (-0.6,1.5) to[out=-70, in = 70,looseness=1] (-0.6,-1.5) --  (1.5,-1.5) -- (1.5,1.5);
\node[scale=1,black] at (0.75,0) {$\M$};
\draw[lightgray,dashed, line width=1pt]  (1.5,1.5)-- (0,0) -- (1.5,-1.5);
\draw[lightgray,dashed, line width=1pt]  (-1.5,1.5)-- (0,0) -- (-1.5,-1.5);
\draw[black, line width=1.5pt] (-0.6,1.5) to[out=-70, in = 70,looseness=1] (-0.6,-1.5) ;
\node[scale=1,black] at (-0.68,1.2) {$\Sigma$};
\node[scale=1,black, rotate=90] at (-1.66,0) {$r=0$};
\draw[black]  (-1.5,1.5)-- (-1.5,-1.5) -- (1.5,-1.5) -- (1.5,1.5)-- (-1.5,1.5)--(-1.5,-1.5);
\node[scale=1,black] at (-1.7,-1.2) {$N$};
\node[scale=1,black] at (1.7,-1.2) {$S$};
\end{tikzpicture}
}
}
\end{tabular}
\caption{
Illustration of the two possible configurations for the defect, with the observer located at the south pole. Left, after introducing the defect, we retain the small portion (shown in grey) of the manifold. We refer to this configuration as the ``small" cut,  corresponding to $\gamma=1$. Right, after introducing the defect, we retain the large portion of the manifold. We call this type of configuration the ``large" cut, which corresponds to $\gamma=-1$.
}
\label{fig: gamma=pm1}
\end{figure}

Here, we introduce the parameter $\Gam=\pm 1$  to keep track of the portion of the manifold retained after introducing the brane as illustrated in Figure~\ref{fig: gamma=pm1}. The small portion of the manifold corresponds to $\Gam=1$ and the large portion to $\Gam=-1$. The coordinates we use to describe the location of the shell are the static patch of the south pole (small cut) or the north pole (large cut). Equivalently, one may think of the manifold as including or not the origin of the $(t,r)$ coordinates.  If the manifold includes the origin, the outward-pointing vector is in the positive radial direction and $\Gam=1$ and if the manifold does not include the origin, the outward pointing is in the negative radial direction and $\Gam=-1$. 
Later on, in Subsection~\ref{sec:puredS}, we will introduce the observer and place it at the south pole (or equivalently on the north pole) of the geometry. The requirement to keep the defect out of the observer's causal diamond translates to $\gamma=-1$.
Moreover, we also use the freedom to choose $\tau$ such that the normal vector be unit normalized:
\be\label{eq: Israel I}
f(R) \, \dot T^2- \frac{\dot R^2}{f(R)} =1~.
\ee
As we will see this also implies the time-time component of the induced metric to be identity.
For later convenience we assign a parameter to radial component of the normal vector and we rewrite it using~\reef{eq: Israel I}
\be\label{eq: beta}
\betaB\equiv \Gam f(R) \, \dot T = \Gam \sqrt{f(R) + \dot R^2}~.
\ee
\paragraph{Induced metric and extrinsic curvature.}
Using the representation of tangent vectors in~\reef{eq:tangent vec} and the definition of induced metric in~\reef{eq: def h}, one can write the induced metric:
\be
ds^2_\Sigma= h_{ij}\, d\xi^i d\xi^j = -d\tau^2+R^2 d\Omega_{d-1}^2~,
\ee
where the condition~\reef{eq: Israel I} sets the time-time component to unity. 
Next, we calculate the extrinsic curvature. Using the explicit expression of the extrinsic curvature in~\reef{eq:pullback of K}, one finds
\begin{equation}
    K_{ij} \,\dd \xi^i \dd \xi^j 
    = \Gam
    \left(
    \frac{3 \dot R^2 \dot T f'}{2f}
    -\frac{1}{2} \dot T^3  f  f' 
    +\dot R \ddot  T 
    -\dot  T \ddot R
    \right) \dd \tau^2 
    +\Gam R \dot{T} f \, d \Omega_{d-1}^2~,
\end{equation}
where to avoid clutter, we keep the radial dependence of $f(R)$ implicit and defined $A'(x)\equiv\partial_{x} A(x)$.
By substituting the derivatives of $T$ using~\reef{eq: beta}, one can rewrite the expression for the extrinsic curvature as a function of $R$ and its derivatives in a concise form:
\be\label{eq:Simple K}
K_{ij} \,\dd \xi^i \dd \xi^j  = - \frac{\dot \betaB}{\dot R} d\tau^2 +  \betaB R \,d\Omega_{d-1}^2~.
\ee
\paragraph{Equation of motion for a perfect fluid.}
Finally, we use the defect equation~\reef{eq: Israel II a} to derive the equation of motion for a thin shell with a spherically symmetric perfect fluid characterized by the stress tensor:
\be
T^{ij}_\Sigma= (\sigma +p) u^i u^j + p \,  h^{ij}~,
\ee
where $\sigma$ is the surface density, $p$ is the pressure and $u^i$ is the proper fluid velocity tangent to the worldvolume of the shell. 
Plugging this stress tensor into second Israel junction condition~\reef{eq: Israel II a}, one can find the density and pressure in terms of $\betaB$:
\ba\label{eq: sigma and p}
\sigma &=\frac{d-1}{8\pi G} \frac{\betaB}{R}~, \\ 
p&= -\frac{1}{8\pi G}\left((d-2)\frac{\betaB}{R}+ \frac{\dot \betaB}{\dot R}\right)=-\frac{1}{(d-1)  \dot R R^{d-2}}\,\partial_\tau \big(\sigma R^{d-1}\big)~.
\ea

In this paper, we consider matter sources with an equation of state given by $p = w \sigma$, where $w$ is a constant. Important values of $w$ include $w = -1$ for brane solutions and $w = 0$ for pressureless dust. One can the solve for $\betaB$ using the equation of state and find:
\be\label{q: beta of R}
\betaB' = \alpha \frac{\betaB}{R}, \quad \text{which leads to} \quad \betaB = {c}\,{R^{\alpha}}~.
\ee
where we used the chain rule $\dot\betaB = \betaB' \dot R$ and introduced $\alpha=-((d-1)w+d-2) $. Here, $c$ is a constant that can be fixed using the total energy of the brane at a given time through the first equation in~\reef{eq: sigma and p}. 
By squaring this equation and substituting $\betaB$ using~\reef{eq: beta}, one can find an equation of motion for the brane 
\be\label{eq: EofM}
\dot R^2 +V(R) =0~,
\ee
with the potential
\be
V_\text{I}(R) = f(R)- {c^2}\,{R^{2\alpha}}~.
\ee
The subscript in the potential refers to scenario I.
It is possible to formally integrate this equation of motion to find the trajectory of the shell in this coordinate system. Using~\reef{eq: Israel I} one finds the trajectory of the shell to satisfy 
\be\label{eq: T of R}
T(\tau)-T_0=\pm |c| \int^{R(\tau)}_{R_0}\frac{dR}{f(R)}\left({c^2-R^{-2\alpha}f(R)}\right)^{-\half}~,
\ee
such that the shell is at radius $R_0$ and time $T_0$ at proper time $\tau_0$. The sign $\pm$ should be picked such that it is consistent with our convention $f(R) \dot T>0$~.

\subsection*{Scenario II}
Most of the analysis of the previous section follows for scenario II as well. More precisely, up to equation~\reef{eq:Simple K} all equations are correct for both left and right manifolds. From now on we use subscripts $\lf$ and $\ri$ respectively for left and right manifolds. The first junction condition \eqref{eq: first Israel} is automatically satisfied by choosing a parametrization where the normal vectors are unit normalized,
\begin{equation}
    f_{\lf}(R)  \dot T^2_{\lf} - \frac{\dot R^2}{f_\lf(R)} = 1 =
    f_{\ri}(R)  \dot T^2_{\ri} - \frac{\dot R^2}{f_\ri(R)}~,
\end{equation}
where $T =  T(\tau)$, $R=R(\tau)$ and derivatives are with respect to $\tau$. Note that, to ensure a continuous gluing of the $(d-1)$-sphere, the left and right manifolds must be joined at the same radial coordinate $r_\text{L}(\tau) = R(\tau) = r_\text{R}(\tau)$. From now on, we always place the observable on the right manifold.

In this parametrization, the second junction condition, equation \eqref{eq: Israel II b}, relates the density and pressure of the shell to the functions $B_\lf$ and $B_\ri$,
\ba\label{eg:sigma and p II}
\sigma &=\frac{d-1}{8\pi G} \frac{\betaB_\lf+\betaB_\ri}{R}~,\\
p&= -\frac{1}{8\pi G}\left((d-2)\frac{\betaB_\lf+\betaB_\ri}{R}+ \frac{\dot\betaB_\lf+\dot\betaB_\ri}{\dot R}\right)~,
\ea
where $B_\lf$ and $B_\ri$ are defined as in \eqref{eq: beta} for the left and right manifolds. Equation \eqref{eg:sigma and p II} is identical to~\reef{eq: sigma and p} by transformation $\betaB\to \betaB_\lf + \betaB_\ri$. 
Again, similar to above, for a perfect fluid, one can solve for $\betaB$ to find:
\be\label{eq: beta lr and c}
\betaB_\lf+\betaB_\ri = {c}\,{R^{\alpha}}~.
\ee
which leads to the equation of motion \eqref{eq: EofM}, but now with a modified potential:
\be
V_{\text{II}}(R) = \half  \left(f_\lf(R)+f_\ri(R)\right)  - \frac{(f_\lf(R)-f_\ri(R))^2} {4c^2 R^{2\alpha}}  - \frac{c^2}{4} R^{2\alpha}~,
\ee
which is independent of $\gamma_\text{L}$ and  $\gamma_\text{R}$, the orientation of the left and right normal vectors.
Note by setting $c\to 2c$ and $f_\lf = f_\ri$ one reproduces the potential $V_{\text{I}}$ from $V_{\text{II}}$. This is a generic feature. One can deduce the equations of motion for the end-of-the-brane from Scenario II by using these redefinitions. Equivalently, one can also set $\betaB_\lf=0$ and $\betaB_\ri=\betaB$.
Similar to~\reef{eq: T of R}, the integrated equation of motion for $T$ is given by
\be
T_{\lf}(\tau)-T_0=\pm  \int^{R(\tau)}_{R_0}\frac{dR}{f_\lf(R)}{\left(\frac{V_\text{II}+ f_{\lf}(R)}{V_\text{II}}\right)^{\half}}~,
\ee
with a similar formula for $T_\ri$. As with Scenario I, the sign $\pm$ of $T_{\text{L}}$ and $T_{\text{R}}$ should be picked such that it is consistent with $f(R) \dot T>0$~.

\paragraph{Orientation of the defect.}
In de Sitter and Schwarzschild–de Sitter we aim to leave the observer's casual diamond  untouched. This implies that, for the portion of the manifold containing the observer, we must consider a large cut with $\gamma = -1$. In scenario II, two types of gluing arise: large–small (left)  and large–large (right),
\begin{center}
\begin{minipage}{0.4\textwidth}
\centering
\begin{tikzpicture}[line width=1. pt, scale=1.4]

\draw[black!50,dashed,line width=1pt] (0,0) -- (1.5,-1.5);
\draw[black!50,dashed,line width=1pt] (0,0) -- (1.5,1.5);

\draw[black!50,dashed,line width=1pt] (-0.665+0.3,-0.665) -- (-1.2,-1.5);
\draw[black!50,dashed,line width=1pt] (-0.665+0.3,0.665) -- (-1.2,1.5);

\draw[black, line width=1.5pt] (-0.6,1.5) to[out=-70, in = 70,looseness=1] (-0.6,-1.5) ;

\draw[black, line width=1pt]  (-1.2,1.5)-- (-1.2,-1.5) -- (1.5,-1.5) -- (1.5,1.5)-- (-1.2,1.5)--(-1.2,-1.5);
\draw[red, line width=1pt](1.5,-1.5) -- (1.5,1.5);
\node[scale=1, black] at (0,-1.7) {large–small gluing};
\node[scale=1,red ,rotate=90] at (1.7,0) {Observer};
\fill[red,opacity=0.2](1.5,-1.5) -- (1.5,1.5) -- (0,0);
\end{tikzpicture}
\end{minipage}
\hspace{0.05\textwidth}
\begin{minipage}{0.4\textwidth}
\centering
\begin{tikzpicture}[line width=1. pt, scale=1.4]

\draw[black!50,dashed,line width=1pt] (0,0) -- (1.5,-1.5);
\draw[black!50,dashed,line width=1pt] (0,0) -- (1.5,1.5);

\draw[black!50,dashed,line width=1pt] (-0.5,0) -- (-2,-1.5);
\draw[black!50,dashed,line width=1pt] (-0.5,0)-- (-2,1.5);

\draw[black, line width=1.5pt] (-0.6,1.5) to[out=-70, in = 70,looseness=1] (-0.6,-1.5) ;

\draw[black, line width=1pt]  (-2,1.5)-- (-2,-1.5) -- (1.5,-1.5) -- (1.5,1.5)-- (-2,1.5)--(-2,-1.5);

\draw[red, line width=1pt](1.5,-1.5) -- (1.5,1.5);
\node[scale=1,red ,rotate=90] at (1.7,0) {Observer};
\fill[red,opacity=0.2](1.5,-1.5) -- (1.5,1.5) -- (0,0);
\node[scale=1, black] at (-0.25,-1.7) {large–large gluing};
\end{tikzpicture}
\end{minipage}
\end{center}
In what follows, we demonstrate that the large–large gluing violates the NEC condition. This will be shown both through explicit calculation and via a more geometric argument based on the Gao–Wald theorem.

\subsection{De Sitter solutions}
\label{sec:puredS}
The equations of motion~\reef{eq: EofM} derived above provide a continuous family of solutions identified by their energy label $c$ and equation-of-state parameter $w$. 
Before presenting explicit solutions  in Subsection~\ref{sec: explicit sol}, here we examine whether the solutions satisfy the matching background condition and the NEC. 

First, let us unfold the options for the possible solutions for both scenarios when the background is de Sitter with metric:
\begin{equation}
ds^2 = -f(r) \dd t^2+ \frac{dr^2}{f(r)} + r^2 \dd\Omega^2_{d-1}~, \quad
f(r)= 1 - \frac{r^2}{\ell^2}~,
\end{equation}
where $\ell$  is de Sitter radius, related to the cosmological constant via $\Lambda=\frac{d(d-1)}{2\ell^2}$.

For \emph{scenario I}, we must restrict to large-cut solutions with $\gamma = -1$. The reason is that only for these solutions is the brane located outside the causal diamond of the observer. The equations of motion \eqref{eq: sigma and p} then yield a brane with a negative energy density and negative tension (corresponding to a positive pressure via $p = -\sigma$ when $w = -1$). For a perfect fluid, the NEC is often expressed as $p + \sigma \geq 0$. Since for brane solutions $p + \sigma = 0$, one could have expected that such objects also satisfy the NEC. However, this is not the case. The reason is that the brane is a codimension-one object, and therefore $T_{ij} k^i k^j\geq0$ is not equivalent to $T_{\mu\nu} k^\mu k^\nu \geq 0$, where $i,j$ are indices in $\Sigma$ and $\mu,\nu$ are indices in de Sitter. In Appendix~\ref{sec: NEC}, we show that for these branes, the NEC holds for null vectors tangent to $\Sigma$ but is violated for null vectors that are not tangent to $\Sigma$. More generally, we find that the NEC is satisfied only when the energy density of the defect $\sigma \geq 0$ and  $w\geq -1$. Later in Subsection \eqref{sec:gaowald}, we will recover this result using the Gao–Wald theorem. We note that although these solutions violate the NEC, the brane lies at the boundary of our manifold. If one imposes boundary conditions such that null geodesics reflect off the brane, then the achronal averaged null energy condition would not be violated. This suggests that such a solution could still be compatible with the requirements of causality.

Let us now discuss  \emph{scenario II}. 
For the moment we relax the matching background condition and allow for the left and right metric have different cosmological constant:
\be
f_{\lf/\ri}(r)= 1 - \frac{r^2}{\ell^2_{\lf/\ri}}~,
\ee
where $\ell_{\lf/\ri}$ is de Sitter radius for the left and right manifold and is given by their corresponding cosmological constants. By putting the observer on the right manifold, again we require $\gamma_\ri=-1$. Therefore, there could be two types of solutions $\gamma_\lf = -1$ (large-large gluing) or $\gamma_\lf = 1$ (small-large gluing). Using~\reef{eg:sigma and p II}, the energy density of the defect for the both cases is given by
\ba\label{eq:sigma Rdot}
\sigma = \frac{d-1}{8\pi G R} \left(\gamma_{\lf}\sqrt{1-\frac{R^2}{\ell^2_\lf}+\dot R^2}-\sqrt{1-\frac{R^2}{\ell^2_\ri}+\dot R^2}\right)~.
\ea
For the large-large gluing, while one can ask for the matching background condition by imposing $\ell_\ri=\ell_\lf$, the energy density is always non-positive and therefore violates the NEC.
This is in agreement with the expectation of the Gao-Wald theorem as it refers to a ``fatter" configuration. 

For the small-large gluing, on the other hand, the positivity of energy density $\sigma > 0$ implies that the de Sitter radius of the left manifold must be larger than the de Sitter radius of the right manifold:
\be
\sigma \geq 0 ~\quad \text{only when} \quad \ell_\lf \geq\ell_\ri~,
\ee
which is incompatible with the matching background condition except for the trivial solution where the left and right manifolds share the same de Sitter radius, $\ell_\lf = \ell_\ri$, leading to $\sigma = 0$. We therefore conclude that, in scenario II, the NEC and matching background condition cannot be simultaneously satisfied within a de Sitter setting.

\subsection{Gao-Wald theorem}
\label{sec:gaowald}
\vspace{-0.25cm}

So far, with explicit calculations, we have seen that the large-large solution is not compatible with the NEC. Here, we review a more generic and elegant perspective to understand this through a corollary of the Gao–Wald theorem~\cite{Gao:2000ga}---which is often interpreted as the exclusion of ``wider'' or ``fatter'' de Sitter spaces. For completeness, we state the relevant result from that paper below:
\emph{
\begin{quote}
    Let $M$ be a spacetime with metric $g_{\mu\nu}$, and let $I^{\pm}(p)$ denote the chronological future/past of given point $p$. Assume that the pair $(M,g_{\mu\nu})$ satisfies the following conditions:
    \begin{itemize}
        \item The spacetime is null geodesically complete, smooth and time-oriented.
        \item It is globally hyperbolic, with a compact Cauchy surface $\Sigma$. 
        \item It satisfies the null energy condition and the null generic condition. 
    \end{itemize}
    Then, there exists Cauchy surfaces $\Sigma_1$ and $\Sigma_2$ with $\Sigma_2 \subset I^{+}(\Sigma_1)$ such that if $p \in I^{+}(\Sigma_2)$ then $\Sigma_1\subset I^{-}(p)$. 
\end{quote}}
Let us begin by reviewing the terminology of the theorem. Time orientability simply means that the spacetime has a non-vanishing timelike vector field. A spacetime is called globally hyperbolic if it contains a Cauchy surface (a spacelike surface that is intersected exactly once by every causal curve in $M$). The null energy condition is the familiar requirement that 
\begin{equation}
    T_{\mu\nu} k^\mu k^\nu \geq 0\,, \quad \text{for all null vectors }k^\mu~. 
\end{equation}
The null generic condition is the additional requirement that each null geodesic in the manifold contains a point for which 
\begin{equation}
    k_{[\eta} R_{\mu]\nu \rho[\sigma}  k_{\tau]} k^{\nu}k^{\rho} \neq 0~,
\end{equation}
where $k^\mu$ is the tangent vector of the null geodesic. This is a technical assumption that is required to apply various results regarding the Raychaudhuri equation. While de Sitter space does not satisfy the null generic condition, the theorem is designed to hold under small perturbations of de Sitter where the condition is satisfied, e.g. through adding matter which in our case is the shell.

The relevant implication of this theorem is that, under a generic perturbation of de Sitter space, the area of an observer’s cosmological horizon at future infinity shrinks to zero at a finite time in the past. Equivalently, at some point in the future the observer (located at $p$) can see the entire Cauchy slice $\Sigma_1$ in de Sitter. In terms of its Penrose diagram, this is often described by saying that de Sitter becomes “taller” rather than “fatter” under a generic perturbation satisfying the null energy condition:
\begin{equation}
\underbrace{
\vcenter{\hbox{
\begin{tikzpicture}[line width=1. pt, scale=1.25]
    \fill[gray!20] (1,1.5-0.1) -- (-0.5,-0.1) -- (-1.5,-1-0.1) -- (-1.5,-1.5) -- (1,-1.5);
    \draw[thick, black] (-1,1.5) -- (-1.5,1.5) -- (-1.5,-1.5) -- (-1,-1.5);
    \draw[thick, black] (-1,1.5) -- (1,1.5) -- (1,-1.5) -- (-1,-1.5);
    \draw[thick, black, line width=1.5pt] (-1,-1.5)  to[out = 75, in = -75] (-1,1.5);
    \draw[line width=1pt,dashed, black](1.,1.3)--(-1.5,1.3);
    \node[scale=1,black] at (-1.75,1.3){$\Sigma_2$};
    \draw[line width=1pt,dashed, black](1.,-1.3)--(-1.5,-1.3);
    \node[scale=1,black] at (-1.75,-1.3){$\Sigma_1$};
    \node[scale=1,black] at (1,1.7){$P$};
    \fill[black] (1,1.4) circle (0.06);
\end{tikzpicture}
}
}}_{{\text{Taller: NEC is not violated}}}
\qquad \qquad
\underbrace{
\vcenter{\hbox{
\begin{tikzpicture}[line width=1. pt, scale=1.25]
     \fill[gray!20](1,1.4) -- (-2+0.1,-1.5) -- (1,-1.5);
    \draw[thick, black] (-2,1.5) -- (-2.7,1.5) -- (-2.7,-1.5) -- (-2,-1.5);
    \draw[thick, black] (-2,1.5) -- (1,1.5) -- (1,-1.5) -- (-2,-1.5);
    \draw[thick, black,line width=1.5pt] (-1,-1.5)  to[out = 75, in = -75] (-1,1.5);
    \draw[line width=1pt,dashed, black](1.,1.3)--(-2.7,1.3);
    \node[scale=1,black] at (-2.95,1.3){$\Sigma_2$};
    \draw[line width=1pt,dashed, black](1.,-1.3)--(-2.7,-1.3);
    \node[scale=1,black] at (-2.95,-1.3){$\Sigma_1$};
    \node[scale=1,black] at (1,1.7){$P$};
    \fill[black] (1,1.4) circle (0.06);
\end{tikzpicture}
}
}}_{{\text{Fatter: NEC is violated}}}.
\end{equation}
In this picture, the grey region represents $I^-(p)$. In the taller spacetime configuration, there exists a pair of Cauchy surfaces $\Sigma_1$ and $\Sigma_2$ such that any point $p$ in the future of $\Sigma_2$ includes $\Sigma_1$ in its past. In contrast, in the fatter configuration, no such pair of Cauchy surfaces can be found. Therefore, upon introducing a perturbation like a shell satisfying the NEC, the spacetime always gets ``taller''. Note that the situation for the Schwarzschild–de Sitter case is different, since one of the conditions of the theorem, namely, null geodesic completeness, is not satisfied.

\subsection{Schwarzschild-de Sitter solutions}\label{sec: SdS}

As discussed in Section~\ref{sec:setup}, the Schwarzschild–de Sitter configuration considered here involves two defects, each satisfying the local field equations. We focus on Scenario~II for brevity; Scenario~I can be obtained from Scenario~II using the steps described in Subsection~\ref{sec:Equation of motion}. Shell $\Sigma_1$ starts from $r=0$ (zigzag line) at global early times, grows to a maximum radius, and collapses back to $r=0$ at global late times. The other shell, $\Sigma_2$, starts from $r=\infty$ (Horizontal solid line), shrinks to a finite minimum radius and grows back to $r=\infty$. Similar to above, the observer causal diamond (in red) is untouched:
\begin{equation}
\vcenter{\hbox{
  \begin{tikzpicture}
    \begin{scope}[scale=1.75]
\draw[black!50,dashed, line width=1.0pt,fill=blueKSV1!70] ({-0.8+0.22},{1-0.22})-- (-0.8,1) -- ({-0.8-0.44},{1-0.44})
        to[out=-82, in = 82] ({-0.8-0.44},{-1+0.44})
        -- (-0.8,-1)
        -- ({-0.8+0.22},{-1+0.22})
          to[out=105, in = -105] ({-0.8+0.22},{1-0.22});
        \draw[thick,decorate,decoration={zigzag,amplitude=0.3mm,segment length=2mm}] (-0.8,1) -- (1,1);
        \draw[thick, decorate,decoration={zigzag,amplitude=0.3mm,segment length=2mm}] (-0.8,-1) -- (1,-1);
        \draw[black!50,dashed,line width=1pt, fill=red!20] (0,0) -- (1,1) -- (2,0) -- (1,-1) -- (0,0);
        \draw[black!50,dashed,line width=1pt] (-0.6,-0.6) -- (0,0) -- (-0.6,0.6);
        \draw[thick] (1,1) -- (2,1) -- (2,-1) -- (1,-1);
        \draw[thick,black] (-0.8,1) -- (-1.6,1) -- (-1.6,-1) -- (-0.8,-1);
        \draw[line width=1.5] (-0.5,-1) to[out = 110, in = -110] (-0.5,1);
        \draw[line width=1.5] (-1.3,-1) to[out = 80, in = -80] (-1.3,1);
        \draw[black!50,dashed,line width=1pt] ({-1.6+0.37},-0.37)--(-1.6,0)--({-1.6+0.37},0.37);
        \draw[thick, ->] (-1.6,-0.55) -- (-1.6,-0.5);
        \draw[thick, ->] (-1.6,0.45) -- (-1.6,0.5);
        \draw[thick, ->] (2,-0.55) -- (2,-0.5);
        \draw[thick, ->] (2,0.45) -- (2,0.5);
        \node at (-0.5,1.2) {$\Sigma_1$};
        \node at (-1.3,1.2) {$\Sigma_2$};
        \end{scope}
    \end{tikzpicture}
        }}
\end{equation}
\\
Here, again, we place the observer on the right hand side and the part of manifold between the shells that does not include the observer (in blue) on the left side. 
The Schwarzschild-de Sitter metric is of the form of~\reef{eq: metric spherical} with:
\be
f_{\lf/\ri}(r)= 1- \frac{\mathcal{M}_{\lf/\ri}} {r^{d-2}} - \frac{r^2}{\ell_{\lf/\ri}^2}~,
\ee
where each side of the shells has the corresponding cosmological constant and the mass parameter:
\be
\mathcal{M}_{\lf/\ri}=\frac{16\pi G m_{\lf/\ri}}{(d-1)\Omega_{d-1}}~,
\ee
where
$m_{\lf/\ri}$ is the mass of the blackhole and $\Omega_{d-1}=\frac{2\pi^{\hd}}{\Gamma(\hd)} $ is the area of the unit $(d-1)$-sphere.

Let us examine the NEC and matching background condition for this case. 
Compared to the de Sitter cases, now the $r=0$ is placed on top and bottom of the Penrose diagram. So, one has to be extra careful with signs of $\gamma$ for each shell. In particular, on $\Sigma_1$, the normal vector with respect to the blue (left) manifold is pointing at the negative $\partial_r$ direction i.e. $\gamma=-1$ and the normal vector  with respect to the red (left) manifold is pointing at the positive $\partial_r$ direction i.e. $\gamma=+1$. 
Taking care of these signs also for $\Sigma_2$, similar to~\reef{eq:sigma Rdot}, the energy density of these shells is given by the equations:
\ba
\sigma_1&= \frac{d-1}{8\pi G R_1} \left(\sqrt{f_\ri(R_1)+\dot R_1^2}- \sqrt{f_\lf(R_1)+\dot R_1^2}\right)~,\\
\sigma_2&= \frac{d-1}{8\pi G R_2} \left(\sqrt{f_\lf(R_2)+\dot R_2^2}- \sqrt{f_\ri(R_2)+\dot R_2^2}\right)~.
\ea

Let us now analyze the NEC, which translates to $\sigma_1, \sigma_2 \geq 0$. This imposes the following conditions on the metric evaluated at the shells:
\be
f_\ri (R_1) \geq f_\lf (R_1)~, \qquad f_\lf (R_2) \geq f_\ri (R_2)~.
\ee
By substituting the explicit expressions for $f_{\lf/\ri}$ into the equations above, one obtains the following inequalities:
\ba\label{eq: cond SdS ineq}
\frac{R_1^d}{\ell_\lf^2} + \mathcal{M}_\lf \geq \frac{R_1^d}{\ell_\ri^2} + \mathcal{M}_\ri~, \qquad
\frac{R_2^d}{\ell_\ri^2} + \mathcal{M}_\ri \geq \frac{R_2^d}{\ell_\lf^2} + \mathcal{M}_\lf~.
\ea
Adding these two inequalities, canceling the mass parameters on both sides, and using the fact that $R_2 \geq R_1$, one finds the first SdS NEC condition 
\be\label{eq:SdS NEC1}
\ell_\lf \geq \ell_\ri~.
\ee
Similarly, by multiplying the first and second inequalities in~\reef{eq: cond SdS ineq} by $R_1^{-d}$ and $R_2^{-d}$, respectively, and eliminating the cosmological constants by adding the resulting expressions, one obtains the second SdS NEC condition
\be\label{eq:SdS NEC2}
\mathcal{M}_\lf \geq \mathcal{M}_\ri~.
\ee
Note that the equalities ($\ell_\lf = \ell_\ri$ and $\mathcal{M}_\lf = \mathcal{M}_\ri$) are simultaneously saturated and that is only for the trivial solution $\sigma_1 = \sigma_2 = 0$. Therefore, the matching background condition cannot be satisfied together with the NEC except in this trivial case.

\paragraph{Imposing the matching background condition.} If we require a single value for the cosmological constant, i.e. $\ell_\text{L} = \ell_{\text{R}}$, we can look for solutions that satisfy the AANEC instead of the NEC.
Here, we consider configurations where the shell escaping to infinity, $\Sigma_2$, satisfies the NEC, while the shell collapsing into the black hole violates it. These configurations correspond to cases where
\begin{equation}
\label{eq:conditionM}
    \mathcal{M}_\text{L} < \mathcal{M}_{\text{R}}~.
\end{equation}

In what follows, we show that this condition on the masses also ensures that the Penrose diagram of SdS becomes taller rather than fatter, compared to the case with no shells. We interpret this as evidence that every complete null geodesic in the spacetime is no longer achronal after the shell insertions.  Recall that a convenient set of Penrose diagram coordinates is given by the redefinitions:
\begin{equation}
    \tanh^{-1}(u\pm v)= t\pm \rho~,
\end{equation}
where $\rho(r)$ is defined via the equation $\text{d}\rho = \text{d}r/f(r)$. The coordinates $(t,\rho)$ span one of static patches in the Penrose diagram, the range of $t$ is from $(-\infty, \infty)$, so is the range of $\rho$, $\rho = - \infty$ and $\rho = \infty$ at the black hole horizon and cosmological horizons, respectively.  The divergences in $\rho$ arise from the behavior of the integral of $1/f(r)$ near the horizons.
If we agree to map $r_0$ to $\rho_0 =0$, with $r_0$ in between the black hole and cosmological horizon, then the line $\rho_0 =0$ maps to the interval $u\in[-1,1]$, while $v=0$. This implies that the height of the SdS Penrose diagram is two. Similarly, the line $t=0$ maps to the interval $v\in[-1,1]$ with fixed $u=0$. Since SdS consists of two copies of the static patch, we conclude that the horizontal size of the Penrose diagram is four. 

We now consider a pair of shells within the static patch. The shells divide the $t=0$ line into three segments: $(r_\text{b}, r_1)$, $(r_1,r_2)$ and  $(r_2,r_{\text{c}})$. Here, $r_1$ and $r_2$  are the locations of the innermost and outermost shells, respectively. The locations of the \emph{right} black hole and cosmological horizons are $r_{\text{b}}$ and $r_\text{c}$ .   The question is whether the width of the spacetime decreases when the pair of shells are introduced in the geometry. The relevant quantity to compute is the amount of length, measured in the  $\rho$ coordinate, lost or gained during the cutting and gluing of the spacetime. For the Penrose diagram to become taller, the net gain should be negative, so we must have that 
\begin{equation}
    \int_{r_1}^{r_2} \frac{\text{d}r}{f_\text{L}(r)} <   
    \int_{r_1}^{r_2} \frac{\text{d}r}{f_\text{R}(r)}~.
\end{equation}
For equal de Sitter radii, $\ell_{\text{R}} = \ell_{\text{L}}$, this equation is equivalent to  \eqref{eq:conditionM}.

A taller Penrose diagram means that \textit{radial} null geodesics (i.e. geodesics with constant angular coordinates) are no longer achronal in the perturbed spacetime. Nonetheless, this is not enough to conclude that the spacetime satisfies the AANEC, as there may be non-radial null geodesics which remain achronal and cross the NEC violating shell. In this paper, we interpret the increased height of the Penrose diagram as evidence that the AANEC is satisfied for these spacetimes, leaving a formal proof of this statement for future work.

Configurations of this type are interesting for two main reasons. First, although $\Sigma_1$ violates the NEC by having a negative energy density, operators that reduce the energy of black hole microstates have been studied in the AdS literature \cite{Papadodimas:2017qit}. Despite being fine-tuned to lower the energy of the specific state $\ket{\psi_0}$, such operators exist in the dual CFT and are also generally present in isolated quantum statistical systems. In the case of SdS, one can consider a setup where the operator preparing the shell $\Sigma_1$ reduces the energy of the spacetime. The fact that $\mathcal{M}_\text{L}$ is smaller than $\mathcal{M}_\text{R}$ can be interpreted as the black hole losing mass as the shell crosses the event horizon (in a process that resembles black hole evaporation through particle pair production). 
The second motivation for considering these shell configurations is that they may help clarify the origin of the entropy of empty de Sitter space at fixed cosmological constant, by taking the limit $\mathcal{M}\to 0$ in which the black hole disappears.

\subsection{Solving the equations of motion}\label{sec: explicit sol}

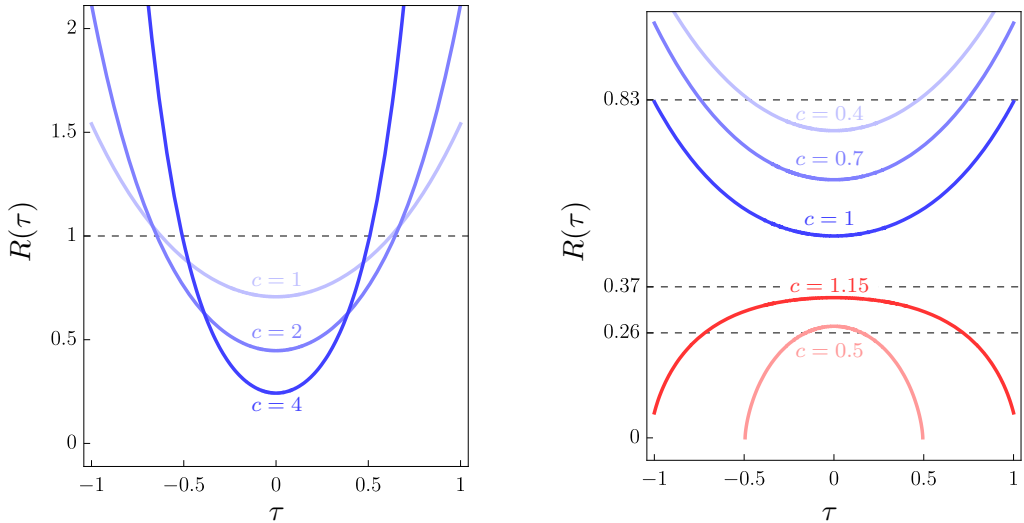
\begin{figure}[t!]
   \centering
\begin{tabular}{cc}
\scalebox{1}{
\raisebox{-40pt}{
\begin{tikzpicture}[line width=1. pt, scale=1.5]
\node[scale=0.66,black] at (0,0) {\includegraphics{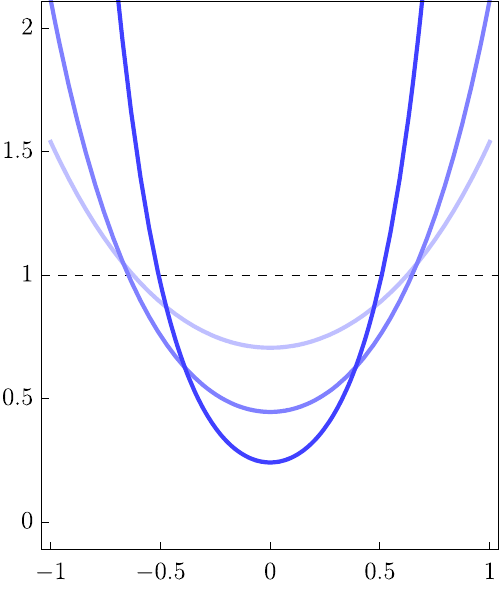}};
\node[scale=1, rotate=0] at (0.15,-0.22) {{\scriptsize
$\textcolor[rgb]{0.75, 0.75, 1}
{c=1}$}};
\node[scale=1, rotate=0] at (0.15,-0.68) {{\scriptsize
$\textcolor[rgb]{0.5, 0.5, 1}
{c=2}$}};
\node[scale=1, rotate=0] at (0.15,-1.33) {{\scriptsize
$\textcolor[rgb]{0.25, 0.25, 1}
{c=4}$}};
\node[scale=1,black, rotate=0] at (0.15,-2.3) {{\small $\tau$}};
\node[scale=1,black, rotate=90] at (-2.1,0.2) {{\small $R(\tau)$}};
\end{tikzpicture}
}
}&
\scalebox{1}{
\raisebox{-40pt}{
\begin{tikzpicture}[line width=1. pt, scale=1.5]
\node[scale=0.66,black] at (0,0) {\includegraphics{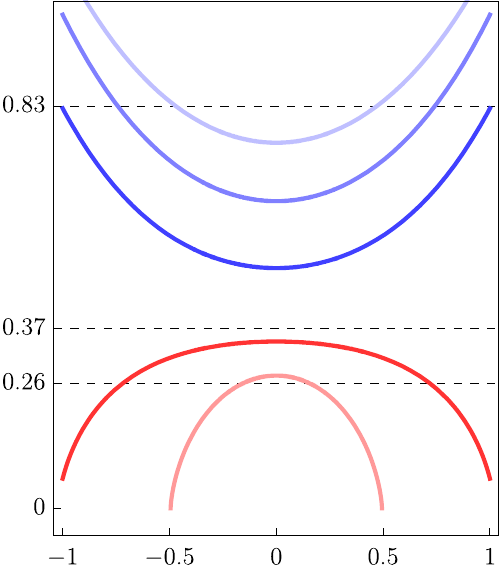}};
\node[scale=1,black, rotate=0] at (0.15,-2.3) {{\small $\tau$}};
\node[scale=1,black, rotate=90] at (-2.1,0.2) {{\small $R(\tau)$}};
\node[scale=1, rotate=0] at (0.15,1.25) {{\scriptsize
$\textcolor[rgb]{0.75, 0.75, 1}
{c=0.4}$}};
\node[scale=1, rotate=0] at (0.15,0.85) {{\scriptsize
$\textcolor[rgb]{0.5, 0.5, 1}
{c=0.7}$}};
\node[scale=1, rotate=0] at (0.15,0.32) {{\scriptsize
$\textcolor[rgb]{0.25, 0.25, 1}
{c=1}$}};
\fill[white](0.15-0.4,-0.3)--(0.15+0.43,-0.3)--(0.15+0.43,-0.25)--(0.15-0.4,-0.25);
\node[scale=1, rotate=0] at (0.15,-0.28) {{\scriptsize
$\textcolor[rgb]{1, 0.2, 0.2}
{c=1.15}$}};
\node[scale=1, rotate=0] at (0.15,-0.85) {{\scriptsize
$\textcolor[rgb]{1, 0.6, 0.6}
{c=0.5}$}};
\end{tikzpicture}
}
}
\end{tabular}
   \vspace{-5pt}
\caption{
Numerical solutions of the equation of motion~\reef{eq: EofM} for the positions of the end-of-the-world brane ($w=-1$) in de Sitter and Schwarzschild–de Sitter spacetimes for various values of the defect energy parameter $c$. Here we set $\ell=1$, and $d=3$. Red and blue curves correspond to collapsing and runaway defects, respectively. The left plot shows brane solutions in de Sitter space where the dotted line indicates $r_\text{c}$, the radius of the cosmological horizon.
The right plot shows end-of-the-world brane solutions in Schwarzschild–de Sitter space for $\mathcal{M}=0.25$. The dotted horizontal lines indicate, in order of increasing radius: the black hole horizon $r_{\text{b}}$, the splitting point between collapsing and runaway shells, and $r_{\text{c}}$. For this set of variables, $c=4\pi G \sigma $.
} 

\label{fig:numerics1}
\end{figure}

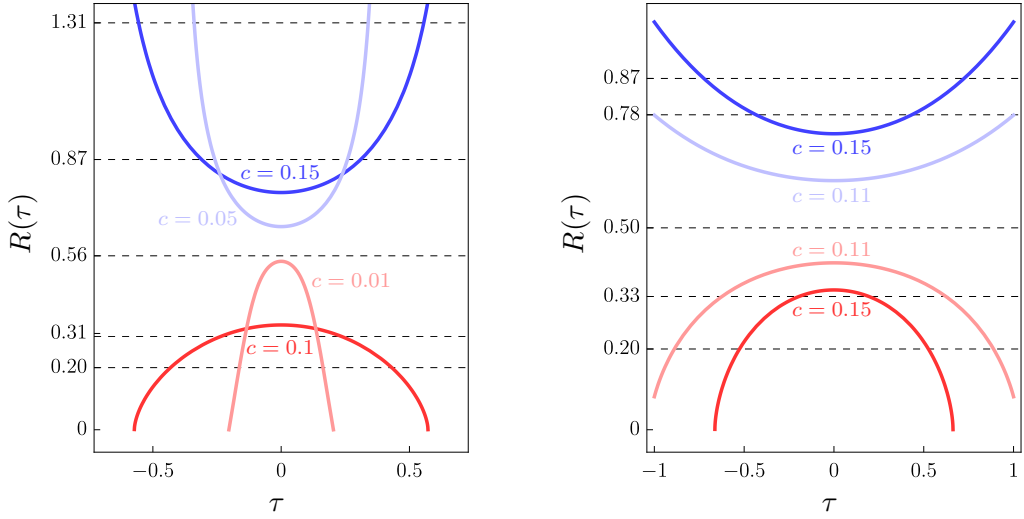
\begin{figure}[h!]
   \centering
\begin{tabular}{cc}
\scalebox{1}{
\raisebox{-33pt}{
\begin{tikzpicture}[line width=1. pt, scale=1.5]
\node[scale=0.66,black] at (0,0) {\includegraphics{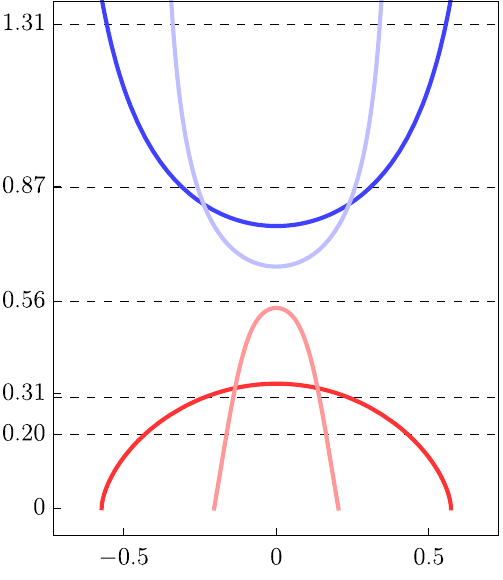}};
\node[scale=1,black, rotate=0] at (0.15,-2.3) {{\small $\tau$}};
\node[scale=1,black, rotate=90] at (-2.1,0.2) {{\small $R(\tau)$}};
\node[scale=1, rotate=0] at (-0.55,0.25) {{\scriptsize
$\textcolor[rgb]{0.75, 0.75, 1}
{c=0.05}$}};
\node[scale=1, rotate=0] at (0.18,0.65) {{\scriptsize
$\textcolor[rgb]{0.25, 0.25, 1}
{c=0.15}$}};

\node[scale=1, rotate=0] at (0.18,-0.9) {{\scriptsize
$\textcolor[rgb]{1, 0.2, 0.2}
{c=0.1}$}};
\node[scale=1, rotate=0] at (0.81,-0.3) {{\scriptsize
$\textcolor[rgb]{1, 0.6, 0.6}
{c=0.01}$}};
\end{tikzpicture}
}
}&
\scalebox{1}{
\raisebox{-33pt}{
\begin{tikzpicture}[line width=1. pt, scale=1.5]
\node[scale=0.66,black] at (0,0) {\includegraphics{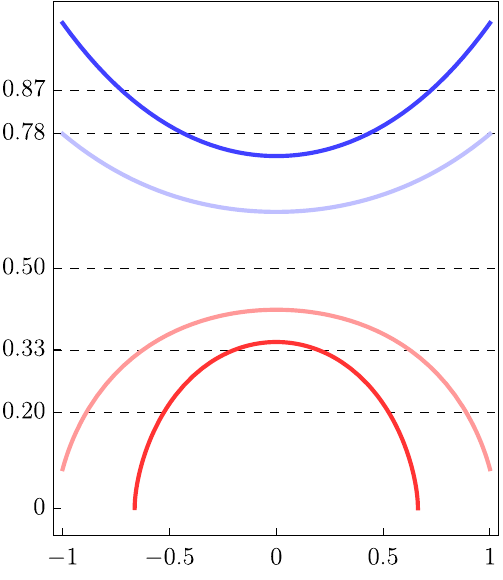}};
\node[scale=1,black, rotate=0] at (0.15,-2.3) {{\small $\tau$}};
\node[scale=1,black, rotate=90] at (-2.1,0.2) {{\small $R(\tau)$}};

\node[scale=1, rotate=0] at (0.17,0.45) {{\scriptsize
$\textcolor[rgb]{0.75, 0.75, 1}
{c=0.11}$}};
\node[scale=1, rotate=0] at (0.17,0.88) {{\scriptsize
$\textcolor[rgb]{0.25, 0.25, 1}
{c=0.15}$}};
\node[scale=1, rotate=0] at (0.17,-0.56) {{\scriptsize
$\textcolor[rgb]{1, 0.2, 0.2}
{c=0.15}$}};
\node[scale=1, rotate=0] at (0.17,-0.03) {{\scriptsize
$\textcolor[rgb]{1, 0.6, 0.6}
{c=0.11}$}};
\end{tikzpicture}
}
}
\end{tabular}
   \vspace{-5pt}
\caption{
Numerical solutions for the positions of shells of dust ($w=0$) gluing two Schwarzschild-de Sitter spaces for several values of the energy parameter $c$ in spacetime dimension $d=3$.
The dotted lines are in increasing order: $r_{\text{b}}^{\text{R}}$, $r_{\text{b}}^{\text{L}}$, the splitting point, $r_{\text{c}}^{\text{R}}$ and $r_{\text{c}}^{\text{L}}$. The left plot shows solutions that  satisfy the NEC.
Here, the corresponding mass parameters are $\mathcal{M}_\lf=0.3$  and $\mathcal{M}_\ri=0.2$. To satisfy the NEC, the cosmological constants of the two sides have to be different: $\ell_\lf=1.5$ and $\ell_\ri=1$. 
In the right plot shells in red violate the NEC, while shells in blue satisfy the NEC. Here, the corresponding mass parameters are $\mathcal{M}_\lf=0.2$  and $\mathcal{M}_\ri=0.3$, and the cosmological constants are identical, $\ell_\lf=\ell_\ri=1$.
} 

\label{fig:numerics2}
\end{figure}

We would like to establish the existence of solutions to the equations describing the location of the defect for a continuous family of defect parameters, namely the equation-of-state parameter $w$ and the energy parameter $c$. The energy parameter $c$ is given by~\reef{q: beta of R} and~\reef{eg:sigma and p II} and is related to energy density $\sigma$ via~\reef{eq: sigma and p} and~\reef{eg:sigma and p II}. Finding an analytical and closed-form solution to the equation of motion~\reef{eq: EofM} is cumbersome. Nevertheless, in this section we verify, through numerical analysis, that such solutions do indeed exist.
We are interested in two classes of solutions. \emph{Collapsing defects} are those that arise from the white hole singularity in the past and collapse into the black hole singularity in the future. These defects only make sense in SdS space and not in de Sitter, where there are no singularities. \emph{Runaway defects}, on the other hand, go from $\mathcal{I}^-$ to $\mathcal{I}^+$. These defects exist in both, SdS and de Sitter spacetimes. For clarity, we first discuss these two classes in the context of end-of-the-world branes; the generalization to shells is discussed later. Numerical solutions are presented in Figures~\ref{fig:numerics1} and~\ref{fig:numerics2}.

\paragraph{Collapsing defects.} 
For this type of defects, there exists proper times $\tau_2>\tau_1$ such that $R(\tau_1) = 0 = R(\tau_2)$. At $\tau_1$ the defect emerges from the white hole and terminates at $\tau_2$ in the black hole. We place the turning point where $\dot R = 0$ to be at $\tau = 0$, we also set $T(0) = 0$ and $\tau_2 = -\tau_1$. Solutions are uniquely determined by the initial condition $R(0) =  R_*$, which must satisfy the algebraic equation  
\begin{equation}
\label{eq:algebraicEq}
    \gamma c \,R_*^{\alpha} =  \sqrt{f(R_*)}~. 
\end{equation}
This formula follows from \eqref{q: beta of R} by requiring $\dot R = 0$. Here, $\alpha = (1-d)w+(2-d)$. The sign of $\gamma = \pm1$ depends on whether we want to keep the large ($\gamma=-1$) or the small ($\gamma=+1$) side of the spacetime. Solutions only exist when $\gamma c \geq 0$. 

For any given values of $M$ and $\ell$ in $f(R)$, we can always find a non-empty range of values for $\gamma c$ where solutions to \eqref{eq:algebraicEq} exist.
When solutions do exist, they always appear in pairs $R_*^{(1)}$, $R_*^{(2)}$ with $R_*^{(2)} \geq R_*^{(1)}$. The choice $R(0) = R_*^{(2)}$ corresponds to a runaway brane, while $R(0) = R_*^{(1)}$ corresponds to a collapsing brane.\footnote{The unique solution where $R_*^{(2)} = R_*^{(1)}$ is a constant function $R(\tau) = R_*^{(1)}$ which sits in between the cosmological and black hole horizons.}
A numerical solution for the collapsing brane can be obtained by integrating the equation:
\begin{equation}
    \dot R(\tau) = -\,\text{sgn}(\tau) \sqrt{c^2 R^{2\alpha} - f(R)}, \quad \text{with} \quad R(0) =  R_*^{(1)}~.
\end{equation}
The sign of the square root is chosen such that $\dot R(\tau) < 0$ for $\tau>0$ and $\dot R(\tau) > 0$ for $\tau<0$; consistent with a brane emerging from the white hole and later entering the black hole.

\paragraph{Runaway defects.} In contrast to collapsing defects, these defects satisfy, $\dot R(\tau) > 0$ for $\tau>0$ and $\dot R(\tau) < 0$ for $\tau <0$, with $\dot R(0)=0$. As discussed above, the initial condition $R(0) = R_*$ must solve~\reef{eq:algebraicEq} and when solutions to this equation exist $R_*$ corresponds to the largest root $R_*^{(2)}$. The location of the runaway brane can be solved numerically by integrating 
\begin{equation}
    \dot R(\tau) =  \text{sgn}(\tau) \sqrt{c^2 R^{2\alpha}-f(R)}, \quad \text{with}\quad R(0) = R_*^{(2)}~.
\end{equation}

By finding the solutions to $R(\tau)$, one can immediately derive $T(\tau)$ using~\reef{eq: beta}.
One might worry that using the static patch coordinates constrains our ability to identify these defects beyond this coordinate system—particularly since $T(\tau)$ diverges as $R(\tau)$ approaches either the cosmological or black hole horizon. However, this is not a real limitation: the equations of motion can always be translated to alternative coordinate systems, such as global de Sitter space.
Below, we present a numerical evaluation of end-of-the-world branes that smoothly cross the cosmological horizon in the Penrose diagram of global de Sitter space. For this evaluation, we use the metric 
\be
ds^2=  \frac{-d\eta^2+d\Omega_d^2}{\cos^2\eta}~,
\ee
where $\eta$ is the global conformal time $\eta\in (-\frac{\pi}{2},\frac{\pi}{2})$ on the vertical axis and the azimuthal angle is on the horizontal axis.  Again, observer is sitting at the south pole on the right side of the diagram:
\begin{equation}
    \vcenter{\hbox{
\begin{tikzpicture}[line width=1. pt, scale=1.5]
\node[scale=0.66,black] at (0,0) {\includegraphics{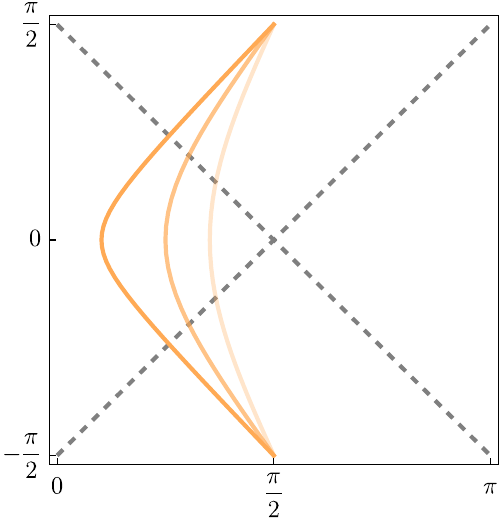}};
\end{tikzpicture}
    }}.
\end{equation}
Here, we set $\ell=1$, $w=-1$ and $d=3$.
The three shell worldvolumes shown in the picture correspond to the energy parameters $c=0.5$, $c=1$ and $c=3$ from lighter to darker orange (right to left).

\paragraph{Shells.} The analysis for shells is very similar to the one for branes. For fixed values of $M_{\text{L}}$, $\ell_\text{L}$ and $M_{\text{R}}$, $\ell_\text{R}$ shell solutions always exist for a range of $c$. The solutions are completely characterized by the roots of the algebraic equation:
\begin{equation}
\label{eq:roots2}
    c R^{\alpha}_* = \gamma_\text{R} \sqrt{f_\text{R}(R_*)} + \gamma_\text{L} \sqrt{f_\text{L}(R_*)}~.
\end{equation}
For collapsing shells satisfying the NEC we are interested in solutions with $\gamma_\text{R} = 1$ and $\gamma_\text{L}=-1$. Runaway shells have the opposite convention. For these values of $\gamma$, when solutions exist there is only one real root $R_*$, which fixes the initial conditions needed to solve the differential equation. When considering shells with equal de Sitter radii $\ell_\text{L} = \ell_\text{R}$, where the collapsing shell violates the NEC but the runaway shell does not, the algebraic equation \reef{eq:roots2} admits two solutions: one corresponding to the collapsing shell and the other to the runaway shell.

\section{Microstate counting from Lorentzian wormholes}
\label{sec:wormholes}

In this paper, we use the gravitational  path integral to define a \emph{coarse-grained} inner product between two quantum states, specified by boundary conditions $i,j$ on the asymptotic boundaries $\mathcal{I}^\pm$ of de Sitter space. Schematically we have that,
\begin{equation}
\label{innerpp}
\overline{\bra{\psi_i}\ket{\psi_j}} = \sum_{M} 
\int \mathcal{D}g_{\mu\nu} \; 
\vcenter{\hbox{
\begin{tikzpicture}
\begin{scope}[scale=1.3]
    \draw[black!50,dashed,line width=1pt,fill = black!10] (-1,1.5) -- (-1.5,1.5) -- (-1.5,-1.5) -- (-1,-1.5);
    \draw[black!50,dashed,line width=1pt,fill = gray!20] (-1,1.5) -- (1,1.5) -- (1,-1.5) -- (-1,-1.5);
    \draw[thick] (-1.5,-1.5) -- (1,-1.5);
    \draw[thick] (-1.5,1.5) -- (1,1.5);
    \fill[black] (-1,-1.5) circle (0.075);
    \fill[black] (-1,1.5) circle (0.075);
    \node at (-0.8,-1.2) {$j$};
    \node at (-0.9,1.2) {$i$};
    \node at (0,-1.8) {$\mathcal{I}^-$};
    \node at (0,1.8) {$\mathcal{I}^+$};
    \node at (-0.25,0) {{\large$e^{iS[g]}$}};
\end{scope}
\end{tikzpicture}}}
\end{equation}
where $\sum_M$ represents a sum over different topologies, and on each topology we do the path integral over all metrics $g$ which can be defined on $M$. Each combination $(M,g)$ is then weighted by the action $S$ described in \eqref{eq: scenario I action} and \eqref{eq: scenario II action}. 
The boundary conditions of interest are those in which the spacetime asymptotically approaches empty de Sitter with either an end-of-the-world brane or a thin shell insertion. 

The path integral \eqref{innerpp} is dominated by the on-shell saddles described in Section~\ref{sec:shells}. These saddles give an inner product in which distinct states are approximately orthogonal. Crucially, the gravitational path integral only provides an approximate evaluation of the inner product, which is why we denote \eqref{innerpp} as an average. The exact inner product between two states is subject to non-pertubative corrections of the form:
\begin{equation}
   \frac{ \bra{\psi_i}\ket{\psi_j}}{
   \sqrt{
   \bra{\psi_i}\ket{\psi_i}
   \bra{\psi_j}\ket{\psi_j}
   }
   } = \delta_{ij} + R_{ij}~,
\end{equation}
where the variables $R_{ij}$ can be thought of as a set of pseudorandom numbers with zero mean. Strictly speaking, these variables are not truly random: in the UV theory, they are in principle computable, but they are  inaccessible at the semiclassical level. Nonetheless, the gravitational path integral has statistical information about their moments, which are encoded in spacetimes with  non-trivial bulk topology. The goal of this section is to construct these topologies, which correspond to Lorentzian wormholes, and compute the moments of these random variables.

The motivation for studying these moments is that they enable one to estimate the dimensionality of the Hilbert space of quantum gravity. Once the moments are determined, they can be used to compute the rank of the Gram matrix
\begin{equation}
G_{ij} = \bra{\psi_i}\ket{\psi_j}~,
\end{equation}
which characterises the dimensionality of the space spanned by the semiclassical vectors $\ket{\psi_i}$. This rank is then interpreted as the dimensionality of the Hilbert space corresponding to de Sitter (and Schwarzschild–de Sitter) spacetime.

This section is organized as follows. In Subsection~\ref{crotches}, we outline how the gravitational path integral can be used to estimate the moments of the variables $R_{ij}$. A crucial element in this computation is the role of topology change in Lorentzian signature, which we briefly review. In Subsection~\ref{sec:Lwormholes}, we then describe the wormhole geometries in de Sitter and Schwarzschild–de Sitter spacetimes and present explicit expressions for their corresponding state overlaps. Finally, in Subsection~\ref{microcounting}, we use these overlaps to determine the dimensionality of the Hilbert space associated with these spacetimes.

\subsection{Topology change in Lorentzian signature}
\label{crotches}

The moments of the random variables $R_{ij}$ can be analyzed through the connected overlap
\begin{equation}
\label{overlap22}
  \overline{
    \bra{\psi_j}\ket{\psi_i}
    \bra{\psi_i}\ket{\psi_j}
    }^{\text{ conn}} \;\;\equiv \;\; \overline{
    \bra{\psi_j}\ket{\psi_i}
    \bra{\psi_i}\ket{\psi_j}
    } -   \overline{
    \bra{\psi_j}\ket{\psi_i}}\;\;\overline{
    \bra{\psi_i}\ket{\psi_j}
    }~.
\end{equation}
This quantity can be computed by evaluating the gravitational path integral with four asymptotic boundaries. Since we are interested in the connected moment, the path integral includes bulk topologies which are fully connected through the bulk:
\begin{equation}
    \vcenter{\hbox{
    \begin{tikzpicture}
    \begin{scope}[scale=1.2]
        \draw[black!50,dashed,line width=1pt, fill=black!10] 
        (-2.5,1.5) --
        (-1,1.5) to[out=-40,in=220] (1,1.5)
        --(2.5,1.5)
        --(2.5,-1.5)
        --(1,-1.5)
        to[out=140,in=40] (-1,-1.5)
        -- (-2.5,-1.5)
        -- (-2.5,1.5);
        \draw[very thick] (-2.5,1.5) -- (-1,1.5);
        \draw[very thick] (1,1.5) -- (2.5,1.5);
        \draw[very thick] (-2.5,-1.5) -- (-1,-1.5);
        \draw[very thick] (1,-1.5) -- (2.5,-1.5);
        \node[anchor=center] at (0,0) {\large ?};
        \fill[black] (-1.75,1.5) circle (0.075);
        \fill[black] (1.75,1.5) circle (0.075);
        \fill[black] (-1.75,-1.5) circle (0.075);
        \fill[black] (1.75,-1.5) circle (0.075);
        \node at (-1.75,1.85) {$j$};
        \node at (-1.75,-1.85) {$i$};
        \node at (1.75,1.85) {$i$};
        \node at (1.75,-1.85) {$j$};
    \end{scope}
    \end{tikzpicture}
    }}.
\end{equation}
As discussed in Section \ref{sec:setup}, disconnected topologies could also contribute to the connected moment. Nevertheless, these configurations are excluded in the presence of an observer, since we require that there exists a curve connecting each ket to its corresponding bra. Hence, the relevant saddles in the presence of an observer are fully connected.
In ``real'' Lorentzian gravity, however, there are no saddles connecting all four boundaries because the nondegeneracy condition on the metric prohibits topology change in Lorentzian signature \cite{Geroch:1967fs,Geroch:1979uc,Hawking:1991nk,Sorkin:1997gi}.  Nevertheless, there are complex geometries in the path integral that realize this connectivity. These geometries are characterized by complex-valued metrics $g_{\mu\nu}$.
Their fundamental building blocks---and, more generally, the building blocks for topology change in Lorentzian signature---are the so-called ``crotch" geometries introduced by Louko and Sorkin \cite{Louko:1995jw}.

\subsection*{The crotch geometry}

Before turning into wormholes in de Sitter, we first review the structure of the crotch geometry and quantify its contribution to the Lorentzian path integral. 

\paragraph{Crotches in Minkowski.}
Louko and Sorkin~\cite{Louko:1995jw} showed that certain smooth complex geometries admit a reinterpretation as Lorentzian spacetimes with ``crotch singularities."
To make this relation concrete, let us explain their argument in the setting of two-dimensional Minkowski spacetime. 
Start with the following coordinate systems on the left and right patches of a double cover of Minkowski space:
\vspace{0.25cm}
\begin{equation}
\label{spacecrtoch}
    \vcenter{
    \hbox{
    \begin{tikzpicture}
    \begin{scope}[scale=1.3]
        \draw[thick] (2,2) -- (-2,2) -- (-2,-2) -- (2,-2) -- (2,2);
       \begin{scope}[xscale=1]
        \draw[thick,decorate, decoration={zigzag, segment length=4pt, amplitude=1pt}] (0,0) -- (-2,0);
\draw[red,line width=1.1pt,->,domain=3.05*pi:5*pi,samples=200,smooth,variable=\t]
        plot ({0.05*\t*cos(\t r)-0.1}, {-0.05*\t*sin(\t r)-0.05});
        \end{scope}
        \fill[black] (0,0) circle (0.075);
        \node at (0,1.5) {$ds_\lf^2 = -\dd t_\lf^2 + \dd x^2_\lf$};
        \node at (2.7,0) {$\phantom{t}$};
        \node at (0,-2.5) {$\phantom{x}$};
    \end{scope}
    \end{tikzpicture}
    }
    }\quad \quad 
        \vcenter{
    \hbox{
    \begin{tikzpicture}
    \begin{scope}[scale=1.3]
        \draw[thick] (2,2) -- (-2,2) -- (-2,-2) -- (2,-2) -- (2,2);
        \begin{scope}[xscale=1]
        \draw[thick,decorate, decoration={zigzag, segment length=4pt, amplitude=1pt}] (0,0) -- (-2,0);
        \draw[red,line width=1.1pt,<-,domain=3.05*pi:5*pi,samples=200,smooth,variable=\t]
        plot ({0.05*\t*cos(\t r)-0.1}, {0.05*\t*sin(\t r)});
        \end{scope}
        \fill[black] (0,0) circle (0.075);
        \node at (0,1.5) {$ds_\ri^2 = -\dd t_\ri^2 + \dd x^2_\ri$};
        \draw[thick, ->] (2.5,-1.5) -- (2.5,1.5);
        \node at (2.7,0) {$t$};
        \draw[thick, ->] (-1.5,-2.3) -- (1.5,-2.3);
        \node at (0,-2.5) {$x$};
    \end{scope}
    \end{tikzpicture}
    }
    }
\end{equation}
The manifold described here is known as a \emph{crotch geometry}. As it is illustrated by the  red arrow, the two covers of Minkowski are identified along a cut which starts at  the origin ($t=x= 0$) and extends out to infinity. Since we are gluing two identical copies along this cut, the resulting spacetime is smooth everywhere except at the origin, where it has a conical singularity with total opening angle $4\pi$. 

To study the effects of this singularity, we need a prescription that regularizes the geometry. Following~\cite{Louko:1995jw}, we implement the coordinate transformation
\begin{equation}
\label{diff}
    t_\lf + i x_\lf  = \frac{1}{2}(u+iv)^2~ \text{ with }~ u\geq-v, \quad \text{and}\quad 
     t_\ri + i x_\ri = \frac{1}{2}(u+iv)^2 ~\text{ with }~ u\leq-v~,
\end{equation}
which describes the two patches in the single $uv$-plane. The metric in this new coordinate system has the form 
\begin{equation}
   ds^2= (u^2+v^2)(\dd u^2 + \dd v^2) -2 (u \dd u - v\dd v )^2~,
\end{equation}
where the singularity is at $u=v=0$. The prescription to remove the singularity at the origin is  to introduce the complex regulator $i\epsilon$,
\begin{equation}
\label{complexreg}
   ds^2=  (u^2+v^2 + i \epsilon)(\dd u^2 + \dd v^2) -2 (u \dd u - v\dd v )^2~.
\end{equation}
The regulator deforms the singular Lorentzian metric into a smooth complex metric that satisfies the Kontsevich–Segal–Witten (KSW) criterion \cite{Witten:2021nzp,Kontsevich:2021dmb}. The idea is to perform all computations with this regulated metric and take the limit $ \epsilon \rightarrow0$ at the end of the computation.\footnote{
Note that these geometries are off-shell, with or without the regulator, since the conical defect introduces a source in the Riemann tensor that is not accounted for by matter. The implications of this regularization procedure will be discussed later in this section. 
} 

In our case, we wish to evaluate the Lorentzian path integral, so we must determine how the singularity at $u=v=0$ affects the Einstein–Hilbert term $\sqrt{-g}R$  in the action. A direct computation shows that the Ricci scalar $R$ times the measure factor $\sqrt{-g}$ is given by:
\begin{equation}
    \sqrt{-g} R \;\dd u \dd v  = 2 i \left[\frac{\epsilon \chi}{(\chi^2+\epsilon^2)^{3/2}}\right] \dd \chi\dd \varphi~,
\end{equation}
where we have defined the radial coordinates 
\begin{equation}
    \label{sphicoord}
u =  \sqrt{\chi}\cos\varphi \quad\text{and}\quad v =\sqrt{\chi}\sin\varphi~,
\end{equation}
with $\chi>0$ and $\varphi \in [0,2\pi)$. In the limit $\epsilon\rightarrow0$, the expression in brackets converges to a Dirac delta function:
\begin{equation}
\label{deltafunction}
    \sqrt{-g} R \;\dd u\dd v \;\;\underset{\epsilon\rightarrow 0}{=}\;\; 2 i \, \delta(\chi) \,\dd \chi \dd \varphi~.
\end{equation}
Upon evaluating the Einstein–Hilbert action $iS_{\text{EH}}$, the conical singularity gives an overall \emph{real} contribution of the form:
\begin{equation}
\label{realcont}
    iS_{\text{EH}} = \frac{i}{16\pi G}\int\sqrt{-g}R = -\frac{1}{4G}~. 
\end{equation}
The important observation is that a real term appears in an otherwise purely imaginary action. Said differently, for any smooth Lorentzian geometry $iS_{\text{EH}}$ is purely imaginary and contributes only to a phase to the Lorentzian path integral. By contrast, with the complex regulated metric $iS_{\text{EH}}$ acquires a real contribution of the form given by \eqref{realcont}.   

As explained in \cite{Blommaert:2023vbz, Marolf:2022ybi}, the result in \eqref{realcont} can be viewed as the analytic continuation of the Gauss–Bonnet theorem, which is valid only in Euclidean signature. In Euclidean signature, surplus angles are sourced by delta functions in the scalar curvature, and for a singularity of opening angle size $2\pi a$ situated at $p$, we have that 
\begin{equation}
    \sqrt{g} R = 4\pi(1-a) \delta(p)~.
\end{equation}
A conical surplus of angle size $4\pi$, i.e. $a=2$, corresponds to a term $-4\pi\delta(p)$ inside the scalar curvature. This is precisely the result obtained from \eqref{deltafunction} after integrating the angular coordinate $\varphi$ and using the relation $\sqrt{g} = i \sqrt{-g}$. 

In addition, a particularly relevant case for this work is when $n$ copies of the base manifold are glued together cyclically. The resulting space is an $n$-fold cover of the original geometry with a conical singularity at the origin and opening angle $2\pi n$. This singularity corresponds to a term in the  Einstein–Hilbert action of the form 
\begin{equation}
     \sqrt{-g} R =  4\pi i (n-1)\delta(p)~.
\end{equation}

\paragraph{Crotches in curved spacetime.} 
In more than two dimensions, the conical singularity lifts to a codimension-two defect $\Sigma$. In \cite{Marolf:2022ybi}, it is shown that this defect contributes to the Einstein-Hilbert action as a Dirac delta function localized at $\Sigma$:
\begin{equation}
\label{eqdelt}
     \sqrt{-g} R =  4\pi i (n-1)\delta(\Sigma) + \text{regular terms}.
\end{equation}
Here, $n$ denotes the number of sheets in the cover. By ``regular terms", we mean the smooth background contributions to the Ricci scalar $R$ coming from the curvature of the spacetime. Evaluating the Einstein–Hilbert action by integrating over the manifold, the conical singularity \eqref{eqdelt} yields the term $4\pi i (n-1)A(\Sigma)$ in the action, where $A(\Sigma)$ is the area of the defect. Accordingly, the Lorentzian path integral picks an exponential suppression of the form
\begin{equation}
     \exp[-(n-1)\frac{A(\Sigma)}{4G}]~. 
\end{equation}

More precisely, let $M$ denote the $n$-fold cover over the geometries $M_j$, $j=1,\dots,n$, along a codimension-two defect $\Sigma$. Then Einstein-Hilbert action of the manifolds $M_j$ and their cover $M$ are related by: 
\begin{equation}
\label{action}
   iS_{\text{EH}}[M] =  -(n-1)\frac{A(\Sigma)}{4G} + i\sum_{j=1}^n S_{\text{EH}}[M_j]~. 
\end{equation}
In appendix \ref{app:crotches}, we reproduce the delta function in \eqref{eqdelt} explicitly for $n=2$, with a base manifold endowed with a metric:
\begin{equation}
    ds^2=-f(r)\dd t^2 + \frac{\dd r^2}{f(r)}+r^2\dd\Omega^2_{d-1}~.
\end{equation} 
matching with the coordinate system we are using in this paper for de Sitter and Schwarzschild-de Sitter.

\paragraph{Evaluation of the Lorentzian path integral.}
Evaluating the path integral on these topologies is subtle because the geometries described above do not satisfy the Einstein equations at the defect $\Sigma$. At this surface, a delta-function source appears in the Riemann tensor which is not accounted for by matter. Despite this issue, these geometries and their role in the path integral have been studied in recent years. In this paper, we follow the prescription proposed in \cite{Marolf:2022ybi}.

The gravitational path integral can be viewed as the infinite-dimensional limit of an integral over a set of complex variables $x_i$, for $i=1,\dots,k$, where $k\rightarrow\infty$. The proposal of \cite{Marolf:2022ybi} is to identify one of these integrals with the integral over the area of the codimension-two defect, $x_1 = A[\Sigma]$. The remaining integrals are then performed over metrics where the area $A[\Sigma]$ of the defect is kept fixed. 
The saddle points of these integrals, referred to as ``fixed-area" saddles,
are not required to satisfy the Einstein equations at the defect but must do so away from it.
These correspond to the crotch geometries discussed earlier. Following this prescription, the path integral over the $n$-fold cover $M$ takes the form:
\begin{equation}
\label{pathint2}
    \int \dd A \;\int_{A[\Sigma] = A} \mathcal{D}g_{\mu\nu} \;\; e^{iS[M]}~,
\end{equation}
where the path integral over the manifold $M$ is restricted to metrics for which the area of the defect is fixed to $A[\Sigma] = A$. Moreover, these metrics are allowed to be singular at $\Sigma$. The next step is to use a saddle-point approximation (which in principle could include loop effects) around the singular crotch geometry.\\[-5pt]

The decomposition of the action \eqref{action}, together with the fact that each $M_j$ in the $n$-fold cover $M$ admits an on-shell geometry satisfying $A[\Sigma] = A$, suggests that the restricted path integral over the $n$-fold cover can be approximated by:
\begin{equation}
\label{PathInt}
\log \;\frac{1}{\mathcal{N}}
    \int_{A[\Sigma] = A} \mathcal{D}g_{\mu\nu} e^{iS[M]} = -(n-1)\frac{A}{4G} + \order{\log G}~, 
\end{equation}
where the normalization factor $\mathcal{N}$ is given by a product of the unrestricted path integrals on the geometries $M_j$
\begin{equation}
\mathcal{N} = \prod_{j=1}^n \int \mathcal{D}g^{(j)}_{\mu\nu} e^{iS[M_j]}~. 
\end{equation}
The final step in evaluating the path integral over the $n$-fold cover $M$ is to use \eqref{PathInt} to perform the integral over the area $A$ in \eqref{pathint2}. 
This integral runs over positive real values of $A$ for which there is a fixed-area saddle. 
In the cases discussed in Subsection \ref{sec:Lwormholes}, the area of these fixed saddles is
parametrized using the radial coordinate $r$ of the static patch, $A = A(r)$, and the integration over $A$ is replaced with an integral over $r$. 
The integration domain for $r$ is compact, with $r \in [r_{\text{min}}, r_{\text{c}}]$, where $r_{\text{min}}$ is set by the geometry, and $r_{\text{c}}$ denotes the radius of the cosmological horizon. 
To evaluate this integral, it is necessary to account for edge effects due to the compact domain of integration. The dominant contribution may arise from $A(r_{\text{min}})$, $A(r_{\text{c}})$, or from an extremal value of $A$ within the integration range. These possibilities will be examined in the next subsection.\\[-5pt]

This prescription for evaluating the path integral has been tested in various two-dimensional AdS setups, where the Lorentzian path integral over geometries with conical singularities has been shown to reproduce familiar results obtained using Euclidean methods. These examples include features like the spectral form factor and the  genus expansion in JT gravity, see e.g. \cite{Held:2024qcl,Blommaert:2023vbz,Colin-Ellerin:2020mva,Colin-Ellerin:2021jev}. It should be noted, however, that \eqref{PathInt} relies strongly on a semiclassical treatment of the path integral. 
In particular, there is an unknown measure factor for the $A$ (or equivalently the $r$) integral that we have not taken into consideration. In what follows, we assume that this measure factor gives only subleading corrections, which is reasonable since it arises from the path integral together with loop corrections around the fixed-area saddle. It would nevertheless be interesting to test these approximations in a more controlled setting, such as in de Sitter JT gravity. We leave a detailed analysis of this point for future work.

\pagebreak

\subsection{Lorentzian wormholes}
\label{sec:Lwormholes}

Having introduced the crotch geometries, we proceed to analyze the wormholes relevant to this paper. In Subsection~\ref{microcounting}, we use these topologies to compute the entropy of de Sitter and Schwarzchild-de Sitter spacetimes. 
The topology for de Sitter shells was already highlighted in \eqref{wormhole}. It is given by  a double cover of de Sitter with a conical surplus located at a radius $r_0$ somewhere in between the outermost shell $r_i$ and the cosmological horizon $r_\text{c}$:
\begin{equation}
\vcenter{
\hbox{
\begin{tikzpicture}
\begin{scope}[ scale=0.5]
\fill[blueKSV1!70] (3,-3) -- (0,0) -- (3,3) -- (3,-3);
\draw[ thick] (-2.3,3) -- (3,3) -- (3,-3) -- (-2.3,-3) -- (-2.3,3);
\node at (1.5,-3.4) {$\mathcal{I}^-$};
\node at (1.5,3.5) {$\mathcal{I}^+$};
\node at (-0.5,-3) {{$\bullet$}};
\draw[ line width=1.5pt] (-0.5,-3) to[out=120, in = 270,looseness=1] (-1.4,0);
\node at (-1.5,3) {{$\bullet$}};
\draw[ line width=1.5pt] (-0.8,0) to[out=90, in = 300,looseness=1] (-1.5,3);
\draw[ thick, black!50,dashed,line width=1pt] (0,0) -- (3,-3);
\draw[ black!50,dashed,line width=1pt] (3,3) -- (0,0);
\draw[thick,blueKSV1!200] (3,-3) -- (3,3);
\node at (3,3) {{\color{blueKSV1!200}$\bullet$}};
\node at (3,-3) {{\color{blueKSV1!200}$\bullet$}};
\node at (-1,-3.5) {{$j$}};
\node at (-1.8,3.5) {{$i$}};
\draw[thick,decorate, decoration={zigzag, segment length=4pt, amplitude=1pt}](-2.3,0)--(-0.2,0);
\node at (-0.2,0) {{$\bullet$}};
\end{scope}
\end{tikzpicture}
}
}
\qquad
\vcenter{
\hbox{
\begin{tikzpicture}
\begin{scope}[ scale=0.5]
\fill[red!20] (3,-3) -- (0,0) -- (3,3) -- (3,-3);
\draw[ thick] (-2.3,3) -- (3,3) -- (3,-3) -- (-2.3,-3) -- (-2.3,3);
\node at (1.5,-3.4) {$\mathcal{I}^-$};
\node at (1.5,3.5) {$\mathcal{I}^+$};
\draw[black!50,dashed,line width=1pt] (0,0) -- (3,-3);
\draw[ black!50,dashed,line width=1pt] (3,3) -- (0,0);
\draw[thick, red] (3,-3) -- (3,3);
\node at (-1.5,-3) {{$\bullet$}};
\draw[ line width=1.5pt] (-1.5,-3) to[out=60, in = 270,looseness=1] (-0.6,0);
\node at (-0.5,3) {{$\bullet$}};
\draw[ line width=1.5pt] (-1.4,0) to[out=90, in = 240,looseness=1] (-0.5,3);
\node at (3,3) {{\color{red}$\bullet$}};
\node at (3,-3) {{\color{red}$\bullet$}};
\node at (-1.8,-3.5) {{$i$}};
\node at (-0.8,3.5) {{$j$}};
\draw[thick,decorate, decoration={zigzag, segment length=4pt, amplitude=1pt}] (-2.3,0)--(-0.2,0);
\node at (-0.2,0) {{$\bullet$}};
\end{scope}
\end{tikzpicture}
}
}
\end{equation}
Note that since the two geometries are smoothly glued along the branch cut, and each satisfies the equations of motion, the resulting spacetime is an on-shell solution---albeit one with a conical singularity in the bulk.
As explained in Subsection~\ref{crotches}, the effect of the defect is to introduce an additional weight of the form
\begin{equation}
    \exp(-\frac{A(r_0)}{4G})~, 
\end{equation}
in the Lorentzian path integral. Here,
$
    A(r) =  r^{d-1}V_{\Omega_{d-1}}
$
is the area of the sphere of radius $r_0$ where the defect is located. When doing the path integral as described in \eqref{pathint2} and \eqref{PathInt}, we find that the area is minimized when $r_0 \rightarrow \text{max}\{r_i,r_j\}$, meaning that the conical singularity attaches to the outermost shell in the geometry
\begin{equation}
\label{overlapsec32}
  \overline{|\bra{\psi_j}\ket{\psi_i}|^2}^{\text{ conn}}   \approx
  \overline{\bra{\psi_i}\ket{\psi_i}}\;
  \overline{\bra{\psi_j}\ket{\psi_j}}\times \exp(-\frac{A({r_0})}{4G}) \quad \text{with} \quad r_0 = \text{max}\{r_i,r_j\}~.
\end{equation}
The approximate sign is because we are ignoring loop corrections. 
Note that we are also minimizing the value of $A(r_0)$ by picking the location of the defect to be at $\text{max}\{r_i,r_j\}$. 

For end-of-the world branes, the setup is very similar, with the topology now taking the form
\begin{equation}
\vcenter{
\hbox{
\begin{tikzpicture}
\begin{scope}[ scale=0.5]
\fill[blueKSV1!70] (3,-3) -- (0,0) -- (3,3) -- (3,-3);
\draw[ thick] (-1.5,3) -- (3,3) -- (3,-3) -- (-0.5,-3);
\node at (-0.5,-3) {{$\bullet$}};
\draw[ line width=1.5pt] (-0.5,-3) to[out=120, in = 270,looseness=1] (-1.4,0);
\node at (-1.5,3) {{$\bullet$}};
\draw[ line width=1.5pt] (-0.8,0) to[out=90, in = 300,looseness=1] (-1.5,3);
\draw[black!50,dashed,line width=1pt] (0,0) -- (3,-3);
\draw[black!50,dashed,line width=1pt] (3,3) -- (0,0);
\draw[thick, blueKSV1!200] (3,-3) -- (3,3);
\node at (3,3) {{\color{blueKSV1!200}$\bullet$}};
\node at (3,-3) {{\color{blueKSV1!200}$\bullet$}};
\node at (-1,-3.5) {{$j$}};
\node at (-1.8,3.5) {{$i$}};
\draw[thick,decorate, decoration={zigzag, segment length=4pt, amplitude=1pt}](-1.4,0)--(-0.2,0);
\node at (-0.2,0) {{$\bullet$}};
\end{scope}
\end{tikzpicture}
}
}
\vcenter{
\hbox{
\begin{tikzpicture}
\begin{scope}[ scale=0.5]
\fill[red!20] (3,-3) -- (0,0) -- (3,3) -- (3,-3);
\draw[ thick] (-0.5,3) -- (3,3) -- (3,-3) -- (-1.5,-3);
\draw[ black!50,dashed,line width=1pt] (0,0) -- (3,-3);
\draw[ black!50,dashed,line width=1pt] (3,3) -- (0,0);
\draw[thick, red] (3,-3) -- (3,3);
\node at (-1.5,-3) {{$\bullet$}};
\draw[ line width=1.5pt] (-1.5,-3) to[out=60, in = 270,looseness=1] (-0.6,0);
\node at (-0.5,3) {{$\bullet$}};
\draw[ line width=1.5pt] (-1.4,0) to[out=90, in = 240,looseness=1] (-0.5,3);
\node at (3,3) {{\color{red}$\bullet$}};
\node at (3,-3) {{\color{red}$\bullet$}};
\node at (-1.8,-3.5) {{$i$}};
\node at (-0.8,3.5) {{$j$}};
\draw[thick,decorate, decoration={zigzag, segment length=4pt, amplitude=1pt}] (-1.4,0)--(-0.2,0);
\node at (-0.2,0) {{$\bullet$}};
\end{scope}
\end{tikzpicture}
}
}
\end{equation}
The only difference is that the spacetime terminates at the brane, so the branch cut ends there as well. The computation of the connected overlap proceeds as in the case of thin shells, and we again recover equation \eqref{overlapsec32}.

\paragraph{Schwarzschild-de Sitter.} As explained in Section~\ref{sec:setup}, for Schwarzschild–de Sitter spacetimes the situation is slightly different because the geometry can be extended to both the left and the right of an observer’s static patch. In this case, we must consider configurations with two thin shells or two end-of-the-world branes. The picture is the most transparent in terms of end-of-the-world branes, where the wormhole topology is given by:
\begin{equation}
\vcenter{\hbox{
    \begin{tikzpicture}
    \begin{scope}[scale=1.6]
        \draw[thick,decorate,decoration={zigzag,amplitude=0.3mm,segment length=2mm}] (-0.45,1) -- (1,1);
        \draw[thick, decorate,decoration={zigzag,amplitude=0.3mm,segment length=2mm}] (-0.45,-1) -- (1,-1);
        \draw[black!50,dashed,line width=1pt, fill=blueKSV1!70] (0,0) -- (1,1) -- (2,0) -- (1,-1) -- (0,0);
        \draw[black!50,dashed,line width=1pt] (-0.6,-0.6) -- (0,0) -- (-0.6,0.6);
        \draw[thick] (1,1) -- (2.3,1);
        \draw[thick] (2.3,-1) -- (1,-1);
        \draw[line width=1.5pt] (-0.5,1) to[out = -110, in = 90] (-0.7,0);
        \draw[line width=1.5pt] ({-1.3+3.6},1) to[out = -80, in = 90] ({-1.2+3.6},0);
        \draw[line width=1.5pt] (-0.8,0) to[out = -90, in = 100] (-0.5,-1);
        \draw[line width=1.5pt] ({-1.15+3.6},0) to[out = -90, in = 90] ({-1.3+3.6},-1);
        \draw[black!50,dashed,line width=1pt] ({2+0.37},-0.37)--(2,0)--({2+0.37},0.37);
        \fill[black] ({-0.5},-1) circle (0.04);
        \fill[black] ({-1.3+3.6},-1) circle (0.04);
        \fill[black] ({-0.5},1) circle (0.04);
        \fill[black] ({-1.3+3.6},1) circle (0.04);
        \draw[thick,decorate, decoration={zigzag, segment length=4pt, amplitude=1pt}](-0,0) -- (-0.8,0);
        \draw[thick,decorate, decoration={zigzag, segment length=4pt, amplitude=1pt}](2.15,0) -- (2.46,0);
        \fill[black] (-0,0) circle (0.04);
        \fill[black] (2.15,0) circle (0.04);
        \node at (1,1.2) {$i$};
        \node at (1,-1.2) {$j$};
        \end{scope}
    \end{tikzpicture}
    }}\qquad 
\vcenter{\hbox{
    \begin{tikzpicture}
    \begin{scope}[scale=1.6, yscale = -1]
        \draw[thick,decorate,decoration={zigzag,amplitude=0.3mm,segment length=2mm}] (-0.45,1) -- (1,1);
        \draw[thick, decorate,decoration={zigzag,amplitude=0.3mm,segment length=2mm}] (-0.45,-1) -- (1,-1);
        \draw[black!50,dashed,line width=1pt, fill=red!20] (0,0) -- (1,1) -- (2,0) -- (1,-1) -- (0,0);
        \draw[black!50,dashed,line width=1pt] (-0.6,-0.6) -- (0,0) -- (-0.6,0.6);
        \draw[thick] (1,1) -- (2.3,1);
        \draw[thick] (2.35,-1) -- (1,-1);
        \draw[line width=1.5pt] (-0.5,1) to[out = -110, in = 90] (-0.7,0);
        \draw[line width=1.5pt] ({-1.3+3.6},1) to[out = -80, in = 90] ({-1.2+3.6},0);
        \draw[line width=1.5pt] (-0.8,0) to[out = -90, in = 100] (-0.5,-1);
        \draw[line width=1.5pt] ({-1.15+3.6},0) to[out = -90, in = 90] ({-1.3+3.6},-1);
        \draw[black!50,dashed,line width=1pt] ({2+0.37},-0.37)--(2,0)--({2+0.37},0.37);
        \fill[black] ({-0.5},-1) circle (0.04);
        \fill[black] ({-1.3+3.6},-1) circle (0.04);
        \fill[black] ({-0.5},1) circle (0.04);
        \fill[black] ({-1.3+3.6},1) circle (0.04);
        \draw[thick,decorate, decoration={zigzag, segment length=4pt, amplitude=1pt}](-0,0) -- (-0.8,0);
        \draw[thick,decorate, decoration={zigzag, segment length=4pt, amplitude=1pt}](2.15,0) -- (2.46,0);
        \fill[black] (-0,0) circle (0.04);
        \fill[black] (2.15,0) circle (0.04);
        \node at (1,1.2) {$i$};
        \node at (1,-1.2) {$j$};
        \end{scope}
    \end{tikzpicture}
    }}
\end{equation}
We remind the reader that the abstract labels $i,j$ denote the different properties of the two shells, i.e. their mass, type of matter, and other internal degrees of freedom. The wormhole geometry is then constructed by gluing along two branch cuts.

Since we have two conical singularities, the Lorentzian path integral acquires a contribution from two area terms:
\begin{equation}
    \exp(-\frac{A(r_{\text{b}})}{4G}) \times  \exp(-\frac{A({r_0})}{4G})~,
\end{equation}
where $r_{\text{b}}$ is the SdS Schwarzschild radius, and $A_{r_0}$ is as in \eqref{overlapsec32}. Here we see a qualitative difference between the black hole and the cosmological horizons. For the cosmological horizon, the conical singularity wants to attach to the outermost defect as to minimize the radial coordinate  $r_i$. For the conical singularity associated to the black hole, the spacetime has a point which minimizes this area: the throat of the Einstein–Rosen bridge at $r = r_{\text{b}}$. Thus, regardless of the specific configuration of branes or thin shells inside the black hole, the conical singularity will attach to this \emph{extremal} surface. The final result is that 
\begin{equation}
\label{overlapsecblack}
  \overline{|\bra{\psi_j}\ket{\psi_i}|^2}^{\text{ conn}}   \approx
  \overline{\bra{\psi_i}\ket{\psi_i}}\;
  \overline{\bra{\psi_j}\ket{\psi_j}}\times \exp(-\frac{A(r_0)+A(r_\text{b})}{4G})~,
\end{equation}

There is an analogous construction of SdS wormholes using thin shells, working with the identified Penrose diagram described in \eqref{identifiedPRS} 
we can consider a branch cut that runs along the right static patch of the diagram. The wormhole that results from this branch cut is given by 
\begin{equation}
\vcenter{\hbox{
    \begin{tikzpicture}
    \begin{scope}[scale=1.6]
        \draw[thick,decorate,decoration={zigzag,amplitude=0.3mm,segment length=2mm}] (-0.8,1) -- (1,1);
        \draw[thick, decorate,decoration={zigzag,amplitude=0.3mm,segment length=2mm}] (-0.8,-1) -- (1,-1);
        \draw[black!50,dashed,line width=1pt, fill=blueKSV1!70] (0,0) -- (1,1) -- (2,0) -- (1,-1) -- (0,0);
        \draw[black!50,dashed,line width=1pt] (-0.6,-0.6) -- (0,0) -- (-0.6,0.6);
        \draw[thick] (1,1) -- (2,1) -- (2,-1) -- (1,-1);
        \draw[thick,black] (-0.8,1) -- (-1.6,1) -- (-1.6,-1) -- (-0.8,-1);
        \draw[line width=1.5pt] (-0.5,1) to[out = -110, in = 90] (-0.7,0);
        \draw[line width=1.5pt] (-1.3,1) to[out = -80, in = 90] (-1.2,0);
        \draw[line width=1.5pt] (-0.8,0) to[out = -90, in = 100] (-0.5,-1);
        \draw[line width=1.5pt] (-1.15,0) to[out = -90, in = 90] (-1.3,-1);
        \draw[black!50,dashed,line width=1pt] ({-1.6+0.37},-0.37)--(-1.6,0)--({-1.6+0.37},0.37);
        \fill[black] (-0.5,-1) circle (0.05);
        \fill[black] (-1.3,-1) circle (0.05);
        \fill[black] (-0.5,1) circle (0.05);
        \fill[black] (-1.3,1) circle (0.05);
        \draw[thick, ->] (-1.6,-0.55) -- (-1.6,-0.5);
        \draw[thick, ->] (-1.6,0.45) -- (-1.6,0.5);
        \draw[thick, ->] (2,-0.55) -- (2,-0.5);
        \draw[thick, ->] (2,0.45) -- (2,0.5);
        \draw[thick,decorate, decoration={zigzag, segment length=4pt, amplitude=1pt}](0,0) -- (-1.4,0);
        \fill[black] (0,0) circle (0.05);
        \fill[black] (-1.4,0) circle (0.05);
        \node at (-0.8,1.2) {$i$};
        \node at (-0.8,-1.2) {$j$};
        \end{scope}
    \end{tikzpicture}
    }}
\quad
\vcenter{\hbox{
    \begin{tikzpicture}
    \begin{scope}[scale=1.6]
        \draw[thick,decorate,decoration={zigzag,amplitude=0.3mm,segment length=2mm}] (-0.8,1) -- (1,1);
        \draw[thick, decorate,decoration={zigzag,amplitude=0.3mm,segment length=2mm}] (-0.8,-1) -- (1,-1);
        \draw[black!50,dashed,line width=1pt, fill=red!20] (0,0) -- (1,1) -- (2,0) -- (1,-1) -- (0,0);
        \draw[black!50,dashed,line width=1pt] (-0.6,-0.6) -- (0,0) -- (-0.6,0.6);
        \draw[thick] (1,1) -- (2,1) -- (2,-1) -- (1,-1);
        \draw[thick,black] (-0.8,1) -- (-1.6,1) -- (-1.6,-1) -- (-0.8,-1);
        \draw[line width=1.5pt] (-0.5,-1) to[out = 110, in = -90] (-0.7,0);
        \draw[line width=1.5pt] (-1.3,-1) to[out = 80, in = -90] (-1.2,0);
        \draw[line width=1.5pt] (-0.8,0) to[out = 90, in = -100] (-0.5,1);
        \draw[line width=1.5pt] (-1.15,0) to[out = 90, in = -90] (-1.3,1);
        \draw[black!50,dashed,line width=1pt] ({-1.6+0.37},-0.37)--(-1.6,0)--({-1.6+0.37},0.37);
        \fill[black] (-0.5,-1) circle (0.05);
        \fill[black] (-1.3,-1) circle (0.05);
        \fill[black] (-0.5,1) circle (0.05);
        \fill[black] (-1.3,1) circle (0.05);
        \draw[thick, ->] (-1.6,-0.55) -- (-1.6,-0.5);
        \draw[thick, ->] (-1.6,0.45) -- (-1.6,0.5);
        \draw[thick, ->] (2,-0.55) -- (2,-0.5);
        \draw[thick, ->] (2,0.45) -- (2,0.5);
        \draw[thick,decorate, decoration={zigzag, segment length=4pt, amplitude=1pt}](0,0) -- (-1.4,0);
        \fill[black] (0,0) circle (0.05);
        \fill[black] (-1.4,0) circle (0.05);
        \node at (-0.8,1.2) {$j$};
        \node at (-0.8,-1.2) {$i$};
        \end{scope}
    \end{tikzpicture}
    }}
\end{equation}
These diagrams share the same features as the case with two end-of-the-world branes. They contain two conical singularities, one attached to the black hole horizon, and the other to the outermost shell in the solution. As a result, equation \eqref{overlapsecblack} also holds for this construction.

\vspace{-7pt}

\subsection{Counting microstates}
\label{microcounting}

Having understood the inner product of states prepared via the gravitational path integral, we now proceed to compute the dimensionality of the Hilbert space spanned by these vectors. This quantity can be obtained by analyzing the rank of the Gram matrix
\begin{equation}
G_{ij} = \bra{\psi_i}\ket{\psi_j}~.
\end{equation}
We begin by considering a finite set of $N$ states that span a Hilbert space defined as
\begin{equation}
\label{Hilbertspace}
    \mathcal{H}_{\text{semicl}} = \text{span}\big\{\ket{\psi_i}, i=1,\dots, N\big\}~.
\end{equation}
As additional vectors are included, the dimension of $\mathcal{H}_{\text{semicl}}$ increases until it eventually saturates. We interpret the corresponding value of $N$ at which this saturation occurs as the dimensionality of the Hilbert space associated with de Sitter or Schwarzschild–de Sitter spacetimes. Equivalently, this corresponds to the smallest value of $N$ for which the Gram matrix acquires its first zero eigenvalue.

We are going to consider states with the property that each shell or brane in $\mathcal{H}_{\text{semicl}}$  is close to the cosmological horizon, i.e. $r_i = r_\text{c} - \delta r$, for a small parameter $\delta r$; we comment on the size of $\delta r$ at the end of this section. 
The reason to consider shells that are close to the cosmological horizon is that these correspond to states which minimize the overlap \eqref{overlapsec32}. Intuitively, because these vectors are as close to mutually orthogonal as possible, the Hilbert space they span attains its maximal possible dimension. 

One way to determine the rank of the Gram matrix is to examine the behavior of the trace of the resolvent,
\begin{equation}
\label{eq:resolvents}
    R(x) = \Tr\left(
    \frac{1}{x-G}
    \right) = 
    \sum_{k=0}^\infty 
    \frac{\Tr(G^k)}{x^{k+1}}~,
\end{equation}
which has a pole whenever $x$ hits an eigenvalue of the Gram matrix. 
In gravity, we do not have access to the exact value of this resolvent. However, we can examine the averaged resolvent $\overline{R}$ by looking at the moments of $\overline{\Tr(G^k)}$, which are encoded in Lorentzian wormholes. The wormholes discussed in \ref{sec:Lwormholes} correspond to the case where $k=2$, but the generalization to higher moments is straightforward, as one can similarly consider the $k$-fold cover of these spacetimes:
\begin{equation}
\label{replicass}
\begin{split}
\overline{\bra{\psi_{i_1}}\ket{\psi_{i_2}}\bra{\psi_{i_2}}\ket{\psi_{i_3}}\dots \bra{\psi_{i_k}}\ket{\psi_{i_1}}} &\approx 
\vcenter{
\hbox{
\begin{tikzpicture}
\begin{scope}[ scale=0.44]
\fill[blueKSV1!70] (3,-3) -- (0,0) -- (3,3) -- (3,-3);
\draw[ thick] (-2.3,3) -- (3,3) -- (3,-3) -- (-2.3,-3) -- (-2.3,3);
\node at (-0.5,-3) {{$\bullet$}};
\draw[ line width=1.5pt] (-0.5,-3) to[out=120, in = 270,looseness=1] (-1.4,0);
\node at (-1.5,3) {{$\bullet$}};
\draw[ line width=1.5pt] (-0.8,0) to[out=90, in = 300,looseness=1] (-1.5,3);
\draw[ black!50,dashed,line width=1pt] (0,0) -- (3,-3);
\draw[ black!50,dashed,line width=1pt] (3,3) -- (0,0);
\node at (-1,-3.5) {{$i_2$}};
\node at (-2,3.5) {{$i_1$}};
\draw[thick,decorate, decoration={zigzag, segment length=4pt, amplitude=1pt}](-2.3,0)--(-0.2,0);
\node at (-0.2,0) {{$\bullet$}};
\end{scope}
\end{tikzpicture}
}
}
\vcenter{
\hbox{
\begin{tikzpicture}
\begin{scope}[ scale=0.44]
\fill[red!20] (3,-3) -- (0,0) -- (3,3) -- (3,-3);
\draw[ thick] (-2.3,3) -- (3,3) -- (3,-3) -- (-2.3,-3) -- (-2.3,3);
\draw[ black!50,dashed,line width=1pt] (0,0) -- (3,-3);
\draw[ black!50,dashed,line width=1pt] (3,3) -- (0,0);
\node at (-1.5,-3) {{$\bullet$}};
\draw[ line width=1.5pt] (-1.5,-3) to[out=60, in = 270,looseness=1] (-0.6,0);
\node at (-0.5,3) {{$\bullet$}};
\draw[ line width=1.5pt] (-1.4,0) to[out=90, in = 240,looseness=1] (-0.5,3);
\node at (-2,-3.5) {{$i_3$}};
\node at (-1,3.5) {{$i_2$}};
\draw[thick,decorate, decoration={zigzag, segment length=4pt, amplitude=1pt}] (-2.3,0)--(-0.2,0);
\node at (-0.2,0) {{$\bullet$}};
\end{scope}
\end{tikzpicture}
}
}\dots \!
\vcenter{
\hbox{
\begin{tikzpicture}
\begin{scope}[ scale=0.44]
\fill[orange!20] (3,-3) -- (0,0) -- (3,3) -- (3,-3);
\draw[ thick] (-2.3,3) -- (3,3) -- (3,-3) -- (-2.3,-3) -- (-2.3,3);
\draw[black!50,dashed,line width=1pt] (0,0) -- (3,-3);
\draw[black!50,dashed,line width=1pt] (3,3) -- (0,0);
\node at (-1.5,-3) {{$\bullet$}};
\draw[ line width=1.5pt] (-1.5,-3) to[out=60, in = 270,looseness=1] (-0.6,0);
\node at (-0.5,3) {{$\bullet$}};
\draw[ line width=1.5pt] (-1.4,0) to[out=90, in = 240,looseness=1] (-0.5,3);
\node at (-2,-3.5) {{$i_1$}};
\node at (-1,3.5) {{$i_k$}};
\draw[thick,decorate, decoration={zigzag, segment length=4pt, amplitude=1pt}] (-2.3,0)--(-0.2,0);
\node at (-0.2,0) {{$\bullet$}};
\end{scope}
\end{tikzpicture}
}
} \\[5pt] &\approx
\overline{\bra{\psi_{i_1}}\ket{\psi_{i_1}}}\;
\overline{\bra{\psi_{i_2}}\ket{\psi_{i_2}}}\dots
\overline{\bra{\psi_{i_k}}\ket{\psi_{i_k}}}\times
\exp[-(k-1)\frac{A({r_0})}{4G}]~,
\end{split}
\end{equation}
where $r_0 = \text{max}\{r_1,\dots, r_k\}$ which by construction is close to $r_{\text{c}}$.\\[-7pt]

We can then use these moments to do the sum of equation \eqref{eq:resolvents} and obtain an explicit expression for $\overline{R(x)}$ by solving a Schwinger–Dyson equation. The steps to formulate and solve the Schwinger–Dyson equation have been extensively discussed in the AdS literature; see, e.g., \cite{Balasubramanian:2024rek, Penington:2019kki,Hsin:2020mfa,Geng:2024jmm} for some recent accounts. Here we only report on the final result, which is that the Gram matrix ceases to be positive definite when 
\begin{equation}
\label{eq:entropy}
    N \gtrsim \exp(\frac{A(r_\text{c})}{4G})~.
\end{equation}
The approximate sign reflects on the various approximations of our computation, such as the loop corrections in the path integral. 
 In formulating the Schwinger–Dyson equations, we also neglected non-planar geometries and nonperturbative contributions from more complicated topologies. The more precise version of \eqref{eq:entropy} is that the Gram matrix ceases to be positive when 
\begin{equation}
\label{entdS}
    \log N  \geq  \frac{A(r_\text{c})}{4G} + \order{\log G}~.
\end{equation}
For the case of Schwarzschild-de Sitter, an analogous computation gives the entropy
\begin{equation}
\label{entSdS}
    \log N = \frac{A(r_{\text{c}})+ A(r_{\text{b}})}{4G} + \order{\log G}~. 
\end{equation}
Note that \eqref{entdS} and \eqref{entSdS} apply to both setups involving end-of-the-world solutions or thin shells.\\[-5pt]

Lastly, we comment on the observation that the conical singularity associated with the cosmological horizon attaches to the brane (or thin shell) rather than to the horizon itself. 
This indicates that the entropy is dominated by states whose shells lie very close to the cosmological horizon. As explained above, this is expected as these are the states which are the the `most orthogonal' to each other since they minimize their inner product. A natural concern, however, is whether there are sufficiently many states within the narrow window $r_i \in (r_\text{c}-\delta r,r_{\text{c}})$ to account for the entropy.\\[-5pt]

As discussed in Section~\ref{sec:shells}, the position of the shells and branes is dynamical and ultimately determined by the parameters of these solutions. So shells and branes which are close to each other correspond to solutions with very similar parameters. 
Since our description is only meant to be effective and semiclassical, one might worry that it breaks down when the parameters describing the shells are too close, e.g. when their masses $|m_i-m_j|\sim e^{-S}$. 
However, this is not a real obstacle, as the effective description has integrated over many of the microscopic degrees of freedom.  Thus, we expect the shells to carry additional internal labels, which may correspond, for example, to matter sources of different flavors or to the specific microscopic arrangement of the dust particles within the shell.
Therefore, at the semiclassical level, we can always prepare a large number of states around $r_i \in (r_\text{c}-\delta r,r_{\text{c}})$, even when $\delta r$ is arbitrarily small.\\[-5pt]

For the case of thin shells, there is another possibility: we can enlarge the state space without introducing different matter flavors by considering configurations with multiple shells.  One shell may be placed arbitrarily close to the cosmological horizon $r_{\text{c}}$, forcing $r_0 \rightarrow r_{\text{c}}$, while the remaining shells lie anywhere else in the bulk. States corresponding to different shell configurations are then approximately orthogonal,
\begin{equation}
\vcenter{\hbox{
\begin{tikzpicture}
\begin{scope}[scale=1.1]
    \draw[thick, black,fill=white] (-1,1.5) -- (-1.6,1.5) -- (-1.6,-1.5) -- (-1,-1.5);
    \draw[thick, black,fill=white] (-1,1.5) -- (1,1.5) -- (1,-1.5) -- (-1,-1.5);
    \draw[black!50,dashed,line width=1pt, fill = red!20] (1,-1.5) -- (-0.5,0) -- (1,1.5);
    \draw[thick, black] (-0.6,-1.5)  --(-0.6,1.5);
    \draw[thick, black] (-0.9,-1.5)  -- (-0.9,1.5);
    \draw[thick, black] (-1.2,-1.5)  -- (-1.2,1.5);
    \node at (-0.6,1.5) {{\color{black}$\bullet$}};
    \node at (-0.6,-1.5) {{\color{black}$\bullet$}};
    \node at (-0.9,1.5) {{\color{black}$\bullet$}};
    \node at (-0.9,-1.5) {{\color{black}$\bullet$}};
    \node at (-1.2,1.5) {{\color{black}$\bullet$}};
    \node at (-1.2,-1.5) {{\color{black}$\bullet$}};
\end{scope}
\end{tikzpicture}}}\,.
\end{equation}

\vspace{-15pt}

\section{Discussion}
\label{sec:discussion}

\vspace{-5pt}

In this paper, we analyzed the inner product of \emph{semiclassical} states constructed using a Lorentzian path integral with additional contributions from branes and shells placed in the spacetime. Within the semiclassical approximation, we find that the variance of the overlaps between such states is determined by complex wormhole geometries, which can be described using spacetimes with Lorentzian conical singularities. By evaluating these overlaps, we found that the microcanonical entropy of de Sitter and Schwarzschild-de Sitter follows the expected Gibbons-Hawking and Bekenstein-Hawking area laws.

In constructing these geometries, we examined two microscopic conditions: the Null Energy Condition (NEC) and the matching background condition. Within the class of configurations explored in this work, we did \emph{not} find solutions satisfying both conditions simultaneously. Instead, we identified solutions that satisfy the NEC but not the matching background condition, and vice versa. Here, we conclude with a discussion of the possible interpretations of this result, and the limitations and potential extensions of our approach.

\vspace{-7pt}

\paragraph{Jump in the cosmological constant.} 
Several of the solutions we considered involve a jump in the cosmological constant across a shell, and one may wonder what the microscopics of such a shell could possibly be. One familiar case where the cosmological constant jumps is across D3-branes sitting inside planar AdS${}_5$. Such a configuration corresponds to a non-trivial point on the Coulomb branch of the gauge theory, and because the RR flux jumps across the brane, so does the cosmological constant. Another well-known case are the Coleman-de Luccia instantons \cite{Coleman:1980aw} which mediate the creation of bubbles of spacetime with a different and smaller cosmological constant inside de Sitter space. 
This requires a landscape of vacua with different vacuum energy à la Bousso and Polchinski \cite{Bousso:2000xa}. In the thin-wall approximation, Coleman-de Luccia instantons effectively look like a tensionfull brane.\footnote{For a recent discussion of these instantons in the context of de Sitter space see e.g. \cite{Irakleous:2025trr}.}

In both examples, the microscopics of the ``defect" are quite well-understood. For D-branes, one could imagine looking at different low-energy excited states on the branes. These would have little impact on the description as an object with only tension, but might provide the required additional ``flavor" label $i$ which appears in the computations in this paper. It is less clear to us whether Coleman-de Luccia instantons can support a similarly large number of internal degrees of freedom without invalidating the tensionfull brane approximation. 

Shells made out of dust particles propagating in a single ambient spacetime do, by construction, not support a jump of the cosmological constant. Therefore, the known objects which allow for such a jump tend to be well-described by positive tension branes. As a result, it is very challenging to find a reliable  microscopic description of a dynamical negative tension (or NEC violating) brane. The standard example of a negative tension object in string theory is an orientifold, but these are not dynamical and carry no internal degrees of freedom.

This suggests that for empty de Sitter solutions, the most natural configurations are those in which a positive-tension brane separates two de Sitter spaces with different cosmological constants. In the context of de Sitter space without additional matter fields, this is reasonable, as the only available degree of freedom that could account for the entropy is in the cosmological constant. In this case, the entropy originates from a diverse set of vacuum solutions that permit the nucleation of multiple de Sitter bubbles in the spacetime. If other forms of matter are included in the theory, then it becomes natural to consider black hole solutions, which we discuss next.

\vspace{-8pt}

\paragraph{NEC in Schwarzschild-de Sitter.}
One way to account for the entropy of de Sitter space, without invoking multiple de Sitter vacua, is to consider small SdS black holes. In Subsection~\ref{sec: SdS}, we discussed a semiclassical configuration involving two shells in a spacetime with a single cosmological constant. Here, it is natural to model the shells as composed of pressureless dust particles. The shell falling into the black hole violates the NEC, while the shell escaping to infinity satisfies the NEC. Despite the NEC violation, we find evidence that suggests that the overall spacetime satisfies the AANEC.
These configurations are interesting because, collapsing matter shells that lower the energy of a ``typical" state have been identified and studied in the AdS literature \cite{Papadodimas:2017qit}. Even though the operators sourcing these shells are fine-tuned, they do exist in the CFT. This suggests that such states may have a genuine microscopic description in the UV theory describing de Sitter. One could then account for the de Sitter entropy by considering the limit in which the SdS black hole vanishes.

\vspace{-8pt}

\paragraph{The entropy of more than one observer.}
An advantage of the framework presented in this paper is that we can use it to count states that are indistinguishable across an entire worldvolume (with multiple observers), rather than only along a single worldline. The causal diamond associated with the worldvolume occupies a larger spacetime region; at $t=0$ it covers more than half of the de Sitter sphere. The associated horizon therefore has a smaller area and hence a smaller entropy. The wormhole geometry is effectively unchanged; the only modification is that the surplus angle of the cut now sits at a smaller radius,
\begin{equation}
\vcenter{
\hbox{
\begin{tikzpicture}
\begin{scope}[scale=0.6]
\draw[black!50,dashed,line width=1pt,fill=blueKSV1!70] (3,-3) -- (2,-3) -- (-1,0) --(2,3) -- (3,3);
\fill[blueKSV1!150] (3,-3) -- (2,-3) -- (2,3) -- (3,3);
\node[rotate = 90] at (2.5,0) {obs. worldvolume};
\draw[ thick] (-2.7,3) -- (3,3) -- (3,-3) -- (-2.7,-3) -- (-2.7,3);
\node at (0.5,-3.4) {$\mathcal{I}^-$};
\node at (0.5,3.5) {$\mathcal{I}^+$};
\node at (-2,3) {{$\bullet$}};
\draw[ line width=1.5pt,black] (-2,3) to[out=-90, in = -270,looseness=1] (-1.7,0);
\node at (-1.2,-3) {{$\bullet$}};
\draw[ line width=1.5pt,black] (-2,0) to[out=-90, in = -250,looseness=1] (-1.2,-3);
\node at (-2,3.5) {{$j$}};
\node at  (-1.2,-3.5) {{$i$}};
\draw[thick,decorate, decoration={zigzag, segment length=4pt, amplitude=1pt}](-2.7,0)--(-1.1,0);
\node at (-1.15,0) {{$\bullet$}};
\end{scope}
\end{tikzpicture}
}
}\quad
\vcenter{
\hbox{
\begin{tikzpicture}
\begin{scope}[scale=0.6]
\draw[black!50,dashed,line width=1pt,fill=red!20] (3,-3) -- (2,-3) -- (-1,0) --(2,3) -- (3,3);
\fill[red!50] (3,-3) -- (2,-3) -- (2,3) -- (3,3);
\node[rotate = 90] at (2.5,0) {obs. worldvolume};
\draw[ thick] (-2.7,3) -- (3,3) -- (3,-3) -- (-2.7,-3) -- (-2.7,3);
\node at (0.5,-3.4) {$\mathcal{I}^-$};
\node at (0.5,3.5) {$\mathcal{I}^+$};
\node at (-2,-3) {{$\bullet$}};
\draw[ line width=1.5pt,black] (-2,-3) to[out=80, in = 270,looseness=1] (-1.7,0);
\node at (-1.2,3) {{$\bullet$}};
\draw[ line width=1.5pt,black] (-2,0) to[out=90, in = 250,looseness=1] (-1.2,3);
\node at (-2,-3.5) {{$j$}};
\node at  (-1.2,3.5) {{$i$}};
\draw[thick,decorate, decoration={zigzag, segment length=4pt, amplitude=1pt}](-2.7,0)--(-1.2,0);
\node at (-1.15,0) {{$\bullet$}};
\end{scope}
\end{tikzpicture}
}
}
\end{equation}

\paragraph{Related constructions.} It would be interesting to explore how our discussion extends to other contexts, for instance by gluing de Sitter space to Anti-de Sitter or Minkowski spacetimes. The embedding of de Sitter spacetimes in AdS and Schwarzschild–AdS backgrounds has been studied in \cite{Freivogel:2005qh,Lowe:2010np}  as a possible way to obtain a controlled holographic framework for analyzing de Sitter physics. However, in these constructions the de Sitter region is usually located inside the black hole horizon, making it difficult to access it directly. In two dimensions, the situation is different as it appears possible to study de Sitter physics through such embeddings (see e.g. \cite{Anninos:2018svg,Espindola:2025wjf}). Most of these analyses have been carried out in Lorentzian or Euclidean signature. It would be interesting to revisit these constructions to see whether complex geometries in these setups could capture de Sitter physics.

Other related constructions involving complex metrics in de Sitter space can be found in \cite{Fumagalli:2024msi,Chen:2020tes}, where the viability of bra-ket wormhole solutions was examined as a possible replacement for the Hartle–Hawking no-boundary state, aiming to address some of its undesirable features.
An alternative approach to the computation of the dS${}_3$ entropy, based on a sum over complex geometries, appears in \cite{Chakravarty:2025sbg}. Those complex geometries are different from the ones we consider and, moreover, appear to violate the KSW criterion. It is unclear to us whether these rather different approaches are in any way connected to the complex geometries discussed in this paper.


\paragraph{Contrast with the QFT Hilbert space.} An important difference between states defined using branes and shells and those defined in context of a quantum field theory (QFT) on a fixed de Sitter background is that the former describe a sector of quantum gravity in which matter backreacts on the geometry. This backreaction captures non-perturbative corrections in inverse powers of Newton’s constant. In contrast, QFT in de Sitter space strictly considers the $G \to 0$,  which by ignoring the backreactions leads to an infinite (and continuous) Hilbert space\cite{10.2307/97833,10.2307/1969129,GelNai47,ThomasSO, Sun:2021thf, Penedones:2023uqc,Anous:2020nxu}.  

\paragraph{Limits of the Euclidean interpretation.} 
 While our emphasis is on the Lorentzian path integral, some of the spacetimes presented in this section also admit an Euclidean preparation. For the geometries shown in Figure~\ref{fig: gamma=pm1}, one can cut the path integral at $t=0$ and perform  a Wick rotation $t \rightarrow i\tau$ to turn the de Sitter metric into a sphere. In global coordinates the explicit transformation reads:
\begin{equation}
    ds^2=-\dd t^2 + \ell^2 \cosh(\frac{t}{\ell})^2 \dd\Omega_d^2 \quad \longrightarrow\quad 
     ds^2= \dd \tau^2 + \ell^2 \cos(\frac{\tau}{\ell})^2 \dd\Omega_d^2.
\end{equation}
The euclidean manifold with an end-of-the-world brane is sphere with a cap removed.  The manifold with the shell is two spheres of radii $\ell_\lf$ and $\ell_\ri$ glued together at the thin shell:
\begin{equation}
\vcenter{\hbox{
\begin{tikzpicture}
\draw[thick, fill = red!20] ({1.5*cos(150)},{1.5*sin(150)}) arc (150:-150:1.5);
\draw[thick, dashed] (0.2,1.45) arc (50:-50:1.9);
\draw[thick,fill=black!10] (-1.3,0) ellipse (0.3 and 0.7);
\node at (-3,0) {EoW brane};
\node at (0.55,-2.15) {$\phantom{Thin shell}$};
\end{tikzpicture}
}} \quad 
\vcenter{\hbox{
\begin{tikzpicture}
\begin{scope}[scale=0.65]
 \draw[thick, fill =  blueKSV1!70] (1.6,0) circle (1.5);
 \draw[thick,fill= red!20] (-1.6,0) circle (2.5);
 \draw[thick,dashed] (1.6,1.5) to[in = 30, out = -30] (1.6,-1.5);
 \draw[thick,dashed] (-1.6,-2.5) to[in = 210, out = 150] (-1.6,2.5);
\fill[blueKSV1!70] (0.69,0) ellipse (0.3 and 1.15);
\draw[thick,dashed] (0.62,-1.14) arc (-90:90:0.2 and 1.14);
\draw[thick] (0.62,1.14) arc (90:270:0.2 and 1.14);
\node at (-1.7,0) {L};
\node at (1.7,0) {R};
\node at (0.55,-3.4) {Thin shell};
\draw[thick] (0.55,-3) -- (0.55,-2);
\end{scope}
\end{tikzpicture}
}}
\end{equation}

As opposed to empty de Sitter, Schwarzschild–de Sitter has no smooth Euclidean continuation. The Euclidean section carries two conical singularities, at the black-hole and cosmological horizons. One can tune the Euclidean time period
\begin{equation}
    \beta = \frac{1}{T_{\text{c}}} \quad\text{or}\quad \beta = \frac{1}{T_{\text{b}}}~,
\end{equation}
to remove one defect, but never both. Here $T_{\text{c,b}}$ correspond to the temperatures of the cosmological and black hole horizon, respectively. Hence there is no completely regular Euclidean manifold. Recent work, see \cite{Draper:2022ofa,Draper:2022xzl,Morvan:2022ybp}, has revisited the interpretation and consequences of these singularities, and  it would be interesting to assess whether a Euclidean preparation of the solutions presented here is sensible when conical singularities are allowed in the Euclidean geometry.

For the wormhole solutions, the Euclidean picture is even less clear. The conical surpluses in the Lorentzian geometries should be viewed as effective descriptions of smooth complex metrics satisfying the KSW criterion. These are complex geometries, and it is not evident if they admit a fully regular Euclidean section.

Finally, we comment on the proposal of \cite{Wang:2025jfd} to describe overlaps using the Euclidean path integral. That work computes the variance of microstate overlaps using Euclidean manifolds which are different from the crotch geometries discussed in this paper.
In their construction, the Euclidean geometries do not have asymptotic boundaries; they only have finite boundaries where an end-of-the-world brane is placed. 
At present, it is an open question to us whether the Euclidean preparation of these shells on finite boundaries makes sense for the computation of the connected moments. 
Unlike Euclidean AdS, Euclidean de Sitter has no asymptotic boundaries, and the closest analog of the construction appears only in Lorentzian signature, where boundary conditions are imposed at $\mathcal{I}^{\pm}$.
If we try to directly map the Lorentzian crotch solutions to Euclidean ones, those appear to also have conical singularities, which are normally not allowed in Euclidean signature.

\paragraph{Acknowledgments}
We would like to thank Victor Gorbenko and Martin Sasieta for discussions and comments on the draft version of this manuscript. 
We also thank Dio Anninos, Vijay Balasubramanian, Pietro Benetti Genolini, Scott Collier, Bartek Czech, Ben Freivogel, Thomas Hartman, Diego Hofman and Julian Sonner for useful suggestions and stimulating discussions.   
 JdB and KSV are supported by the European Research Council under the European Unions Seventh Framework Programme (FP7/2007-2013), ERC Grant agreement ADG 834878. KSV is also supported  by the Dutch Research Council (NWO  \raisebox{-3pt}{\includegraphics[height=0.9\baselineskip]{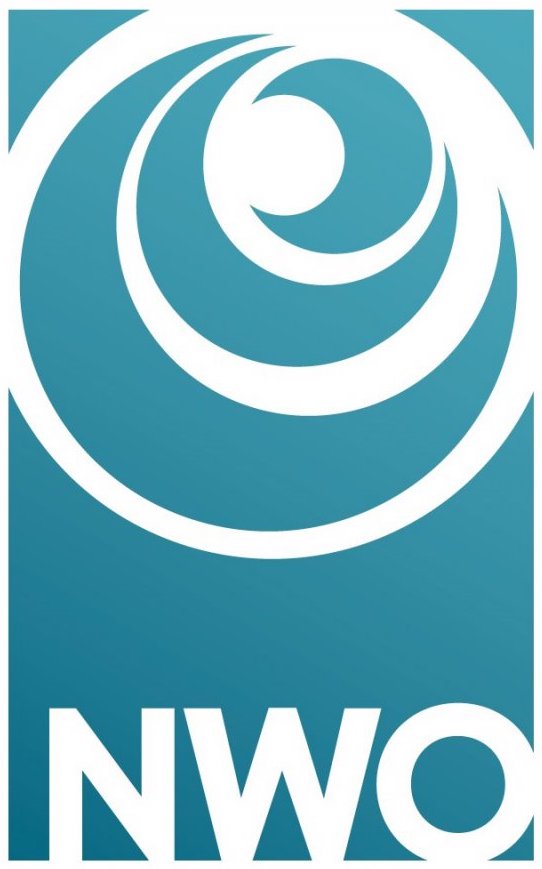}}\,) under {\hypersetup{urlcolor=black}\href {https://doi.org/10.61686/PRTOQ68396}{the grant}} for the project \textit{Constraining the Space of Cosmological Theories} with file number VI.Veni.242.438 of the research programme NWO Talent Programme, Veni. DL is supported by the SNF starting grant “The Fundamental Description of the Expanding Universe”. DL is also supported by the Fonds National Suisse de la Recherche Scientifique (Schweizerischer Nationalfonds zur Förderung der wissenschaftlichen Forschung) through the Project Grant 200021\_215300 and the NCCR51NF40-141869 The Mathematics of Physics (SwissMAP).

\vskip 4pt

\pagebreak

\appendix
\section{Notation and conventions}\label{eq:conventions}
In this section, we summarize the notation and conventions used in this paper. The results are valid for both Euclidean and Lorentzian manifolds, while we focus mainly on Lorentzian manifolds $M$ of $(d+1)$-dimensions with a metric $g_{\mu\nu}$ of signature mostly plus, that is, $(-,+,+,\cdots,+)$. Greek indices are used for tensors defined on $(d+1)$-dimensional manifolds, while Latin indices are used for tensors living on hypersurfaces defined on $M$.

Let $\Sigma$ be a codimension-one hypersurface  ($d$-dimensional) on the manifold $M$ that is defined as
\be
\Sigma= \{x^\mu \in \M: f(x^\mu) = C\}~,
\ee
for some constant $C$ or equivalently 
\be
\Sigma= \{x^\mu \in \M: x^\mu (\xi^a)\}~,
\ee
where $\xi^a$ with $a=\{1,\cdots,d\}$ are coordinates intrinsic to the hypersurface $\Sigma$. In the case that $\Sigma$ is the boundary of the manifold $M$, we fix the sign of $f$ and $C$ by defining
\be
\Sigma= \partial M~:\qquad M=\{x:f(x)<C\}~.
\ee

The normal vector to $\Sigma$ is given by
\be\label{eq:normal def}
n_\mu=   \frac{\e\,\partial_\mu f}{\sqrt{|g^{\mu\nu}\partial_\mu f\partial_\nu f}|}=  \frac{\e\, \partial_\mu f}{\sqrt{\e\,g^{\mu\nu}\partial_\mu f\partial_\nu f}}~,
\ee
which is unit-normalized 
\be
n_\mu n^\mu = \epsilon~,\qquad \text{where} \quad \epsilon=\pm1~.
\ee
The overall sign/orientation of the normal vector is picked such that $n^\alpha \partial_\alpha f>0$ which is equivalent to the convention that $n^\mu$ is outward-pointing when $\Sigma=\partial M$.
In a Lorentzian manifold $M$, $\Sigma$ is called spacelike if the normal vector is timelike i.e. $\epsilon=-1$ and $\Sigma$ is timelike if the normal vector is spacelike i.e. $\epsilon=1$. For Euclidean manifolds $M$, the normal vector is trivially spacelike and so is $\Sigma$. We also define the set of vectors: 
\be
e_i^\mu=\frac{\partial x^\mu}{\partial \xi^a}~,
\ee
which are tangent to curves contained in $\Sigma$: $n_\mu e_i^\mu =0$~.

The projection tensor to the tangent space of $\Sigma $ is therefore given by
\be\label{eq:project tensor}
h_{\mu\nu} = g_{\mu \nu} -\e \, n_{\mu}n_{\nu}~,
\ee
and the induced metric is defined as the pullback operation
\be\label{eq: def h}
h_{a b} = e_a^\mu e_b^\nu g_{\mu \nu}~.
\ee
 By following the  convention for the orientation of the normal vector defined as well as taking the hypersurface to be the boundary of the manifold $M$ i.e. $\Sigma=\partial M$ we can write Stokes's theorem as\cite{Poisson:2009pwt}
\be\label{eq:Stokes}
\int_{M} \sqrt{|g|}\,  \nabla_\mu V^\mu = \int_{\Sigma} \, \sqrt{|h|} \,  \e \,n_\mu V^\mu~,
\ee
where $V^\mu$ is an arbitrary vector on $M$.
Finally, the extrinsic curvature is defined as 
\be\label{eq: K def}
K_{\mu \nu} = \nabla_\mu n_\nu -\e  \, n_\mu n^\alpha \nabla_\alpha n_\nu~.
\ee
Notice that the sign of the extrinsic curvature is dictated based on the choice of sign in the normal vector definition in~\reef{eq:normal def}. Finally, let us spell out the pull-back of the extrinsic curvature onto $\Sigma$:
\be\label{eq:pullback of K}
K_{ab} =  e_a^\mu e_b^\nu K_{\mu \nu} = -n_\alpha \left(\partial_{a} e^\alpha_{b}+\Gamma^\alpha_{\mu \nu}e^\mu_{a}e^\nu_{b} \right)~,
\ee
in which $\Gamma_{\mu \nu}^\alpha$ is the Christoffel symbol and we used $\partial_{a}\equiv\partial/\partial\xi^a$ for brevity. 
It is easy to verify the above expression using the fact that $e_a^\mu n^\nu=0$ and $\nabla_\mu n_\nu = \partial_\mu n_\nu - \Gamma_{\mu \nu}^\alpha n_\alpha $.

\section{Gibbons-Hawking-York Action}\label{sec:GHY}
 The Einstein-Hilbert term would produce the correct Einstein field equations for a manifold with no boundary. In presence of a boundary, one needs to add the so-called Gibbons-Hawking-York term to the action to derive the correct Einstein field equations. In this section, starting from the Einstein-Hilbert action, we review the Gibbons-Hawking-York action in both the Dirichlet and Neumann boundary conditions. These are standard results in general relativity, which we review here for completeness and to establish our conventions. For a more detailed discussion see e.g. \cite{Poisson:2009pwt, Visser:1995cc, Bavera:2018boundary}.

\subsection{Variation of Einstein-Hilbert action}
The Einstein-Hilbert action is given by
\be
S_{\text{EH}}= \frac{1}{16 \pi G} \int_M \sqrt{|g|} \left(R-2\Lambda\right)~ + \int_{M} \sqrt{|g|}\, \mathcal{L}_{\text{matter}}~,
\ee
where the integral is over manifold $M$ with boundary $\partial M =\Sigma$,  $g$ is the determinant of metric $g_{\mu\nu}$ on $M$, $R$ is the Ricci scalar and the normalization is picked such that it gives the canonical definition of Newton constant $G$. Here $M$ could be both Lorentzian and Euclidean manifold. By varying the action one finds
\be\label{eq:EH var}
\delta S_{\text{EH}}=  \frac{1}{16\pi G}\int_\M \sqrt{|g|}\left[
    \left(R_{\mu\nu} - \frac{1}{2}(R - 2\Lambda) g_{\mu\nu}\right)\delta g^{\mu\nu} + \nabla_\alpha V^{\alpha}\right] + \int_{M} \delta\left(\sqrt{|g|}\, \mathcal{L}_{\text{matter}}\right)~,
\ee
where we used identities of $\nabla_\alpha g_{\mu \nu}=0$, the variation of determinant $\delta\sqrt{|g|} = -\frac{1}{2}\sqrt{|g|}g_{\mu\nu}\delta g^{\mu\nu}$ and the Palatini identity
\be
\delta R_{\mu \nu }=\delta {R^{\rho }}_{\mu \alpha \nu }=\nabla _{\rho }\delta \Gamma _{\mu \nu}^{\rho }-\nabla _{\nu }\delta \Gamma _{\mu \rho }^{\rho }~,
\ee
and we defined
\be
V^\alpha \equiv g^{\mu\nu}\delta \Gamma^{\alpha}_{\mu\nu}
    -g^{\mu\alpha}\delta \Gamma_{\mu\nu}^{\nu}~.
\ee
Using the definition of stress-tensor
\be
 T_{\mu \nu }\equiv\frac {-2}{\sqrt{|g|}}\frac {\delta ({\sqrt{|g|}}{\mathcal {L}}_{\text {matter} })}{\delta g^{\mu \nu }}~,
\ee
and by rearranging equation~\reef{eq:EH var} one finds
\ba\label{eq:EH var2}
\delta S_{\text{EH}}=  \frac{1}{16\pi G}\int_\M \sqrt{|g|}
    \left(R_{\mu\nu} - \frac{1}{2}(R - 2\Lambda) g_{\mu\nu} - 8\pi G T_{\mu \nu}\right)\delta g^{\mu\nu} + \frac{1}{16\pi G}\int_\M \sqrt{|g|} \nabla_\alpha V^{\alpha}
\ea
The first term would reproduce the desired Einstein field equations. The second term is a total derivative that vanishes for a manifold without a boundary. 
In what follows, we show that to cancel this term one has to add the Gibbons-Hawking-York boundary term.

\subsection{Boundary term of Einstein-Hilbert action}
We start by rewriting the second term in~\reef{eq:EH var2} using Stock's theorem in~\reef{eq:Stokes}. More concretely, we have
\be
\int_\M \sqrt{|g|} \,\nabla_\alpha V^{\alpha} = \int_{\Sigma} \sqrt{|h|}\,  \e \,n_{\alpha} V^{\alpha}~.
\ee 
Let us rewrite this boundary term in terms of the variation of normal vector and metric. 
Using the identities
\ba
\delta\left(\nabla_\mu n_\nu \right) = \nabla_\mu \delta n_\nu -n_\lambda \delta \Gamma_{\mu\nu}^\lambda ~,\\
\delta\left(\nabla_\mu n^\mu \right) = \nabla_\mu \delta n^\mu + n^\nu\delta \Gamma_{\mu\nu}^\mu ~,
\ea
one can replace the variation of the connections in terms of the variation of the normal vector as
\ba\label{eq:n dot V}
 n_\alpha V^\alpha &= g^{\mu \nu} n_\alpha \delta\Gamma^{\alpha}_{\mu\nu} - n^\mu \delta \Gamma_{\mu\nu}^{\nu}\\
 &= g^{\mu \nu} \left[\nabla_\mu \delta n_\nu-\delta \left(\nabla_\mu n_\nu\right)\right] + \nabla_\mu \delta n^\mu - \delta\left(\nabla_\mu n^\mu\right)\\
 &= \nabla_\mu \left( g^{\mu \nu} \delta n_\nu\right) - \delta\left(g^{\mu \nu}\nabla_\mu n_\nu\right) +  \nabla_\mu n_\nu \delta g^{\mu \nu}+ \nabla_\mu \delta n^\mu - \delta\left(\nabla_\mu n^\mu\right)\\
 &= \nabla_\mu n_\nu \delta g^{\mu \nu} - 2 \delta (\nabla_\mu n^\mu) + \nabla_\mu c^\mu
\ea
where we again used the identity $\nabla_\alpha \, g_{\mu \nu}=0$ and defined $c^\mu\equiv \delta n^\mu + g^{\mu \nu} \delta n_\nu = \delta g^{\mu \nu} n_\nu + 2g^{\mu \nu} \delta n_\nu$. Next, we rewrite the last term in~\reef{eq:n dot V} to find a pure boundary term on the hypersurface $\Sigma$ itself. To do so, we introduce the induced covariant derivative on the hypersurface as $\mathcal{D}_\mu \equiv P^\alpha_\mu \nabla_\alpha  =(\delta^\alpha_\mu - \e \,n_\mu n^\alpha ) \nabla_\alpha  $. Then, it is straightforward to verify 
\ba\label{eq: nabla c}
\nabla_\mu c^\mu&= \mathcal{D}_\mu c^\mu+ \e \, n_\mu n^\alpha \nabla_\alpha c^\mu\\
&= \mathcal{D}_\mu c^\mu+  \e \,  n^\alpha \nabla_\alpha \left(n_\mu c^\mu\right) - \e \,  c^\mu n^\alpha  \nabla_\alpha n_\mu \\ 
&= \mathcal{D}_\mu c^\mu- \e \,  c^\mu n^\alpha  \nabla_\alpha n_\mu \\
&= \mathcal{D}_\mu c^\mu- \e \, n^\alpha  \nabla_\alpha n_\mu \, n_\nu \, \delta g^{\mu \nu} - 2 \e g^{\mu \nu} \delta n_\nu n^\alpha  \nabla_\alpha n_\mu~,\\
\ea
where, in the second line, we used the fact that $c^\mu$, following its definition, is orthogonal to the normal vector: $n_\mu c^\mu= n_\mu \delta n^\mu + n^\mu \delta n_{\mu} = \delta \e =0$. 

Now we show that the last term in the last line of~\reef{eq: nabla c} vanishes. 
Using the definition of the normal vector, it is easy to prove that its variation is proportional to itself. More precisely:
\be
\delta n_\alpha = -\half (n_\mu n_\nu \delta g^{\mu \nu}) n_\alpha~.
\ee
Since the normal vector is orthogonal to the acceleration vector (i.e. $n^\mu \nabla_\nu n_\mu=0$), the last term in the last line of~\reef{eq: nabla c} vanishes and we find
\be
\nabla_\mu c^\mu = \mathcal{D}_\mu c^\mu- \e \, n^\alpha  \nabla_\alpha n_\nu \, n_\mu \, \delta g^{\mu \nu}~.
\ee
Putting everything together we find
\ba
 n_\alpha V^\alpha &= \left( \nabla_\mu n_\nu - \e \, n_\mu n^\alpha  \nabla_\alpha n_\nu  \right)\delta g^{\mu \nu} - 2 \delta (\nabla_\mu n^\mu)  +\mathcal{D}_\mu c^\mu\\
 &= K_{\mu \nu} \delta g^{\mu \nu} -2 \delta K  +\mathcal{D}_\mu c^\mu~.
\ea
At the final stage of our derivation, we simply rewrite the boundary term in~\reef{eq:EH var2}, in terms of the variation of extrinsic curvature  on the boundary:
\ba\label{eq: boundary term GHY}
\int_\M \sqrt{|g|} \,\nabla_\alpha V^{\alpha} &= \int_{\Sigma} \sqrt{|h|}\, \e\, n_{\alpha} V^{\alpha}\\
&= \int_{\Sigma} \sqrt{|h|}\,\e\, \left(K_{\mu \nu} \delta g^{\mu \nu} -2 \delta K  +\mathcal{D}_\mu c^\mu\right)\\
&=-2 \e \int_{\Sigma} \delta  \left(\sqrt{|h|}K\right) + \e\int_{\Sigma} \sqrt{|h|}    \left(K_{\mu \nu}-K h_{\mu \nu}\right)  \delta h^{\mu \nu}\, 
\ea
where we used the fact that $\mathcal{D}_\mu c^\mu$ is a total derivative on the boundary and would vanish on $\Sigma$ using Stock's theorem. We also used $K_{\mu \nu} \delta g^{\mu \nu} = K_{\mu\nu} \delta h^{\mu \nu}$ easily verifiable from the definition of the extrinsic curvature and induced metric.

In summary, the variation of Einstein-Hilbert action in the presence of a boundary is given by
\ba\label{EH var end}
\delta S_{\text{EH}}&=  \frac{1}{16\pi G}\int_\M \sqrt{|g|}
    \left(R_{\mu\nu} - \frac{1}{2}(R - 2\Lambda) g_{\mu\nu} - 8\pi G T_{\mu \nu}\right)\delta g^{\mu\nu} \\
    &- \frac{\e}{8\pi G} \int_{\Sigma} \delta  \left(\sqrt{|h|}K\right)\\
    &+ \frac{\e}{16\pi G} \int_{\Sigma} \sqrt{|h|}    \left(K_{\mu \nu}-K h_{\mu \nu}\right)  \delta h^{\mu \nu}~.\\
    ~&
\ea

\subsection{Gibbons-Hawking-York term}
To derive the Einstein equations in the bulk, one must add an additional term to the Einstein–Hilbert action to cancel the boundary contribution $S_{\text{GHY}}$ in \reef{EH var end}:
\ba
S = S_{\text{EH}} + S_{\text{GHY}}~,
\ea
This boundary term was first introduced by York~\cite{York:1972sj} and Gibbons–Hawking~\cite{Gibbons:1976ue}, but it is subject to the specific imposed boundary conditions.
In what follows, we discuss three possible boundary conditions and their corresponding Gibbons–Hawking term contributions:
\begin{itemize}
\item \textbf{Dirichlet} boundary condition corresponds to freezing of the metric at the boundary. More precisely, it refers to vanishing variation of the metric  at the boundary  i.e. $\delta g_{\mu \nu}|_{\Sigma}= \delta h_{\mu \nu} =0$. As a result, the second term in~\reef{eq: boundary term GHY} vanishes resulting in the absence of the boundary energy tensor $T^{\Sigma}_{\mu \nu} =0$. Therefore, the total action is given by:
\be
S_{\text{GHY}} =\frac{\e}{8 \pi G} \int_{\Sigma} \sqrt{|h|} \,K~.
\ee

\item \textbf{Neumann} boundary condition implies existence of a matter at the boundary with a divergence-less stress tensor given by the extrinsic curvature. In this boundary condition Gibbons-Hawking-York boundary term  includes a matter field localized at the boundary:
\be\label{eq: Neumann GHY}
S_{\text{GHY}}= \frac{\e}{8 \pi G} \int_{\Sigma} \sqrt{|h|} \,K~ + \int_{\Sigma} \sqrt{|h|}\, \mathcal{L}^\Sigma_{\text{matter}}~.
\ee
This leads to two sets of field equations,  the classic Einstein equations in the bulk plus a boundary field equation by requiring the $\delta h^{\mu\nu}$ coefficient to vanish:
\be
K_{\mu\nu}- K h_{\mu\nu}=8\pi\e G \, T^{\Sigma}_{\mu \nu}~,
\ee
where we define the boundary stress tensor as
\be
 T^{\Sigma}_{\mu \nu}\equiv\frac {-2}{\sqrt{|h|}}\frac {\delta ({\sqrt{|h|}}\mathcal {L}^{\Sigma}_{\text {matter} })}{\delta h^{\mu \nu }}~.
\ee
Using Gauss–Codazzi equations, it is straightforward to show that by requiring zero flux on the boundary, the boundary stress tensor is divergence-less. 
\item Lastly, we  consider two manifolds $M_1$ and $M_2$ that are glued at the shell at $\Sigma=\partial M_1=\partial M_2$ where the action is sum of the action of each manifold with a source matter at the intersection:
\ba
S_{\text{total}} &= S_1 +S_2+ S^{\Sigma}_{\text{matter}}\\
&= S_{1,\text{EH}}+S_{1,\text{GHY}}+S_{2,\text{EH}}+S_{2,\text{GHY}}+ S^{\Sigma}_{\text{matter}}~.
\ea
Following the procedure in~\reef{eq: Neumann GHY}, the above action results in Einstein equation in the bulk plus matter field with the following stress tensor
\be
\kappa_{\mu\nu}- \kappa\, h_{\mu\nu}=8\pi\e G \, T^{\Sigma}_{\mu \nu}~,
\ee
where $\kappa_{\mu\nu}$ is defined sum of extrinsic curvatures of the manifold $M_1$ and $M_2$ at the brane: 
\be
\kappa^{\mu\nu} = K_{1}^{\mu \nu} +K_{2}^{\mu \nu}~,
\ee
which is equivalent to Israel junction conditions. 
Note that in some sources $\kappa^{\mu\nu}$ is defined as sum of two extrinsic curvature, this is a result of different conventions for normal vector and consequently extrinsic curvature. Here, we take our normal vector for both manifolds to be outward-pointing (hence $n_1=-n_2$) while some sources take the first manifold normal vector outward-pointing and second inward-pointing.  
\end{itemize}
\section{NEC for a shell}\label{sec: NEC}
The Null Energy Condition (NEC) states that for every future-pointing null vector field $k^\mu$ the contraction with the stress tensor $T^{\mu \nu}$ is non-negative:
\be\label{eq: nec}
k^\mu k^\nu T_{\mu \nu} \geq 0~.
\ee
When dealing with a perfect fluid in the bulk with proper velocity $u^\mu$, energy density $\rho$ and pressure $P$, the stress tensor is given by 
\be
T_{\mu\nu}= (\rho+P)\, u_\mu u_\nu + P \,g_{\mu \nu}~.
\ee
Therefore, the NEC in the case of a perfect fluid in the bulk would translate to
\be
k^\mu k^\nu T_{\mu \nu} =  (\rho+P) (u\cdot k)^2 \geq 0~\qquad \Longrightarrow\quad \text{NEC:}\quad\rho + P\geq 0~,
\ee
where we used the shorthand notation $u\cdot k \equiv u_\mu k^\mu$. 
Now we want to derive a similar relation for the case of a lower-dimensional perfect fluid \textit{constrained} to a codimension-one hypersurface $\Sigma$ with stress tensor
\be
T_{ij} = (\sigma+p)\, u_i u_j\,  + p\,h_{ij}~,
\ee
corresponding to Scenario II in the main text. Here, $\sigma$ is the surface density, $p$ is the lower-dimensional pressure and again we used the Latin letters for the lower-dimensional indices. 
After pulling back the stress tensor to a higher-dimensional tensor with bulk indices and using~\reef{eq:project tensor}, one finds
\ba
T^{\Sigma}_{\mu \nu} = (\sigma+p)\, u_\mu u_\nu \,  + p\,(g_{\mu \nu}-\e \,n_\mu n_\nu)~,
\ea
which compared to a bulk perfect fluid has an additional $p\,n_\mu n_\nu$ term which can be interpreted as the shear pressure. The NEC condition in~\reef{eq: nec} would translate to 
\be\label{eq: NEC shell}
k^\mu k^\nu T^\Sigma_{\mu \nu} =  (\sigma+p) (u\cdot k)^2 - \e \, p\, (n\cdot k)^2\geq 0~.
\ee

In what follows, we will examine  the implications of the equation above for the case of interest, with spacelike normal vector ($\e=1$) and timelike proper velocity ($u_\mu u^\mu \leq 0$)~.
First of all, for pressureless dust ($p=0$), similar to the bulk fluid, the NEC is equivalent to the positivity of the energy density:
\be
\text{NEC for pressureless dust iff:} \qquad \sigma \geq 0~.
\ee
This is in agreement with intuition. Dust particles constrained  to a thin shell can also be considered as dust particles moving in higher dimensional manifolds. 
Similarly, for timelike brane solution with $w=-1$:
\begin{equation}
    \text{NEC for branes iff: }\qquad \sigma = -p \geq 0~,
\end{equation}
which also aligns with intuition as stable brane configurations have a positive tension. 

Next, we move to a more generic perfect fluid with equation of state $p= w \sigma$. With a change of coordinates such that the $u^{\hat \mu}= (\mathcal{U},0,0,\cdots)$ and $n^{\hat \mu}=(0,\mathcal{N},0,\cdots)$, for some $\mathcal{U}$ and $\mathcal{V}$ so that $u$ and $n$ are unit-normalized,  the null vector can be decomposed as:
\be
k^{\hat \mu} = a_u u^{\hat \mu} + a_n n^{\hat \mu} + \sum_{i=1}^{d-1} a_i t^{\hat \mu}_i~,
\ee
with orthonormal set of spacelike vectors $t^\mu_i=(0,\cdots,1,\cdots,0)$ and for real $a$'s. 
Using this parametrization, the NEC relation in~\reef{eq: NEC shell} becomes of the form:
\be
\sigma (1+w) a_u^2 \geq \sigma w a^2_n~.
\ee
From the fact that $k$ is a null vector, we know that there exists a constraint on the set of $a$'s given by
$a_u^2= a_n^2 + \sum_{i=1}^{d-1}  a_i^2~,
$ which implies that $a_u^2\geq a_n^2\geq 0$.
So the NEC is satisfied if:~\footnote{Here we assumed that the matter equation of state parameter satisfies $w\geq -1$, i.e. it is not phantom.} 
\be
\sigma \geq 0~.
\ee
Note that so far we only showed that this is a sufficient condition. To show it is also a necessary condition, note that~\reef{eq: NEC shell} should be true for any $k^{\hat \mu}$. In particular, it should be satisfied for $k^{\hat \mu}=\half \left( u + n\right )$ which makes the condition above iff, as we have $a_i^2=0$ for this choice of the null vector.

\section{Crotches in de Sitter}
\label{app:crotches}

In this appendix, we explicitly reproduce the delta function in \eqref{eqdelt} for the wormhole geometries discussed in subsection \eqref{sec:Lwormholes}. This is achieved by analyzing the crotch singularities arising in the double cover of de Sitter endowed with the metric:
\begin{equation}
\label{metrics}
   ds^2= -f(r)\dd t^2 + \frac{\dd r^2}{f(r)}+r^2\dd\Omega^2_{d-1}~,
\end{equation}
where the conical singularity is located at the point $t=0$ and $r=r_0$. The idea of the argument is that, within an infinitesimally small neighborhood around the conical singularity $\Sigma$ (here $\Sigma$ is the sphere of radius $r_0$), one can work with a coordinate system with a flat metric. For~\eqref{metrics}, we can do this by redefining the radial coordinate $r\rightarrow \rho$, such that 
\begin{equation}
\label{flatmetric}
    ds^2= f(r) \left[
    -\dd t^2 + \dd \rho^2
    \right]+r^2 \dd \Omega_{d-1}^2~. 
\end{equation}
Here, $r = r(\rho)$ is a function of $\rho$ satisfying differential equation $r'(\rho) =  f(r)$. The next step is to repeat the analysis \eqref{spacecrtoch} for the coordinates $t$ and $\rho$. The mapping to the $uv$-plane is given by the transformation
\begin{equation}
    t + i (\rho-\rho_0) = \frac{1}{2}(u+ i v)^2~,
\end{equation}
where $r(\rho_0) =r_0$. Introducing the same $i\epsilon$ regulator as in \eqref{complexreg}, we find a complex, but everywhere smooth, metric 
\begin{equation}
    ds^2= f(r) \left[
   (u^2+v^2+i\epsilon)(\dd u^2 + \dd v^2)-2(u\dd u-v \dd v)^2
    \right]+r^2 \dd \Omega_{d-1}^2~.
\end{equation}
Again, $r$, and consequently $f(r)$, are implicitly functions of $u$ and $v$.
We are interested in the behavior of this metric in a small neighborhood $U_\e$ around the conical singularity located at $u= v=0$.  That is, we want to study this metric in the limit $\chi,\epsilon \rightarrow0$ with fixed ratio $\chi/\epsilon$. Here $\chi$ is the radial coordinated defined as
\begin{equation}
    u =  \sqrt{\chi}\cos\varphi \quad\text{and}\quad v =\sqrt{\chi}\sin\varphi~.
\end{equation}
A direct computation shows that, at a small neighborhood $U_\e$ around this point, the Einstein-Hilbert term of this metric behaves like
\begin{equation}
    \sqrt{-g}R = 2 i \left[\frac{\epsilon \chi}{(\chi^2+\epsilon^2)^{3/2}}\right]\times r_0^{d-1}\sqrt{-g_{\Omega}} \quad \text{ as $\chi,\epsilon\rightarrow 0$}~,
\end{equation}
where $g_{\Omega}$ denotes the determinant of the metric on the $(d-1)$-sphere. Note that again in $\e\to 0$ limit this only has support on $\chi\to 0$. Integrating over the small neighborhood $U_\e$ (i.e. around $\chi=0$ and $0\leq\varphi<2\pi$) then gives the desired result:
\begin{equation}
\label{eqcurvature}
   \frac{1}{16\pi G} \int_{U_\e}\dd \chi\; \dd \varphi \; \dd V_{\Omega}\;
   \sqrt{-g}R  =i \frac{r_0^{d-1} V_{\Omega_{d-1}}}{4G} = i \frac{A(\Sigma)}{4G}~,
\end{equation}
where $\Sigma$ is the area of the $(d-1)$-sphere of radius $r_0$. Note that since we are integrating only along an infinitesimally small neighborhood of the singularity, we do not have to include smooth curvature corrections inside \eqref{eqcurvature}.

\bibliographystyle{ytphys}
\bibliography{ref}

@article{Anninos:2018svg,
    author = "Anninos, Dionysios and Galante, Dami{\'a}n A. and Hofman, Diego M.",
    title = "{De Sitter horizons {\&} holographic liquids}",
    eprint = "1811.08153",
    archivePrefix = "arXiv",
    primaryClass = "hep-th",
    doi = "10.1007/JHEP07(2019)038",
    journal = "JHEP",
    volume = "07",
    pages = "038",
    year = "2019"
}

@article{Lowe:2010np,
    author = "Lowe, David A. and Roy, Shubho",
    title = "{Punctuated eternal inflation via AdS/CFT}",
    eprint = "1004.1402",
    archivePrefix = "arXiv",
    primaryClass = "hep-th",
    reportNumber = "BROWN-HET-1593",
    doi = "10.1103/PhysRevD.82.063508",
    journal = "Phys. Rev. D",
    volume = "82",
    pages = "063508",
    year = "2010"
}

@article{Freivogel:2005qh,
    author = "Freivogel, Ben and Hubeny, Veronika E. and Maloney, Alexander and Myers, Robert C. and Rangamani, Mukund and Shenker, Stephen",
    title = "{Inflation in AdS/CFT}",
    eprint = "hep-th/0510046",
    archivePrefix = "arXiv",
    reportNumber = "SLAC-PUB-11505, SU-ITP-05-27, UCB-PTH-05-30, LBNL-58913, DCPT-05-45",
    doi = "10.1088/1126-6708/2006/03/007",
    journal = "JHEP",
    volume = "03",
    pages = "007",
    year = "2006"
}

@article{Coleman:1980aw,
    author = "Coleman, Sidney R. and De Luccia, Frank",
    title = "{Gravitational Effects on and of Vacuum Decay}",
    reportNumber = "SLAC-PUB-2463",
    doi = "10.1103/PhysRevD.21.3305",
    journal = "Phys. Rev. D",
    volume = "21",
    pages = "3305",
    year = "1980"
}

@article{Irakleous:2025trr,
    author = "Irakleous, Anastasios and Rondeau, Fran{\c{c}}ois and Toumbas, Nicolaos",
    title = "{Holography for de Sitter bubble geometries}",
    eprint = "2509.19423",
    archivePrefix = "arXiv",
    primaryClass = "hep-th",
    month = "9",
    year = "2025"
}

@article{York:1972sj,
    author = "York, Jr., James W.",
    title = "{Role of conformal three geometry in the dynamics of gravitation}",
    doi = "10.1103/PhysRevLett.28.1082",
    journal = "Phys. Rev. Lett.",
    volume = "28",
    pages = "1082--1085",
    year = "1972"
}

@article{Israel:1966rt,
    author = "Israel, W.",
    title = "{Singular hypersurfaces and thin shells in general relativity}",
    doi = "10.1007/BF02710419",
    journal = "Nuovo Cim. B",
    volume = "44S10",
    pages = "1",
    year = "1966",
    note = "[Erratum: Nuovo Cim.B 48, 463 (1967)]"
}

@article{Graham:2007va,
    author = "Graham, Noah and Olum, Ken D.",
    title = "{Achronal averaged null energy condition}",
    eprint = "0705.3193",
    archivePrefix = "arXiv",
    primaryClass = "gr-qc",
    doi = "10.1103/PhysRevD.76.064001",
    journal = "Phys. Rev. D",
    volume = "76",
    pages = "064001",
    year = "2007"
}

@article{Hartman:2016lgu,
    author = "Hartman, Thomas and Kundu, Sandipan and Tajdini, Amirhossein",
    title = "{Averaged Null Energy Condition from Causality}",
    eprint = "1610.05308",
    archivePrefix = "arXiv",
    primaryClass = "hep-th",
    doi = "10.1007/JHEP07(2017)066",
    journal = "JHEP",
    volume = "07",
    pages = "066",
    year = "2017"
}

@article{Kelly:2014mra,
    author = "Kelly, William R. and Wall, Aron C.",
    title = "{Holographic proof of the averaged null energy condition}",
    eprint = "1408.3566",
    archivePrefix = "arXiv",
    primaryClass = "gr-qc",
    doi = "10.1103/PhysRevD.90.106003",
    journal = "Phys. Rev. D",
    volume = "90",
    number = "10",
    pages = "106003",
    year = "2014",
    note = "[Erratum: Phys.Rev.D 91, 069902 (2015)]"
}

@book{Bavera:2018boundary,
  author= "Bavera, Simone S.",
  title = "{The Boundary Terms of the Einstein-Hilbert Action}",
  address = "Zurich, Switzerland",
  month = "March",
  year = "2018"
}

@book{Poisson:2009pwt,
    author = "Poisson, Eric",
    title = "{A Relativist's Toolkit: The Mathematics of Black-Hole Mechanics}",
    doi = "10.1017/CBO9780511606601",
    publisher = "Cambridge University Press",
    month = "12",
    year = "2009"
}

@book{Visser:1995cc,
    author = "Visser, Matt",
    title = "{Lorentzian wormholes: From Einstein to Hawking}",
    isbn = "978-1-56396-653-8",
    year = "1995"
}

@article{Chandra:2022bqq,
    author = "Chandra, Jeevan and Collier, Scott and Hartman, Thomas and Maloney, Alexander",
    title = "{Semiclassical 3D gravity as an average of large-c CFTs}",
    eprint = "2203.06511",
    archivePrefix = "arXiv",
    primaryClass = "hep-th",
    doi = "10.1007/JHEP12(2022)069",
    journal = "JHEP",
    volume = "12",
    pages = "069",
    year = "2022"
}

@article{10.2307/97833,
 ISSN = {00804630},
 URL = {http://www.jstor.org/stable/97833},
 author = {Harish-Chandra},
 journal = {Proceedings of the Royal Society of London. Series A, Mathematical and Physical Sciences},
 number = {1018},
 pages = {372--401},
 publisher = {The Royal Society},
 title = {Infinite Irreducible Representations of the Lorentz Group},
 volume = {189},
 year = {1947}
}

@article{Sun:2021thf,
    author = "Sun, Zimo",
    title = "{A note on the representations of SO(1,d + 1)}",
    eprint = "2111.04591",
    archivePrefix = "arXiv",
    primaryClass = "hep-th",
    doi = "10.1142/S0129055X24300073",
    journal = "Rev. Math. Phys.",
    volume = "37",
    number = "01",
    pages = "2430007",
    year = "2025"
}

@article{Anous:2020nxu,
    author = "Anous, Tarek and Skulte, Jim",
    title = "{An invitation to the principal series}",
    eprint = "2007.04975",
    archivePrefix = "arXiv",
    primaryClass = "hep-th",
    doi = "10.21468/SciPostPhys.9.3.028",
    journal = "SciPost Phys.",
    volume = "9",
    number = "3",
    pages = "028",
    year = "2020"
}

@article{10.2307/1969129,
 ISSN = {0003486X},
 URL = {http://www.jstor.org/stable/1969129},
 author = {V. Bargmann},
 journal = {Annals of Mathematics},
 number = {3},
 pages = {568--640},
 publisher = {Annals of Mathematics},
 title = {Irreducible Unitary Representations of the Lorentz Group},
 volume = {48},
 year = {1947}
}

@article{Penedones:2023uqc,
    author = "Penedones, Joao and Salehi Vaziri, Kamran and Sun, Zimo",
    title = "{Hilbert space of quantum field theory in de Sitter spacetime}",
    eprint = "2301.04146",
    archivePrefix = "arXiv",
    primaryClass = "hep-th",
    doi = "10.1103/PhysRevD.111.045001",
    journal = "Phys. Rev. D",
    volume = "111",
    number = "4",
    pages = "045001",
    year = "2025"
}

@article{GelNai47,
author = " I.~M.~Gel'fand, M.~A.~Naimark.",
title="{ Unitary representations of the Lorentz group}",
journal=" Izv. Akad. Nauk SSSR Ser. Mat ",
pages=" 411--504",
volume=" 11",
year= "1947"
}

@article{ThomasSO,
 ISSN = {0003486X},
 URL = {http://www.jstor.org/stable/1968990},
 author = {L. H. Thomas},
 journal = {Annals of Mathematics},
 number = {1},
 pages = {113--126},
 publisher = {Annals of Mathematics},
 title = {On Unitary Representations of the Group of De Sitter Space},
 volume = {42},
 year = {1941}
}

@article{Chandra:2022fwi,
    author = "Chandra, Jeevan and Hartman, Thomas",
    title = "{Coarse graining pure states in AdS/CFT}",
    eprint = "2206.03414",
    archivePrefix = "arXiv",
    primaryClass = "hep-th",
    month = "6",
    year = "2022"
}

@article{DiUbaldo:2023qli,
    author = "Di Ubaldo, Gabriele and Perlmutter, Eric",
    title = "{AdS$_3$/RMT$_2$ Duality}",
    eprint = "2307.03707",
    archivePrefix = "arXiv",
    primaryClass = "hep-th",
    month = "7",
    year = "2023"
}

@article{Balasubramanian:2022gmo,
    author = "Balasubramanian, Vijay and Lawrence, Albion and Magan, Javier M. and Sasieta, Martin",
    title = "{Microscopic origin of the entropy of black holes in general relativity}",
    eprint = "2212.02447",
    archivePrefix = "arXiv",
    primaryClass = "hep-th",
    month = "12",
    year = "2022"
}

@article{Balasubramanian:2022lnw,
    author = "Balasubramanian, Vijay and Lawrence, Albion and Magan, Javier M. and Sasieta, Martin",
    title = "{Microscopic origin of the entropy of astrophysical black holes}",
    eprint = "2212.08623",
    archivePrefix = "arXiv",
    primaryClass = "hep-th",
    month = "12",
    year = "2022"
}

@article{Sasieta:2022ksu,
    author = "Sasieta, Martin",
    title = "{Wormholes from heavy operator statistics in AdS/CFT}",
    eprint = "2211.11794",
    archivePrefix = "arXiv",
    primaryClass = "hep-th",
    doi = "10.1007/JHEP03(2023)158",
    journal = "JHEP",
    volume = "03",
    pages = "158",
    year = "2023"
}

@article{Hartle:1983ai,
    author = "Hartle, J. B. and Hawking, S. W.",
    editor = "Fang, Li-Zhi and Ruffini, R.",
    title = "{Wave Function of the Universe}",
    reportNumber = "PRINT-83-0937 (CAMBRIDGE)",
    doi = "10.1103/PhysRevD.28.2960",
    journal = "Phys. Rev. D",
    volume = "28",
    pages = "2960--2975",
    year = "1983"
}

@article{Climent:2024trz,
    author = "Climent, Ana and Emparan, Roberto and Magan, Javier M. and Sasieta, Martin and Vilar L{\'o}pez, Alejandro",
    title = "{Universal construction of black hole microstates}",
    eprint = "2401.08775",
    archivePrefix = "arXiv",
    primaryClass = "hep-th",
    doi = "10.1103/PhysRevD.109.086024",
    journal = "Phys. Rev. D",
    volume = "109",
    number = "8",
    pages = "086024",
    year = "2024"
}

@article{Hsin:2020mfa,
    author = "Hsin, Po-Shen and Iliesiu, Luca V. and Yang, Zhenbin",
    title = "{A violation of global symmetries from replica wormholes and the fate of black hole remnants}",
    eprint = "2011.09444",
    archivePrefix = "arXiv",
    primaryClass = "hep-th",
    reportNumber = "CALT-TH-2020-051",
    doi = "10.1088/1361-6382/ac2134",
    journal = "Class. Quant. Grav.",
    volume = "38",
    number = "19",
    pages = "194004",
    year = "2021"
}

@article{Penington:2019kki,
    author = "Penington, Geoff and Shenker, Stephen H. and Stanford, Douglas and Yang, Zhenbin",
    title = "{Replica wormholes and the black hole interior}",
    eprint = "1911.11977",
    archivePrefix = "arXiv",
    primaryClass = "hep-th",
    doi = "10.1007/JHEP03(2022)205",
    journal = "JHEP",
    volume = "03",
    pages = "205",
    year = "2022"
}

@article{Collier:2023fwi,
    author = "Collier, Scott and Eberhardt, Lorenz and Zhang, Mengyang",
    title = "{Solving 3d Gravity with Virasoro TQFT}",
    eprint = "2304.13650",
    archivePrefix = "arXiv",
    primaryClass = "hep-th",
    month = "4",
    year = "2023"
}

@article{Wang:2025jfd,
    author = "Wang, Zhi",
    title = "{Microscopic origin of the entropy of de Sitter spacetime}",
    eprint = "2506.03058",
    archivePrefix = "arXiv",
    primaryClass = "hep-th",
    month = "6",
    year = "2025"
}

@article{Witten:2021nzp,
    author = "Witten, Edward",
    title = "{A Note On Complex Spacetime Metrics}",
    eprint = "2111.06514",
    archivePrefix = "arXiv",
    primaryClass = "hep-th",
    month = "11",
    year = "2021"
}

@article{Blommaert:2025bgd,
    author = "Blommaert, Andreas and Kudler-Flam, Jonah and Urbach, Erez Y.",
    title = "{Absolute entropy and the observer's no-boundary state}",
    eprint = "2505.14771",
    archivePrefix = "arXiv",
    primaryClass = "hep-th",
    month = "5",
    year = "2025"
}

@article{Lanczos:1924bgi,
    author = "Lanczos, Kornel",
    title = {{Fl{\"a}chenhafte Verteilung der Materie in der Einsteinschen Gravitationstheorie}},
    doi = "10.1002/andp.19243791403",
    journal = "Annalen Phys.",
    volume = "379",
    number = "14",
    pages = "518--540",
    year = "1924"
}

@article{Morvan:2022ybp,
    author = "Morvan, Edward K. and van der Schaar, Jan Pieter and Visser, Manus R.",
    title = "{On the Euclidean action of de Sitter black holes and constrained instantons}",
    eprint = "2203.06155",
    archivePrefix = "arXiv",
    primaryClass = "hep-th",
    doi = "10.21468/SciPostPhys.14.2.022",
    journal = "SciPost Phys.",
    volume = "14",
    number = "2",
    pages = "022",
    year = "2023"
}

@article{Gao:2000ga,
    author = "Gao, Sijie and Wald, Robert M.",
    title = "{Theorems on gravitational time delay and related issues}",
    eprint = "gr-qc/0007021",
    archivePrefix = "arXiv",
    doi = "10.1088/0264-9381/17/24/305",
    journal = "Class. Quant. Grav.",
    volume = "17",
    pages = "4999--5008",
    year = "2000"
}

@article{Harlow:2025pvj,
    author = "Harlow, Daniel and Usatyuk, Mykhaylo and Zhao, Ying",
    title = "{Quantum mechanics and observers for gravity in a closed universe}",
    eprint = "2501.02359",
    archivePrefix = "arXiv",
    primaryClass = "hep-th",
    reportNumber = "MIT-CTP/5824",
    month = "1",
    year = "2025"
}

@article{Abdalla:2025gzn,
    author = "Abdalla, Ahmed I. and Antonini, Stefano and Iliesiu, Luca V. and Levine, Adam",
    title = "{The gravitational path integral from an observer{\textquoteright}s point of view}",
    eprint = "2501.02632",
    archivePrefix = "arXiv",
    primaryClass = "hep-th",
    doi = "10.1007/JHEP05(2025)059",
    journal = "JHEP",
    volume = "05",
    pages = "059",
    year = "2025"
}

@article{Kontsevich:2021dmb,
    author = "Kontsevich, Maxim and Segal, Graeme",
    title = "{Wick Rotation and the Positivity of Energy in Quantum Field Theory}",
    eprint = "2105.10161",
    archivePrefix = "arXiv",
    primaryClass = "hep-th",
    doi = "10.1093/qmath/haab027",
    journal = "Quart. J. Math. Oxford Ser.",
    volume = "72",
    number = "1-2",
    pages = "673--699",
    year = "2021"
}

@article{Sorkin:1997gi,
    author = "Sorkin, Rafael D.",
    title = "{Forks in the road, on the way to quantum gravity}",
    eprint = "gr-qc/9706002",
    archivePrefix = "arXiv",
    reportNumber = "SU-GP-93-12-2, SU-GP-93-12-2",
    doi = "10.1007/BF02435709",
    journal = "Int. J. Theor. Phys.",
    volume = "36",
    pages = "2759--2781",
    year = "1997"
}

@article{Geroch:1967fs,
    author = "Geroch, Robert P.",
    title = "{Topology in general relativity}",
    doi = "10.1063/1.1705276",
    journal = "J. Math. Phys.",
    volume = "8",
    pages = "782--786",
    year = "1967"
}

@inbook{Geroch:1979uc,
    author = "Geroch, Robert P. and Horowitz, G. T.",
    title = "{Global structure of spacetimes}",
    booktitle = "{General Relativity}: {An Einstein Centenary Survey}",
    pages = "212--293",
    year = "1979"
}

@article{Hawking:1991nk,
    author = "Hawking, S. W.",
    title = "{The Chronology protection conjecture}",
    reportNumber = "DAMTP-R-91-15",
    doi = "10.1103/PhysRevD.46.603",
    journal = "Phys. Rev. D",
    volume = "46",
    pages = "603--611",
    year = "1992"
}

@article{Colin-Ellerin:2020mva,
    author = "Colin-Ellerin, Sean and Dong, Xi and Marolf, Donald and Rangamani, Mukund and Wang, Zhencheng",
    title = "{Real-time gravitational replicas: Formalism and a variational principle}",
    eprint = "2012.00828",
    archivePrefix = "arXiv",
    primaryClass = "hep-th",
    doi = "10.1007/JHEP05(2021)117",
    journal = "JHEP",
    volume = "05",
    pages = "117",
    year = "2021"
}

@article{Draper:2022ofa,
    author = "Draper, Patrick and Farkas, Szilard",
    title = "{Euclidean de Sitter black holes and microcanonical equilibrium}",
    eprint = "2203.01871",
    archivePrefix = "arXiv",
    primaryClass = "hep-th",
    doi = "10.1103/PhysRevD.105.126021",
    journal = "Phys. Rev. D",
    volume = "105",
    number = "12",
    pages = "126021",
    year = "2022"
}

@article{Draper:2022xzl,
    author = "Draper, Patrick and Farkas, Szilard",
    title = "{de Sitter black holes as constrained states in the Euclidean path integral}",
    eprint = "2203.02426",
    archivePrefix = "arXiv",
    primaryClass = "hep-th",
    doi = "10.1103/PhysRevD.105.126022",
    journal = "Phys. Rev. D",
    volume = "105",
    number = "12",
    pages = "126022",
    year = "2022"
}

@article{Arias:2019pzy,
    author = "Arias, Cesar and Diaz, Felipe and Sundell, Per",
    title = "{De Sitter Space and Entanglement}",
    eprint = "1901.04554",
    archivePrefix = "arXiv",
    primaryClass = "hep-th",
    doi = "10.1088/1361-6382/ab5b78",
    journal = "Class. Quant. Grav.",
    volume = "37",
    number = "1",
    pages = "015009",
    year = "2020"
}

@article{Geng:2024jmm,
    author = "Geng, Hao and Jiang, Yikun",
    title = "{Microscopic origin of the entropy of single-sided black holes}",
    eprint = "2409.12219",
    archivePrefix = "arXiv",
    primaryClass = "hep-th",
    doi = "10.1007/JHEP04(2025)133",
    journal = "JHEP",
    volume = "04",
    pages = "133",
    year = "2025"
}

@article{Chandrasekaran:2022cip,
    author = "Chandrasekaran, Venkatesa and Longo, Roberto and Penington, Geoff and Witten, Edward",
    title = "{An algebra of observables for de Sitter space}",
    eprint = "2206.10780",
    archivePrefix = "arXiv",
    primaryClass = "hep-th",
    doi = "10.1007/JHEP02(2023)082",
    journal = "JHEP",
    volume = "02",
    pages = "082",
    year = "2023"
}

@article{Witten:2023xze,
    author = "Witten, Edward",
    title = "{A background-independent algebra in quantum gravity}",
    eprint = "2308.03663",
    archivePrefix = "arXiv",
    primaryClass = "hep-th",
    doi = "10.1007/JHEP03(2024)077",
    journal = "JHEP",
    volume = "03",
    pages = "077",
    year = "2024"
}

@article{Balasubramanian:2025hns,
    author = "Balasubramanian, Vijay and Yildirim, Tom",
    title = "{How to Count States in Gravity}",
    eprint = "2506.15767",
    archivePrefix = "arXiv",
    primaryClass = "hep-th",
    month = "6",
    year = "2025"
}

@article{Bousso:2000xa,
    author = "Bousso, Raphael and Polchinski, Joseph",
    title = "{Quantization of four form fluxes and dynamical neutralization of the cosmological constant}",
    eprint = "hep-th/0004134",
    archivePrefix = "arXiv",
    reportNumber = "SU-ITP-00-12, NSF-ITP-00-40",
    doi = "10.1088/1126-6708/2000/06/006",
    journal = "JHEP",
    volume = "06",
    pages = "006",
    year = "2000"
}

@article{Colin-Ellerin:2021jev,
    author = "Colin-Ellerin, Sean and Dong, Xi and Marolf, Donald and Rangamani, Mukund and Wang, Zhencheng",
    title = "{Real-time gravitational replicas: low dimensional examples}",
    eprint = "2105.07002",
    archivePrefix = "arXiv",
    primaryClass = "hep-th",
    doi = "10.1007/JHEP08(2021)171",
    journal = "JHEP",
    volume = "08",
    pages = "171",
    year = "2021"
}

@article{Blommaert:2023vbz,
    author = "Blommaert, Andreas and Kruthoff, Jorrit and Yao, Shunyu",
    title = "{The power of Lorentzian wormholes}",
    eprint = "2302.01360",
    archivePrefix = "arXiv",
    primaryClass = "hep-th",
    doi = "10.1007/JHEP10(2023)005",
    journal = "JHEP",
    volume = "10",
    pages = "005",
    year = "2023"
}

@article{deBoer:2025rct,
    author = "de Boer, Jan and Kames-King, Joshua and Post, Boris",
    title = "{Surgery and statistics in 3d gravity}",
    eprint = "2506.04151",
    archivePrefix = "arXiv",
    primaryClass = "hep-th",
    month = "6",
    year = "2025"
}

@article{deBoer:2023vsm,
    author = "de Boer, Jan and Liska, Diego and Post, Boris and Sasieta, Martin",
    title = "{A principle of maximum ignorance for semiclassical gravity}",
    eprint = "2311.08132",
    archivePrefix = "arXiv",
    primaryClass = "hep-th",
    month = "11",
    year = "2023"
}

@article{Balasubramanian:2024rek,
    author = "Balasubramanian, Vijay and Craps, Ben and Hernandez, Juan and Khramtsov, Mikhail and Knysh, Maria",
    title = "{Counting microstates of out-of-equilibrium black hole fluctuations}",
    eprint = "2412.06884",
    archivePrefix = "arXiv",
    primaryClass = "hep-th",
    doi = "10.1007/JHEP06(2025)083",
    journal = "JHEP",
    volume = "06",
    pages = "083",
    year = "2025"
}

@article{Belin:2020hea,
    author = "Belin, Alexandre and de Boer, Jan",
    title = "{Random statistics of OPE coefficients and Euclidean wormholes}",
    eprint = "2006.05499",
    archivePrefix = "arXiv",
    primaryClass = "hep-th",
    reportNumber = "CERN-TH-2020-096",
    doi = "10.1088/1361-6382/ac1082",
    journal = "Class. Quant. Grav.",
    volume = "38",
    number = "16",
    pages = "164001",
    year = "2021"
}

@article{Altland:2020ccq,
    author = "Altland, Alexander and Sonner, Julian",
    title = "{Late time physics of holographic quantum chaos}",
    eprint = "2008.02271",
    archivePrefix = "arXiv",
    primaryClass = "hep-th",
    doi = "10.21468/SciPostPhys.11.2.034",
    journal = "SciPost Phys.",
    volume = "11",
    pages = "034",
    year = "2021"
}

@article{Altland:2022xqx,
    author = "Altland, Alexander and Post, Boris and Sonner, Julian and van der Heijden, Jeremy and Verlinde, Erik",
    title = "{Quantum chaos in 2D gravity}",
    eprint = "2204.07583",
    archivePrefix = "arXiv",
    primaryClass = "hep-th",
    month = "4",
    year = "2022"
}

@article{Saad:2019lba,
    author = "Saad, Phil and Shenker, Stephen H. and Stanford, Douglas",
    title = "{JT gravity as a matrix integral}",
    eprint = "1903.11115",
    archivePrefix = "arXiv",
    primaryClass = "hep-th",
    month = "3",
    year = "2019"
}

@article{Louko:1995jw,
    author = "Louko, Jorma and Sorkin, Rafael D.",
    title = "{Complex actions in two-dimensional topology change}",
    eprint = "gr-qc/9511023",
    archivePrefix = "arXiv",
    reportNumber = "SU-GP-95-5-1, WISC-MILW-95-TH-16, MDDP-PP-96-40",
    doi = "10.1088/0264-9381/14/1/018",
    journal = "Class. Quant. Grav.",
    volume = "14",
    pages = "179--204",
    year = "1997"
}

@article{Freivogel:2021ivu,
    author = "Freivogel, Ben and Nikolakopoulou, Dora and Rotundo, Antonio F.",
    title = "{Wormholes from Averaging over States}",
    eprint = "2105.12771",
    archivePrefix = "arXiv",
    primaryClass = "hep-th",
    month = "5",
    year = "2021"
}

@article{Espindola:2025wjf,
    author = "Esp{\'\i}ndola, Ricardo and Miyashita, Shoichiro",
    title = "{Flow-geometry microstates}",
    eprint = "2510.18901",
    archivePrefix = "arXiv",
    primaryClass = "hep-th",
    month = "10",
    year = "2025"
}

@article{Held:2024qcl,
    author = "Held, Jesse and Liu, Xiaoyi and Marolf, Donald and Wang, Zhencheng",
    title = "{Euclidean and complex geometries from real-time computations of gravitational R{\'e}nyi entropies}",
    eprint = "2409.17428",
    archivePrefix = "arXiv",
    primaryClass = "hep-th",
    doi = "10.1007/JHEP02(2025)136",
    journal = "JHEP",
    volume = "02",
    pages = "136",
    year = "2025"
}

@article{Papadodimas:2017qit,
    author = "Papadodimas, Kyriakos",
    title = "{A class of non-equilibrium states and the black hole interior}",
    eprint = "1708.06328",
    archivePrefix = "arXiv",
    primaryClass = "hep-th",
    reportNumber = "CERN-TH-2017-160",
    month = "8",
    year = "2017"
}

@article{Almheiri:2019qdq,
    author = "Almheiri, Ahmed and Hartman, Thomas and Maldacena, Juan and Shaghoulian, Edgar and Tajdini, Amirhossein",
    title = "{Replica Wormholes and the Entropy of Hawking Radiation}",
    eprint = "1911.12333",
    archivePrefix = "arXiv",
    primaryClass = "hep-th",
    doi = "10.1007/JHEP05(2020)013",
    journal = "JHEP",
    volume = "05",
    pages = "013",
    year = "2020"
}

@article{Gibbons:1977mu,
    author = "Gibbons, G. W. and Hawking, S. W.",
    title = "{Cosmological Event Horizons, Thermodynamics, and Particle Creation}",
    doi = "10.1103/PhysRevD.15.2738",
    journal = "Phys. Rev. D",
    volume = "15",
    pages = "2738--2751",
    year = "1977"
}

@article{Marolf:2022ybi,
    author = "Marolf, Donald",
    title = "{Gravitational thermodynamics without the conformal factor problem: partition functions and Euclidean saddles from Lorentzian path integrals}",
    eprint = "2203.07421",
    archivePrefix = "arXiv",
    primaryClass = "hep-th",
    doi = "10.1007/JHEP07(2022)108",
    journal = "JHEP",
    volume = "07",
    pages = "108",
    year = "2022"
}

@article{Gibbons:1976ue,
    author = "Gibbons, G. W. and Hawking, S. W.",
    title = "{Action Integrals and Partition Functions in Quantum Gravity}",
    reportNumber = "PRINT-76-0995 (CAMBRIDGE)",
    doi = "10.1103/PhysRevD.15.2752",
    journal = "Phys. Rev. D",
    volume = "15",
    pages = "2752--2756",
    year = "1977"
}

@article{Chen:2020tes,
    author = "Chen, Yiming and Gorbenko, Victor and Maldacena, Juan",
    title = "{Bra-ket wormholes in gravitationally prepared states}",
    eprint = "2007.16091",
    archivePrefix = "arXiv",
    primaryClass = "hep-th",
    doi = "10.1007/JHEP02(2021)009",
    journal = "JHEP",
    volume = "02",
    pages = "009",
    year = "2021"
}

@article{Fumagalli:2024msi,
    author = "Fumagalli, Alessandro and Gorbenko, Victor and Kames-King, Joshua",
    title = "{De Sitter Bra-Ket wormholes}",
    eprint = "2408.08351",
    archivePrefix = "arXiv",
    primaryClass = "hep-th",
    doi = "10.1007/JHEP05(2025)074",
    journal = "JHEP",
    volume = "05",
    pages = "074",
    year = "2025"
}

@article{Gibbons:1978ac,
    author = "Gibbons, G. W. and Hawking, S. W. and Perry, M. J.",
    title = "{Path Integrals and the Indefiniteness of the Gravitational Action}",
    reportNumber = "PRINT-78-0375 (CAMBRIDGE)",
    doi = "10.1016/0550-3213(78)90161-X",
    journal = "Nucl. Phys. B",
    volume = "138",
    pages = "141--150",
    year = "1978"
}

@article{Narovlansky:2023lfz,
    author = "Narovlansky, Vladimir and Verlinde, Herman",
    title = "{Double-scaled SYK and de Sitter holography}",
    eprint = "2310.16994",
    archivePrefix = "arXiv",
    primaryClass = "hep-th",
    doi = "10.1007/JHEP05(2025)032",
    journal = "JHEP",
    volume = "05",
    pages = "032",
    year = "2025"
}

@article{Boruch:2025ilr,
    author = "Boruch, Jan and Di Ubaldo, Gabriele and Haehl, Felix M. and Perlmutter, Eric and Rozali, Moshe",
    title = "{Modular-Invariant Random Matrix Theory and AdS3 Wormholes}",
    eprint = "2503.00101",
    archivePrefix = "arXiv",
    primaryClass = "hep-th",
    reportNumber = "RIKEN-iTHEMS-Report-25",
    doi = "10.1103/4hhn-c6mp",
    journal = "Phys. Rev. Lett.",
    volume = "135",
    number = "12",
    pages = "121602",
    year = "2025"
}

@article{Collier:2025lux,
    author = {Collier, Scott and Eberhardt, Lorenz and M{\"u}hlmann, Beatrix},
    title = "{A microscopic realization of dS$_3$}",
    eprint = "2501.01486",
    archivePrefix = "arXiv",
    primaryClass = "hep-th",
    doi = "10.21468/SciPostPhys.18.4.131",
    journal = "SciPost Phys.",
    volume = "18",
    number = "4",
    pages = "131",
    year = "2025"
}

@article{Chakravarty:2025sbg,
    author = "Chakravarty, Joydeep and Maloney, Alexander and Namjou, Keivan and Ross, Simon F.",
    title = "{The spectrum of pure dS$_{3}$ gravity in the static patch}",
    eprint = "2505.06420",
    archivePrefix = "arXiv",
    primaryClass = "hep-th",
    doi = "10.1007/JHEP10(2025)021",
    journal = "JHEP",
    volume = "10",
    pages = "021",
    year = "2025"
}
\end{document}